\documentstyle[12pt,graphicx]{article}
\begin{document}
\baselineskip 18pt
\def\today{\ifcase\month\or
 January\or February\or March\or April\or May\or June\or
 July\or August\or September\or October\or November\or December\fi
 \space\number\day, \number\year}

%
\def\thebibliography#1{\section*{References\markboth
 {References}{References}}\list
 {[\arabic{enumi}]}{\settowidth\labelwidth{[#1]}
 \leftmargin\labelwidth
 \advance\leftmargin\labelsep
 \usecounter{enumi}}
 \def\newblock{\hskip .11em plus .33em minus .07em}
 \sloppy
 \sfcode`\.=1000\relax}
\let\endthebibliography=\endlist
\def\beq{\begin{equation}}
\def\eeq{\end{equation}}
\def\beqn{\begin{eqnarray}}
\def\eeqn{\end{eqnarray}}
\def\rmuu{\gamma^{\mu}}
\def\rmud{\gamma_{\mu}}
\def\PL{{1-\gamma_5\over 2}}
\def\PR{{1+\gamma_5\over 2}}
\def\sinW2{\sin^2\theta_W}
\def\AEM{\alpha_{EM}}
\def\mul{M_{\tilde{u} L}^2}
\def\mur{M_{\tilde{u} R}^2}
\def\mdl{M_{\tilde{d} L}^2}
\def\mdr{M_{\tilde{d} R}^2}
\def\mz2{M_{z}^2}
\def\c2b{\cos 2\beta}
\def\au{A_u}
\def\ad{A_d}
\def\cob{\cot \beta}
\def\v#1{v_#1}
\def\tb{\tan\beta}
\def\epem{$e^+e^-$}
\def\KK{$K^0$-$\overline{K^0}$}
\def\wi{\omega_i}
\def\xj{\Upsilon_j}
\def\Wmu{W_\mu}
\def\Wnu{W_\nu}
\def\m#1{{\tilde m}_#1}
\def\mH{m_H}
\def\mw#1{{\tilde m}_{\omega #1}}
\def\mx#1{{\tilde m}_{\Upsilon^{0}_#1}}
\def\mc#1{{\tilde m}_{\Upsilon^{+}_#1}}
\def\mwi{{\tilde m}_{\omega i}}
\def\mxi{{\tilde m}_{\Upsilon^{0}_i}}
\def\mci{{\tilde m}_{\Upsilon^{+}_i}}
\def\mz{M_z}
\def\sw{\sin\theta_W}
\def\cw{\cos\theta_W}
\def\cb{\cos\beta}
\def\sb{\sin\beta}
\def\rwi{r_{\omega i}}
\def\rxj{r_{\Upsilon j}}
\def\rfp{r_f'}
\def\Kik{K_{ik}}
\def\Fq2{F_{2}(q^2)}
\def\f{\({\cal F}\)}
\def\d1{{\f(\tilde c;\tilde s;\tilde W)+ \f(\tilde c;\tilde \mu;\tilde W)}}
\def\tw{\tan\theta_W}
\def\sec2w{sec^2\theta_W}

\begin{titlepage}

\begin{center}
{\Large {\bf{Couplings of Vector-Spinor Representation for $\bf{SO(10)}$ Model Building}}}\\
\vskip 0.5 true cm \vspace{2cm}
\renewcommand{\thefootnote}
{\fnsymbol{footnote}}
 Pran Nath and Raza M. Syed
\vskip 0.5 true cm
\end{center}

\noindent { Department of Physics, Northeastern University,
Boston, MA 02115-5000, USA} \\
\vskip 1.0 true cm \centerline{\bf Abstract}
\medskip
Higgs multiplet in the vector-spinor representations of $SO(10)$,
i.e., the $144+\overline{144}$ multiplet can  break the $SO(10)$
gauge symmetry spontaneously  in  one step down to the Standard
Model gauge group symmetry $SU(3)_C\times SU(2)_L\times U(1)_Y$
and a recent analysis has used such vector-spinors for building  a
new class of $SO(10)$ grand unification models (hep-ph/0506312) .
Here we discuss the  techniques for the computation of  several classes of  
vector-spinor couplings using the Basic Theorem on the SO(2N) vertex
 expansion developed by the  authors.  The computations  include the  cubic couplings of the
vector-spinors with $SO(10)$ tensors, quartic self-couplings of the vector-spinors,
and couplings  of the vector-spinors with spinor representations of $SO(10)$.
 The last set  include couplings of vector-spinors with the 16-plets of quarks and lepton
 and with the 16 and $\overline{16}$ of Higgs.
 These couplings provide a crucial tool for further development of the $SO(10)$ grand unification
 using vector-spinor representations. These  include study of quark-lepton masses,
 analysis of dimension five operators including baryon and lepton number violating operators,
 and study of neutrino masses and  mixings. Illustrative examples are given for
 their computation using a sample of vector-spinor couplings.

\end{titlepage}

\tableofcontents
\section{Introduction}
 $SO(10)$ is a favored group for the unification of the electro-weak and the strong
 interactions\cite{georgi,early}. However, there is a wide array of possibilities for model building
 within the gauge group. Thus while the remarkable feature of $SO(10)$ is that it
 unifies one generation of quarks and leptons within one irreducible representation,
 i.e., the 16 plet representation, the Higgs sector of the theory is largely
 unconstrained and thus there exist a wide variety of models which differ by the
 choice of the Higgs sector of the theory.  In most models the  Higgs sector is generally
 quite elaborate involving several Higgs multiplets necessary for the breaking
  of  $SO(10)$ symmetry in steps down to the Standard Model gauge group
 $SU(3)_C\times SU(2)_L\times U(1)_Y$.  An interesting recent proposal  made by Babu,
 Gogoladze and  the  authors  is to use a
 single pair of $144+\overline{144}$ multiplet to break the $SO(10)$ gauge group in one step down
 to the Standard Model gauge symmetry\cite{bgns}. The couplings involving the $144$ and
 $\overline{144}$ are rather intricate and not easily computable.  However, significant
 progress has occurred recently in how one may compute couplings involving spinor and
 tensor representations of SO(10)\cite{ns,ns1}. An important result in such constructions is
 the so called Basic Theorem deduced in Ref.\cite{ns} using oscillator  techniques\cite{ms,wz}
   which facilitates the computations
 of vertices involving spinor and tensor $SO(10)$ representations. Thus using the
 basic theorem, couplings of the type $16\times 16\times 10$, $16\times 10\times 120$,
 $16\times 16\times\overline{126}$ and $16^{\dagger}\times 16\times 1$, $16^{\dagger}\times 16\times 1$
 were computed in Ref.\cite{ns} and further applications of the technique were made in
 Ref.\cite{Wiesenfeldt:2005zx}.
 Now the couplings of the $144$ and $\overline{144}$
 are more involved. This is so because of two factors: first we are dealing with a
 vector-spinor rather than just a spinor representation of $SO(10)$. Second the
 vector-spinor is constrained in order that it correspond to the irreducible $144$ or
 $\overline{144}$  representation of $SO(10)$. Nonetheless, we will find that
 the techniques of Ref.\cite{ns} appropriately adopted to this case will prove
 very useful in the analysis of $SO(10)$ vertices: cubic, quartic  or of
 higher order. In this paper we will limit ourselves to the analysis of cubic
 and quartic interactions where the  $144$ and $\overline{144}$ are involved.
The detailed  knowledge of the couplings of a gauge group are useful in model building\cite{recent},
and in extracting the
 implications of the models for spontaneous symmetry breaking,  neutrino oscillations\cite{Mohapatra:2005wg}
 proton decay\cite{pdecay1,pdecay2,pdecay3}, computation of the mass spectra and a variety of other applications.
This provides the motivation for a detalied analysis of the couplings discussed below.

The outline of the rest of the paper is as follows:  In Sec.2 we
give a brief summary of previous results  which are essential for
the developments of the succeeding sections. Here we discuss the
generators of $SO(10)$ in the $SU(5)\times U(1)$ basis using the
oscillator approach.  We then state the so called Basic  Theorem
that significantly facilitates the computation of couplings for
spinor and tensor representations in $SO(10)$. In Sec.3 we address
the question of how one may treat the  144 irreducible
representation through the use of a constrained vector-spinor.
This is so because, the vector-spinor in $SO(10)$ has $16\times
10=160$ components, and we need  a constraint to eliminate sixteen
components to get the irreducible $144$-plet tensor. In this
section we also decompose  the 144 in representations of
$SU(5)\times U(1)$ and define their normalizations.  An analysis
of the cubic  couplings  of $144$ and $\overline{144}$ with the
10, 120 and $\overline{126}$ tensor representations is given in
Sec.4. Here we also discuss the cubic couplings of
$\overline{144}$ with 1, 45 and 210 tensor representations. The
corresponding cubic couplings involving just the 144 plets can be
gotten by from these in straightforward fashion and are not
explicitly exhibited. In Sec.5 we discuss the self couplings  of
the vector-spinor representations. These couplings cannot be cubic
and the allowed couplings must at least be quartic or higher. We
compute the quartic couplings. These can be of several types. Thus
$144\times \overline{144}$ can couple  with $144\times
\overline{144}$ by mediation by 1, 45 and 210. Additionally, there
are couplings where $144\times 144$ and $\overline{144}\times
\overline{144}$ can couple with $144\times 144$ and
$\overline{144}\times \overline{144}$ either by mediation by 10,
120 or $126+\overline{126}$. Thus there  are  a  variety of
quartic self-couplings  involving spinors. In Sec.6 we discuss the
couplings of vector-spinors with the 16-plet of matter. Here we
consider couplings where $144\times 144$ and $\overline{144}\times
\overline{144}$
 couple with  $16\times 16$ plets of quark-lepton matter multiplets via mediation by
 10, 120 and $126+\overline{126}$.
 In Sec.7 we discuss the gauge couplings of the 144 and $\overline{144}$ with the
 singlet gauge field and with a 45 plet of gauge field  belonging to the adjoint
 representation of $SO(10)$.
  In Sec.8 we give some illustrative examples of how the vector-spinor couplings
  are to be used in model building.
 Conclusions are given in Sec.9. Some further details of the
 quartic couplings from 10-plet mediation are given in Appendix A, and similar
 details for 120-plet mediation are given in Appendix B, and from
 $126+\overline{126}$ are given in Appendix C. Field normalizations for fields
 other than the 144-plet are given in Appendix D, and in Appendix E we discuss
 how one may limit to the case of one generation of $144+\overline{144}$-plet of
 vector-spinor fields. In Appendix F we illustrate to the reader,
 by means of an example, the technique used for the actual computation of a
 vector-spinor  coupling.

\section{Preliminaries}
\newcommand{\Nu}{\mbox{\boldmath$\nu$}}
An efficient decomposition of the $SO(10)$ vertices is in the
$SU(5)\times U(1)$ basis. In Sec.(2.1) we give the basic formulae
for the decomposition of the $SO(10)$ generators in this basis and
in Sec.(2.2) we give the Basis Theorem for the computation of the
SO(10) vertices.
\subsection{SO(10) generators in $\bf{SU(5)\times U(1)}$ basis}

We begin by defining the Clifford elements, $\Gamma_{\mu}$
($\mu=1,2,...,10$) in terms of creation and destruction operators,
$b_i$ and $b_i^{\dagger}$ ($i=1,2,...,5$)\cite{ms,wz}
\begin{equation}
\Gamma_{2i}= (b_i+ b_i^{\dagger});~~~ \Gamma_{2i-1}= -i(b_i-
b_i^{\dagger})
\end{equation}
so that
\begin{equation}
\{\Gamma_{\mu},\Gamma_{\nu}\}=2\delta_{\mu\nu}.
\end{equation}
where
\begin{equation}
\{b_i,b_j^{\dagger}\}=\delta_{i}^j;~~~\{b_i,b_j\}=0;
~~~\{b_i^{\dagger},b_j^{\dagger}\}=0
\end{equation}
and that the $SU(5)$ singlet state $|0>$ satisfies $b_i|0>=0$.

 The 45 generators of $SO(10)$ in the spinor representation are
 \begin{equation}
 \Sigma_{\rho\sigma}=\frac{1}{2i}[\Gamma_\rho,\Gamma_\sigma]
 \end{equation}

In the analysis of $SO(10)$ invariant interactions one also needs
the equivalent of  charge conjugation operator given by
\begin{equation}
B=\prod_{\mu =odd}\Gamma_{\mu}= -i\prod_{k=1}^5
(b_k-b_k^{\dagger})
\end{equation}

 The
semi-spinors $\Psi_{(\pm)\acute{a}}$ ($\acute{a}=1,2,3$)
transforms as a 16($\overline{16}$)-dimensional irreducible
representation of $SO(10)$ and contains
$1+\overline{5}+10$($1+5+\overline{10}$) in its $SU(5)$
decomposition. They are given by
\begin{equation}
|\Psi_{(+)\acute{a}}>=|0>{\bf
M}_{\acute{a}}+\frac{1}{2}b_i^{\dagger}b_j^{\dagger}|0>{\bf
M}_{\acute{a}}^{ij} +\frac{1}{24}\epsilon^{ijklm}b_j^{\dagger}
b_k^{\dagger}b_l^{\dagger}b_m^{\dagger}|0>{\bf M}_{\acute{a}i}
\end{equation}
\begin{equation}
|\Psi_{(-)\acute{a}}>=b_1^{\dagger}b_2^{\dagger}b_3^{\dagger}b_4^{\dagger}b_5^{\dagger}|0>{\bf
N}_{\acute{a}}+\frac{1}{12}\epsilon^{ijklm}b_k^{\dagger}b_l^{\dagger}b_m^{\dagger}|0>{\bf
N}_{\acute{a}ij}+b_i^{\dagger}|0>{\bf N}_{\acute{a}}^i
\end{equation}

\subsection{The Basic Theorem for computation of $\bf{SO(10)}$ vertices}

We now review the recently developed technique\cite{ns}
 for the analysis of $SO(2N)$ invariant
couplings which allows a full exhibition of the $SU(N)$ invariant
 content of the spinor and tensor
representations. The technique utilizes a basis consisting of a
specific set of reducible $SU(N)$ tensors in terms of which the
$SO(2N)$ invariant couplings have a simple expansion.
To that end, we note that the natural basis for the expansion of
the $SO(2N)$ vertex is in terms of a specific set of $SU(N)$
reducible tensors, ${\Phi}_{c_k}$ and ${\Phi}_{\overline c_k}$
which we define as
$A^k\equiv{\Phi}_{c_k}\equiv{\Phi}_{2k}+i{\Phi}_{2k-1},~
A_k\equiv{\Phi}_{\overline c_k}\equiv{\Phi}_{2k}-i{\Phi}_{2k-1}$.
This is extended immediately
 to define the quantity $\Phi_{c_ic_j\bar c_k..}$
with an arbitrary number of unbarred and barred indices where each
 $c$ index can be expanded out so that
$A^iA^jA_k...={\Phi}_{c_ic_j\overline
c_k...}={\Phi}_{2ic_j\overline c_k...}+i{\Phi}_{2i-1c_j\overline
 c_k...}~$etc.. Thus, for example, the quantity  $\Phi_{c_ic_j\overline
c_k...c_N}$ is a sum of
 $2^N$ terms gotten by expanding all the c indices.
$\Phi_{c_ic_j\overline c_k...c_n}$ is completely anti-symmetric in
the interchange of its c indices whether unbarred or barred:
 ${\Phi}_{c_i\overline c_jc_k...\overline c_n}=-{\Phi}_{c_k\overline c_jc_i...\overline c_n}$.
Further, $ {\Phi}^*_{c_i\overline c_jc_k...\overline
c_n}={\Phi}_{\overline c_ic_j\overline
  c_k...c_n}$ etc.. We now make the observation\cite{ns1} that the object
$\Phi_{c_ic_j\overline c_k...c_n}$ transforms like a reducible
representation of $SU(N)$. Thus if we are able to compute the $SO(2N)$
invariant couplings in terms of these reducible tensors of $SU(N)$
then there remains only the further step of decomposing the
reducible
 tensors into their irreducible parts. These results are
codified in the so called The Basic Theorem which we discuss next.

 The vertex $\Gamma_{\mu}\Gamma_{\nu}\Gamma_
{\lambda}..\Gamma_{\sigma}$ $\Phi_{\mu\nu\lambda ..\sigma}$ where
$\Phi_{\mu\nu\lambda ..\sigma}$ is a Higgs tensor, appears often
in $SO(2N)$ invariant couplings  and can be expanded in the
following form
\begin{eqnarray}
\Gamma_{\mu}\Gamma_{\nu}\Gamma_{\lambda}..\Gamma_{\sigma}
\Phi_{\mu\nu\lambda ...\sigma}= b_i^{\dagger}
b_j^{\dagger}b_k^{\dagger}...b_n^{\dagger} \Phi_{c_ic_jc_k...c_n}
+\left(b_i b_j^{\dagger}b_k^{\dagger}...b_n^{\dagger}
\Phi_{\overline c_ic_jc_k...c_n}+~perms\right)\nonumber\\
+\left(b_i b_jb_k^{\dagger}...b_n^{\dagger} \Phi_{\overline
c_i\overline c_jc_{k}...c_n}+~perms\right)+ ...
+\left(b_ib_jb_k...b_{n-1}b_n^{\dagger}\Phi_{\overline
c_i\overline c_j
\overline c_k...\overline c_{n-1}c_n}+~perms\right)\nonumber\\
+ b_ib_jb_k...b_n \Phi_{\overline c_i\overline c_j\overline
c_k...\overline c_n}
\end{eqnarray}
 As mentioned above,
the object $\Phi_{c_ic_j\overline c_k...c_n}$ transforms like a
reducible representation of $SU(N)$ which can be further decomposed
in its irreducible parts.

\section{144 and $\bf{\overline {144}}$ as Constrained
Vector-Spinor Mutiplets}

In this section we discuss the $SU(5)$ particle content of the 144
and ${\overline {144}}$ vector-spinors and their expansion  in
terms of oscillator modes. We also normalize the fields in the decomposition
of vector-spinors. Finally, we define the notation
that is used in the paper.

\subsection{Field content in $\bf{SU(5)\times U(1)}$ decomposition}

We begin by discussing first the field content of the reducible
vector-spinor  160 and $\overline {160}$ multiplets\cite {bgns}:
\begin{equation}\label{bar160spinor}
|\Psi_{(+)\acute{a}\mu}>=|0>{\bf
P}_{\acute{a}\mu}+\frac{1}{2}b_i^{\dagger}b_j^{\dagger}|0>{\bf
P}_{\acute{a}\mu}^{ij} +\frac{1}{24}\epsilon^{ijklm}b_j^{\dagger}
b_k^{\dagger}b_l^{\dagger}b_m^{\dagger}|0>{\bf P}_{\acute{a}i\mu}
\end{equation}
\begin{equation}\label{160spinor}
|\Psi_{(-)\acute{b}\mu}>=b_1^{\dagger}b_2^{\dagger}
b_3^{\dagger}b_4^{\dagger}b_5^{\dagger}|0>{\bf Q}_{\acute{b}\mu}
+\frac{1}{12}\epsilon^{ijklm}b_k^{\dagger}b_l^{\dagger}
b_m^{\dagger}|0>{\bf Q}_{\acute{b}ij\mu}+b_i^{\dagger}|0>{\bf
Q}_{\acute{b}\mu}^i
\end{equation}
where the lower case Latin letters $i,j,k,l,m,...=1,2,...,5$ are
$SU(5)$ indices, the lower case Greek letters
$\mu,\nu,\rho,...=1,2,...,10$ represent $SO(10)$ indices, while
the lower case Latin letters with accent $\acute{a}, \acute{b},
\acute{c}, \acute{d}=1,2,3$ are generation indices. The $SU(5)$
field content of $160+\overline{160}$ multiplet is
\begin{eqnarray}\label{su5decompositionofbar160,160}
\overline{160} (\Psi_{(+)\mu})=1({\bf {\widehat P}})+\bar 5({\bf
P}_{i})+5 ({\bf P}^i)+5({\bf {\widehat P}}^i)+{\overline
{10}}({\bf P}_{ij}) +{\overline {10}}({\bf {\widehat
P}}_{ij})+{\overline {15}}({\bf P}_{ij}^{(S)})\nonumber\\
+24 ({\bf P}^i_j)+{\overline {40}}({\bf { P}}_{jkl}^i)+45 ({\bf
P}^{ij}_k)
\\
160 (\Psi_{(-)\mu})=1({\bf {\widehat Q}})+5({\bf Q}^{i})+\bar 5
({\bf Q}_i)+\bar 5({\bf {\widehat Q}}_i)+10({\bf Q}^{ij}) +10({\bf
{\widehat
Q}}^{ij})+15({\bf Q}^{ij}_{(S)})\nonumber\\
+24 ({\bf Q}^i_j)+40({\bf { Q}}^{ijk}_l)+{\overline {45}} ({\bf
Q}_{jk}^i)
\end{eqnarray}
These $SU(5)$ fields are extracted from the reducible fields
appearing in Eqs.(\ref{bar160spinor}) and (\ref{160spinor}) as
follows:
\begin{eqnarray}
100={\overline {50}}+50:~~~{\bf P}_{\mu}^{ij}=\left({\bf
P}_{c_k}^{ij},{\bf P}_{\overline c_{k}}^{ij}\right)\equiv
          \left({\bf R}^{[ij]k},{\bf R}_{k}^{[ij]}\right)\nonumber\\
{\overline {100}}={\overline {50}}+50:~~~{\bf Q}_{\mu
ij}=\left({\bf Q}_{ijc_k},{\bf Q}_{ij\overline c_{k}}\right)\equiv
          \left({\bf S}^k_{[ij]},{\bf S}_{[ij]k}\right)\nonumber\\
{\overline {50}}=25+{\overline {25}}:~~~{\bf P}_{i\mu }=\left({\bf
P}_{ic_k},{\bf P}_{i\overline c_{k}}\right)\equiv
          \left({\bf R}^{k}_i,{\bf R}_{ik}\right)\nonumber\\
50=25+25:~~~{\bf Q}_{\mu }^i=\left({\bf Q}_{\overline
c_{k}}^i,{\bf Q}_{c_k}^i\right)\equiv \left({\bf S}^{i}_k,{\bf
S}^{ik}\right)\nonumber\\
10={\overline {5}}+5:~~~{\bf P}_{\mu}=\left({\bf P}_{c_k},{\bf
P}_{\overline c_{k}}\right)\equiv
          \left({\bf P}^{k},{\bf P}_{k}\right)\nonumber\\
{ \overline {10}}={\overline {5}}+5:~~~{\bf Q}_{\mu}=\left({\bf
Q}_{c_k},{\bf Q}_{\overline c_{k}}\right)\equiv
          \left({\bf Q}^{k},{\bf Q}_{k}\right)\nonumber\\
\nonumber\\
 50=45+5:~~~{\bf R}_k^{[ij]}={\bf
P}_k^{ij}+\frac{1}{4}\left(\delta^j_k\widehat { {\bf
P}}^i-\delta^i_k\widehat {{\bf P}}^j\right)\nonumber\\
{\overline {50}}={\overline {45}}+{\overline {5}}:~~~{\bf
S}^i_{[jk]}={\bf Q}^i_{jk}+\frac{1}{4}\left({\delta^i_k\widehat
{\bf
Q}}_j-{\delta^i_{j}\widehat  {\bf Q}}_k\right)\nonumber\\
{\overline {50}}={\overline {40}}+{\overline {10}}:~~~{\bf
R}^{[ij]k}=\epsilon^{ijlmn}{\bf
P}_{lmn}^{k}+\epsilon^{ijklm}\widehat {
{\bf P}}_{lm}\nonumber\\
50=40+10:~~~{\bf S}_{[ij]k}=\epsilon_{ijlmn}{\bf
Q}^{lmn}_{k}+\epsilon_{ijklm}\widehat { {\bf
Q}}^{lm}\nonumber\\
25=24+1:~~~{\bf R}^i_j={\bf P}^i_j+\frac{1}{5}\delta^i_j{\widehat
{\bf P}}~~~~~~~~~~~~~~~~~~\nonumber\\
25=24+1:~~~{\bf S}^i_j={\bf Q}^i_j+\frac{1}{5}\delta^i_j{\widehat
{\bf Q}}~~~~~~~~~~~~~~~~~~\nonumber\\
{\overline {25}}={\overline {10}}+{\overline {15}}:~~~{\bf
R}_{ij}=\frac{1}{2}\left({\bf P}_{ij}+{\bf
P}_{ij}^{(S)}\right)~~~~~~~~~~~~\nonumber\\
25=10+15:~~~{\bf S}^{ij}=\frac{1}{2}\left({\bf Q}^{ij}+{\bf
Q}^{ij}_{(S)}\right)~~~~~~~~~~~~
\end{eqnarray}

\subsection{Oscillator mode expansion of irreducible $\bf{144}$ and
$\bf{\overline{144}}$ multiplets }

 The vector-spinor
$|\Psi_{(+)\mu}>$ is unconstrained, has $160$ components and is
reducible. To see how the $160$ plet can  be reduced, we note that
$\Gamma_{\mu} |\Psi_{(+)\mu}>$ is a 16 dimensional $SO(10)$
spinor. Thus one way to define an irreducible $144$
($\overline{144}$) dimensional vector-spinor is to impose the
constraint \beqn \Gamma_{\mu} |{\Upsilon}_{(\pm)\mu}>=0
\label{144constraint} \eeqn We explore now the implication of the
above constraint. The contraction of
 $\Gamma_{\mu}$ with the 160+$\overline {160}$ multiplet $|\Psi_{(\pm)\mu}>$ gives
\begin{eqnarray}
\Gamma_{\mu}|\Psi_{(+)\mu}>=b_1^{\dagger}b_2^{\dagger}
b_3^{\dagger}b_4^{\dagger}b_5^{\dagger}|0>{\widehat {\bf P}}
+\frac{1}{12}\epsilon^{ijklm}b_k^{\dagger}b_l^{\dagger}
b_m^{\dagger}|0>\left({\bf P}_{ij}+6{\widehat {\bf
P}}_{ij}\right)\nonumber\\
+b_i^{\dagger}|0>\left({\bf P}^{i}+{\widehat {\bf P}}^{i}\right)
\nonumber\\
 \Gamma_{\mu}|\Psi_{(-)\mu}>=|0>{\widehat {\bf
P}}+\frac{1}{2}b_i^{\dagger}b_j^{\dagger}|0>\left({\bf
Q}^{ij}+6{\widehat {\bf Q}}^{ij}\right)\nonumber\\
+\frac{1}{24}\epsilon^{ijklm}b_j^{\dagger}
b_k^{\dagger}b_l^{\dagger}b_m^{\dagger}|0>\left({\bf
Q}_{i}+{\widehat {\bf Q}}_{i}\right)
\end{eqnarray}
Thus to get the $144$ and $\overline {144}$ spinor,
$|\Upsilon_{(\pm)\mu}>$, we need to impose the following
conditions:
\begin{eqnarray}\label{constraints}
{\widehat {\bf P}}=0,~~~{\widehat {\bf P}}^i=-{\bf
P}^i,~~~{\widehat
{\bf P}}_{ij}=-\frac{1}{6}{\bf P}_{ij}\nonumber\\
{\widehat {\bf Q}}=0,~~~{\widehat {\bf Q}}_i=-{\bf
Q}_i,~~~{\widehat {\bf Q}}^{ij}=-\frac{1}{6}{\bf Q}^{ij}
\end{eqnarray}
Hence, we have following relation
 \begin{equation}
 |{\Upsilon}_{(\pm)\mu}>=\left(|\Psi_{(\pm)\mu}>\right)_{constraint~of~Eq.(\ref{constraints})}
\end{equation}
The above implies that certain components of the 160 and
$\overline{160}$ multiplets are either zero or are related thus
reducing the number of independent components from 160 to 144. For
completeness, we  give the expansion of the constrained $144$ and
$\overline{144}$ vector-spinors in its oscillator modes
\begin{eqnarray}
{{\overline{144}}\choose 144}:~~~
|\Upsilon_{(\pm)\mu}>=\left(|\Upsilon_{(\pm)c_n}>,|\Upsilon_{(\pm)\bar
c_n}>\right)\nonumber\\
|\Upsilon_{(+)c_n}>=|0>{\bf
P}^n+\frac{1}{2}b_i^{\dagger}b_j^{\dagger}|0>\left[\epsilon^{ijklm}{\bf
P}^n_{klm}-\frac{1}{6}\epsilon^{ijnlm}{\bf P}_{lm}\right]\nonumber\\
+\frac{1}{24}\epsilon^{ijklm}b_j^{\dagger}
b_k^{\dagger}b_l^{\dagger}b_m^{\dagger}|0>{\bf P}_{i}^n\nonumber\\
|\Upsilon_{(+)\bar c_n}>=|0>{\bf
P}_n+\frac{1}{2}b_i^{\dagger}b_j^{\dagger}|0>\left[{\bf
P}^{ij}_{n}+\frac{1}{4}\left(\delta^i_n{\bf P}^j-\delta^j_n{\bf P}^i\right)\right]\nonumber\\
+\frac{1}{24}\epsilon^{ijklm}b_j^{\dagger}
b_k^{\dagger}b_l^{\dagger}b_m^{\dagger}|0>\left[\frac{1}{2}{\bf P}_{in}+\frac{1}{2}{\bf P}_{in}^{(S)}\right]\nonumber\\
|\Upsilon_{(-)c_n}> =b_1^{\dagger}b_2^{\dagger}
b_3^{\dagger}b_4^{\dagger}b_5^{\dagger}|0>{\bf Q}^n
+\frac{1}{12}\epsilon^{ijklm}b_k^{\dagger}b_l^{\dagger}
b_m^{\dagger}|0>\left[{\bf
Q}_{ij}^{n}+\frac{1}{4}\left(\delta^n_i{\bf Q}_j-\delta^n_j{\bf Q}_i\right)\right]\nonumber\\
+b_i^{\dagger}|0>\left[\frac{1}{2}{\bf Q}^{in}+\frac{1}{2}{\bf
Q}^{in}_{(S)}\right]\nonumber\\
|\Upsilon_{(-)\bar c_n}> =b_1^{\dagger}b_2^{\dagger}
b_3^{\dagger}b_4^{\dagger}b_5^{\dagger}|0>{\bf Q}_n
+\frac{1}{12}\epsilon^{ijklm}b_k^{\dagger}b_l^{\dagger}
b_m^{\dagger}|0>\left[\epsilon_{ijklm}{\bf
Q}_n^{klm}-\frac{1}{6}\epsilon_{ijnlm}{\bf Q}^{lm}\right]
\nonumber\\
+b_i^{\dagger}|0>{\bf Q}_n^i~~~
\end{eqnarray}

\subsection{Normalization conditions and notation}

To normalize the $SU(5)$ fields contained in the tensor,
$|\Upsilon_{(\pm)\mu}>$, we carry out a field redefinition
\begin{eqnarray}
\{{\overline 5}\}:~~{\bf P}_i={\cal P}_i,~~~~\{5\}:~~{\bf
P}^i=\frac{2}{\sqrt 5}{\cal P}^i,~~~~\{\overline {10}\}:~~{\bf
P}_{ij}=\sqrt{\frac{6}{5}}{\cal P}_{ij}\nonumber\\
\{\overline {15}\}:~~{\bf P}^{(S)}_{ij}=\sqrt{2}{\cal
P}^{(S)}_{ij},~~~~\{24\}:~~{\bf P}^i_j={\cal P}^i_j,~~~~
\{\overline {40}\}:~~{\bf P}^{l}_{ijk}=\frac{1}{6}{\cal
P}^{l}_{ijk}\nonumber\\
\{45\}:~~{\bf P}^{ij}_{k}={\cal
P}^{ij}_{k}~~~~~~~~~~~~~~~~~~~~~~~~~~~~~~~~~~
\end{eqnarray}
\begin{eqnarray}
\{5\}:~~{\bf Q}^i={\cal Q}^i,~~~~\{{\overline 5}\}:~~{\bf
Q}_i=\frac{2}{\sqrt 5}{\cal Q}_i,~~~~\{10\}:~~{\bf
Q}^{ij}=\sqrt{\frac{6}{5}}{\cal Q}^{ij}\nonumber\\
\{15\}:~~{\bf Q}_{(S)}^{ij}=\sqrt{2}{\cal
Q}_{(S)}^{ij},~~~~\{24\}:~~{\bf Q}^i_j={\cal Q}^i_j,~~~~
\{40\}:~~{\bf Q}^{ijk}_{l}=\frac{1}{6}{\cal
Q}^{ijk}_{l}\nonumber\\
\{\overline {45}\}:~~{\bf Q}^{k}_{ij}={\cal
Q}^{k}_{ij}~~~~~~~~~~~~~~~~~~~~~~~~~~~~~~~~
\end{eqnarray}
In terms of the normalized fields, the kinetic energy of the $144$
and $\overline {144}$:\\
$-<\partial_A\Upsilon_{(\pm)\mu}|\partial^A\Upsilon_{(\pm)\mu}>$
takes the form
\begin{eqnarray}
{\mathsf L}^{144}_{kin}=-\partial_A{\cal
P}^{i\dagger}\partial^A{\cal P}_i-\partial_A{\cal
P}_i^{\dagger}\partial^A{\cal P}_i-\frac{1}{2!}\partial_A{\cal
P}^{ij\dagger}\partial^A{\cal
P}_{ij}\nonumber\\
-\frac{1}{2!}\partial_A{\cal P}_{(S)}^{ij\dagger}\partial^A{\cal
P}^{(S)}_{ij}-\partial_A{\cal P}^{i\dagger}_j\partial^A{\cal
P}_{i}^j-\frac{1}{3!}\partial_A{\cal
P}^{ijk\dagger}_l\partial^A{\cal
P}_{ijk}^l\nonumber\\
-\frac{1}{2!}\partial_A{\cal P}^{k\dagger}_{ij}\partial^A{\cal
P}^{ij}_k
\end{eqnarray}
\begin{eqnarray}
{\mathsf L}^{\overline{144}}_{kin}=-\partial_A{\cal
Q}_i^{\dagger}\partial^A{\cal Q}^i-\partial_A{\cal
Q}^{i\dagger}\partial^A{\cal Q}_i-\frac{1}{2!}\partial_A{\cal
Q}^{\dagger}_{ij}\partial^A{\cal
Q}^{ij}\nonumber\\
-\frac{1}{2!}\partial_A{\cal Q}_{ij}^{(S)\dagger}\partial^A{\cal
Q}^{ij}_{(S)}-\partial_A{\cal Q}^{i\dagger}_j\partial^A{\cal
Q}_{i}^j-\frac{1}{3!}\partial_A{\cal
Q}^{l\dagger}_{ijk}\partial^A{\cal
Q}_{l}^{ijk}\nonumber\\
-\frac{1}{2!}\partial_A{\cal Q}^{ij\dagger}_{k}\partial^A{\cal
Q}^{k}_{ij}
\end{eqnarray}
where $A=0,1,2,3$ represents the Lorentz index.\\

 For ease of reference we give below the notations that will be used in
 much of the paper.
\begin{enumerate}
\item The set of indices ($\cal U$,$\cal U'$)...($\cal Z$,$\cal
Z'$) run over several Higgs representations of the same kind
 \item
$\cal M^{(.)}$ represents mass matrices \item $h^{(.)}$, ${\bar
h}^{(.)}$, $f^{(.)}$, ${\bar f}^{(.)}$, $g^{(.)}$, ${\bar
g}^{(.)}$; $k^{(.)}$, ${\bar k}^{(.)}$, $l^{(.)}$, ${\bar
l}^{(.)}$ are constants

\item An antisymmetric product of four $\Gamma $'s for example, is
represented by
\begin{equation}
\Gamma_{[\mu}\Gamma_{\nu}\Gamma_{\rho}
\Gamma_{\lambda]}=\frac{1}{4!}\sum_P(-1)^{\delta_P}
\Gamma_{\mu_{P(1)}}\Gamma_{\nu_{P(2)}}\Gamma_{\rho_{P(3)}}
\Gamma_{\lambda_{P(4)}}\nonumber
\end{equation}
with $\sum_P$ denoting the sum over all permutations and
$\delta_P$ takes on the value $0$ and $1$ for even and odd
permutations respectively.
\end{enumerate}

\section{Higgs Sector Cubic Couplings}
In this section we compute the cubic couplings in the superpotential
 involving two  vector-spinors
and one each of the tensors 1, 10, 45, 120, 210, and ${\overline
{126}}$ plet of Higgs. We will discuss their $SU(5)\times U(1)$
decomposed form below.

\subsection{The $\bf{(144\times \overline{144}\times 1)}$ couplings}

The $(144\times \overline{144}\times 1)$ coupling structure in the superpotential is

\begin{equation}
{\mathsf
W}^{(1)}=h^{^{(1)}}_{\acute{a}\acute{b}}<\Upsilon^{*}_{(-)\acute{a}\mu}|B|\Upsilon_{(+)\acute{b}\mu}>\Phi
\end{equation}
Where $\Phi$ is the 1-plet of Higgs field. The coupling structure in
the $SU(5)\times U(1)$ decomposed form is
\begin{eqnarray}
{\mathsf W}^{(1)}=ih^{^{(1)}}_{\acute{a}\acute{b}}
\left[\frac{3}{5}{\cal Q}_{\acute{a}i}^{\bf T}{\cal
P}^i_{\acute{b}}+{\cal Q}_{\acute{a}}^{i\bf T}{\cal
P}_{\acute{b}i}+\frac{1}{10}{\cal Q}_{\acute{a}}^{ij\bf T}{\cal
P}_{\acute{b}ij}+\frac{1}{2}{\cal Q}_{(S)\acute{a}}^{ij\bf
T}{\cal P}^{(S)}_{\acute{b}ij}\right.\nonumber\\
\left.+{\cal Q}_{\acute{a}j}^{i\bf T}{\cal
P}_{\acute{b}i}^j-\frac{1}{6}{\cal Q}_{\acute{a}l}^{ijk\bf T}{\cal
P}_{\acute{b}ijk}^l-\frac{1}{2}{\cal Q}_{\acute{a}ij}^{k\bf
T}{\cal P}_{\acute{b}k}^{ij}\right]{\mathsf H}
\end{eqnarray}

\subsection{ The $\bf{(144\times \overline{144}\times 45)}$ couplings}

The $(144\times \overline{144}\times 45)$ couplings in the
superpotential is

\begin{equation}
{\mathsf
W}^{(45)}=\frac{1}{2!}h^{^{(45)}}_{\acute{a}\acute{b}}<\Upsilon^{*}_{(-)\acute{a}\mu}|B\Sigma_{\rho\sigma}|\Upsilon_{(+)\acute{b}
\mu}>\Phi_{\rho\sigma}
\end{equation}
where $\Phi_{\rho\sigma}$ represents the 45-plet of Higgs field.
The couplings in their $SU(5)\times U(1)$ decomposed form are given by

\begin{eqnarray}
{\mathsf
W}^{(45)}=h^{^{(45)}}_{\acute{a}\acute{b}}\left\{\left[\frac{3}{\sqrt
{10}}{\cal Q}_{\acute{a}j}^{i\bf T}{\cal
P}_{\acute{b}i}^j+\frac{11}{10\sqrt {10}}{\cal
Q}_{\acute{a}}^{ij\bf T}{\cal P}_{\acute{b}ij}+\frac{3}{\sqrt
{10}}{\cal Q}_{(S)\acute{a}}^{ij\bf T}{\cal
P}^{(S)}_{\acute{b}ij}+\frac{1}{2\sqrt {10}}{\cal
Q}_{\acute{a}ij}^{k\bf T}{\cal
P}_{\acute{b}k}^{ij}\right.\right.\nonumber\\
\left.\left.-\frac{19}{5\sqrt {10}}{\cal Q}_{\acute{a}i}^{\bf
T}{\cal P}^i_{\acute{b}}-\sqrt{\frac{5}{2}}{\cal
Q}_{\acute{a}}^{i\bf T}{\cal
P}_{\acute{b}i}+\frac{1}{6\sqrt{10}}{\cal Q}_{\acute{a}l}^{ijk\bf
T}{\cal
P}_{\acute{b}ijk}^l\right]{\mathsf H}\right.\nonumber\\
\left.+\left[-\frac{1}{\sqrt 2}{\cal Q}_{\acute{a}}^{k\bf T}{\cal
P}_{\acute{b}k}^{lm}-\frac{1}{\sqrt {10}}{\cal
Q}_{\acute{a}}^{l\bf T}{\cal P}^m_{\acute{b}} + \frac{2}{\sqrt
{15}}{\cal Q}_{\acute{a}}^{lk\bf T}{\cal
P}_{\acute{b}k}^m+\frac{1}{\sqrt 2}{\cal Q}_{\acute{a}n}^{klm\bf
T}{\cal
P}_{\acute{b}k}^n\right.\right.\nonumber\\
\left.\left.+\frac{7}{20\sqrt {3}}\epsilon^{ijklm}{\cal
Q}_{\acute{a}i}^{\bf T}{\cal P}_{\acute{b}jk} -\frac{1}{3\sqrt
{10}}\epsilon^{ijklm}{\cal Q}_{\acute{a}n}^{\bf T}{\cal
P}_{\acute{b}ijk}^{n} -
\frac{1}{4}\sqrt{\frac{3}{5}}\epsilon^{ijklm}{\cal
Q}_{\acute{a}ij}^{n\bf T}{\cal P}_{\acute{b}nk}\right.\right.\nonumber\\
\left.\left.+\frac{1}{4}\epsilon^{ijklm}{\cal
Q}_{\acute{a}ij}^{n\bf T}{\cal
P}^{(S)}_{\acute{b}nk}\right]{\mathsf H}_{lm}\right.\nonumber\\
\left.+\left[-\frac{1}{\sqrt 2}{\cal Q}_{\acute{a}lm}^{k\bf
T}{\cal P}_{\acute{b}k}+\frac{1}{\sqrt {10}}{\cal
Q}_{\acute{a}l}^{\bf T}{\cal P}_{\acute{b}m} + \frac{2}{\sqrt
{15}}{\cal Q}_{\acute{a}l}^{k\bf T}{\cal
P}_{\acute{b}km}+\frac{1}{\sqrt 2}{\cal Q}_{\acute{a}n}^{k\bf
T}{\cal
P}_{\acute{b}klm}^n\right.\right.\nonumber\\
\left.\left.+\frac{7}{20\sqrt {3}}\epsilon_{ijklm}{\cal
Q}_{\acute{a}}^{ij\bf T}{\cal P}^{k}_{\acute{b}} -\frac{1}{3\sqrt
{10}}\epsilon_{ijklm}{\cal Q}_{\acute{a}n}^{ijk\bf T}{\cal
P}_{\acute{b}}^{n} +\frac{1}{4}\sqrt{\frac{3}{5}}
\epsilon_{ijklm}{\cal
Q}_{\acute{a}}^{in\bf T}{\cal P}_{\acute{b}n}^{jk}\right.\right.\nonumber\\
\left.\left.+\frac{1}{4}\epsilon_{ijklm}{\cal
Q}_{(S)\acute{a}}^{in\bf T}{\cal
P}^{jk}_{\acute{b}n}\right]{\mathsf H}^{lm}\right.\nonumber\\
\left.+\left[\sqrt 2{\cal Q}_{\acute{a}ik}^{l\bf T}{\cal
P}_{\acute{b}l}^{kj} -\frac{1}{\sqrt {10}}{\cal
Q}_{\acute{a}ik}^{j\bf T}{\cal P}_{\acute{b}}^k+\frac{1}{\sqrt
{10}}{\cal Q}_{\acute{a}k}^{\bf T}{\cal
P}_{\acute{b}i}^{kj}-\frac{3}{10\sqrt
{2}}{\cal Q}_{ai}^{\bf T}{\cal P}_{\acute{b}}^j\right.\right.\nonumber\\
\left.\left.+\frac{1}{\sqrt {2}}{\cal Q}_{\acute{a}m}^{jkl\bf
T}{\cal P}_{\acute{b}kli}^m -\frac{1}{\sqrt {15}}{\cal
Q}_{\acute{a}i}^{jkl\bf T}{\cal P}_{\acute{b}kl} -\frac{1}{\sqrt
{15}}{\cal Q}_{\acute{a}}^{kl\bf T}{\cal P}_{\acute{b}ikl}^j
+\frac{1}{15\sqrt {2}}{\cal Q}_{\acute{a}}^{jk\bf T}{\cal
P}_{\acute{b}ki}\right.\right.\nonumber\\
\left.\left.-\sqrt{\frac{3}{10}}{\cal Q}_{\acute{a}}^{jk\bf
T}{\cal P}_{\acute{b}ki}^{(S)} +\sqrt{\frac{3}{10}}{\cal
Q}_{(S)\acute{a}}^{jk\bf T}{\cal P}_{\acute{b}ki}-\frac{1}{\sqrt 2
}{\cal Q}_{(S)\acute{a}}^{jk\bf T}{\cal
P}_{\acute{b}ki}^{(S)}-\sqrt{2}{\cal Q}_{\acute{a}k}^{j\bf T}{\cal
P}_{\acute{b}i}^{k}\right]{\mathsf H}^{i}_j\right\}\nonumber\\
\end{eqnarray}

\subsection{The $\bf{(144\times \overline{144}\times 210)}$ couplings}

The $(144\times \overline{144}\times 210)$ coupling structure is

\begin{equation}
{\mathsf
W}^{(210)}=\frac{1}{4!}h^{^{(210)}}_{\acute{a}\acute{b}}<\Upsilon^{*}_{(-)\acute{a}\mu}|B\Gamma_{[\nu}\Gamma_{\rho}\Gamma_{\sigma}
\Gamma_{\lambda]}|\Upsilon_{(+)\acute{b}\mu}>\Phi_{\nu\rho\sigma\lambda}
\end{equation}
where $\Phi_{\nu\rho\sigma\lambda}$ represents the 210-plet of
Higgs field.  The superpotential in the $SU(5)\times U(1)$
decomposed form is

\begin{eqnarray}
{\mathsf
W}^{(210)}=ih^{^{(210)}}_{\acute{a}\acute{b}}\left\{\left[\frac{1}{2\sqrt
{15}}{\cal Q}_{\acute{a}j}^{i\bf T}{\cal
P}_{\acute{b}i}^j+\frac{1}{4\sqrt {15}}{\cal Q}_{\acute{a}}^{ij\bf
T}{\cal P}_{\acute{b}ij}+\frac{1}{\sqrt {15}}{\cal
Q}_{(S)\acute{a}}^{ij\bf T}{\cal
P}^{(S)}_{\acute{b}ij}+\frac{1}{4\sqrt {15}}{\cal
Q}_{\acute{a}ij}^{k\bf T}{\cal
P}_{\acute{b}k}^{ij}\right.\right.\nonumber\\
\left.\left.+\frac{7}{10}\sqrt{\frac{3}{5}}{\cal
Q}_{\acute{a}i}^{\bf T}{\cal
P}^i_{\acute{b}}+\frac{1}{2}\sqrt{\frac{5}{3}}{\cal
Q}_{\acute{a}}^{i\bf T}{\cal
P}_{\acute{b}i}+\frac{1}{12\sqrt{15}}{\cal Q}_{\acute{a}l}^{ijk\bf
T}{\cal
P}_{\acute{b}ijk}^l\right]{\mathsf H}\right.\nonumber\\
\left.+\left[-\frac{1}{2\sqrt 2}{\cal Q}_{\acute{a}}^{k\bf T}{\cal
P}_{\acute{b}k}^{lm}-\frac{1}{2\sqrt {10}}{\cal
Q}_{\acute{a}}^{l\bf T}{\cal P}^m_{\acute{b}} + \frac{1}{3\sqrt
{15}}{\cal Q}_{\acute{a}}^{lk\bf T}{\cal
P}_{\acute{b}k}^m-\frac{1}{6\sqrt 2}{\cal Q}_{\acute{a}n}^{klm\bf
T}{\cal
P}_{\acute{b}k}^n\right.\right.\nonumber\\
\left.\left.+\frac{1}{4}\sqrt{\frac{3}{10}}\epsilon^{ijklm}{\cal
Q}_{\acute{a}i}^{\bf T}{\cal P}_{\acute{b}jk} -\frac{1}{6\sqrt
{10}}\epsilon^{ijklm}{\cal Q}_{\acute{a}n}^{\bf T}{\cal
P}_{\acute{b}ijk}^{n}+ \frac{1}{8\sqrt{15}}\epsilon^{ijklm}{\cal
Q}_{\acute{a}ij}^{n\bf T}{\cal P}_{\acute{b}nk}\right.\right.\nonumber\\
\left.\left.-\frac{1}{24}\epsilon^{ijklm}{\cal
Q}_{\acute{a}ij}^{n\bf T}{\cal
P}^{(S)}_{\acute{b}nk}\right]{\mathsf H}_{lm}\right.\nonumber\\
\left.+\left[-\frac{1}{2\sqrt 2}{\cal Q}_{\acute{a}lm}^{k\bf
T}{\cal P}_{\acute{b}k}-\frac{1}{2\sqrt {10}}{\cal
Q}_{\acute{a}l}^{\bf T}{\cal P}_{\acute{b}m}+ \frac{1}{3\sqrt
{15}}{\cal Q}_{\acute{a}l}^{k\bf T}{\cal
P}_{\acute{b}km}+\frac{1}{6\sqrt 2}{\cal Q}_{\acute{a}n}^{k\bf
T}{\cal
P}_{\acute{b}klm}^n\right.\right.\nonumber\\
\left.\left.-\frac{1}{4}\sqrt{\frac{3}{10}}\epsilon_{ijklm}{\cal
Q}_{\acute{a}}^{ij\bf T}{\cal P}^{k}_{\acute{b}} +\frac{1}{6\sqrt
{10}}\epsilon_{ijklm}{\cal Q}_{\acute{a}n}^{ijk\bf T}{\cal
P}_{\acute{b}}^{n} +\frac{1}{8\sqrt{15}} \epsilon_{ijklm}{\cal
Q}_{\acute{a}}^{in\bf T}{\cal P}_{\acute{b}n}^{jk}\right.\right.\nonumber\\
\left.\left.+\frac{1}{24}\epsilon_{ijklm}{\cal
Q}_{(S)\acute{a}}^{in\bf T}{\cal
P}^{jk}_{\acute{b}n}\right]{\mathsf H}^{lm}\right.\nonumber\\
\left.+\left[-\frac{1}{3\sqrt 2}{\cal Q}_{\acute{a}ik}^{l\bf
T}{\cal P}_{\acute{b}l}^{kj} +\frac{1}{6\sqrt {10}}{\cal
Q}_{\acute{a}ik}^{j\bf T}{\cal P}_{\acute{b}}^k-\frac{1}{6\sqrt
{10}}{\cal Q}_{\acute{a}k}^{\bf T}{\cal
P}_{\acute{b}i}^{kj}+\frac{1}{20\sqrt{2}}{\cal Q}_{\acute{a}i}^{\bf T}{\cal P}_{\acute{b}}^j\right.\right.\nonumber\\
\left.\left.-\frac{1}{6\sqrt {2}}{\cal Q}_{\acute{a}m}^{jkl\bf
T}{\cal P}_{\acute{b}ikl}^m +\frac{1}{6\sqrt {15}}{\cal
Q}_{\acute{a}i}^{jkl\bf T}{\cal P}_{\acute{b}}^{kl}
+\frac{1}{6\sqrt {15}}{\cal Q}_{\acute{a}}^{kl\bf T}{\cal
P}_{\acute{b}ikl}^j +\frac{5\sqrt {2}}{9}{\cal
Q}_{\acute{a}}^{jk\bf T}{\cal
P}_{\acute{b}ki}\right.\right.\nonumber\\
\left.\left.-\frac{1}{2}\sqrt{\frac{3}{10}}{\cal
Q}_{\acute{a}}^{jk\bf T}{\cal
P}_{\acute{b}ki}^{(S)}+\frac{1}{2}\sqrt{\frac{3}{10}}{\cal
Q}_{(S)\acute{a}}^{jk\bf T}{\cal P}_{\acute{b}ki}-\frac{1}{2\sqrt
2}{\cal Q}_{(S)\acute{a}}^{jk\bf T}{\cal
P}_{\acute{b}ki}^{(S)}-\frac{1}{\sqrt{2}}{\cal
Q}_{\acute{a}k}^{j\bf T}{\cal P}_{\acute{b}i}^{k}\right]{\mathsf H}^{i}_j\right.\nonumber\\
\left.+\left[-\frac{1}{\sqrt{5}}{\cal Q}_{\acute{a}}^{j\bf T}{\cal
P}_{\acute{b}ji} +\frac{1}{\sqrt{3}}{\cal Q}_{\acute{a}}^{j\bf
T}{\cal P}_{\acute{b}ji}^{(S)} +2\sqrt{\frac{2}{15}}{\cal
Q}_{\acute{a}j}^{\bf T}{\cal P}_{\acute{b}i}^j\right]{\mathsf
H}^i\right.\nonumber\\
\left.+\left[\frac{1}{\sqrt{5}}{\cal Q}_{\acute{a}}^{ij\bf T}{\cal
P}_{\acute{b}j} +\frac{1}{\sqrt{3}}{\cal Q}_{(S)\acute{a}}^{ij\bf
T}{\cal P}_{\acute{b}j} +2\sqrt{\frac{2}{15}}{\cal
Q}_{\acute{a}j}^{i\bf T}{\cal P}_{\acute{b}}^j\right]{\mathsf
H}_i\right.\nonumber\\
\left.+\left[\frac{1}{2\sqrt 6}{\cal Q}_{\acute{a}ij}^{m\bf
T}{\cal P}_{\acute{b}m}^{kl} +\frac{1}{2\sqrt {30}}{\cal
Q}_{\acute{a}ij}^{k\bf T}{\cal P}_{\acute{b}}^{l} -\frac{1}{2\sqrt
{30}}{\cal Q}_{\acute{a}i}^{\bf T}{\cal P}_{\acute{b}j}^{kl}
+\frac{1}{2\sqrt 6}{\cal Q}_{\acute{a}n}^{klm\bf T}{\cal
P}_{\acute{b}mij}^{n}
\right.\right.\nonumber\\
\left.\left.+\frac{1}{3\sqrt 5}{\cal Q}_{\acute{a}i}^{klm\bf
T}{\cal P}_{\acute{b}mj} +\frac{1}{3\sqrt 5}{\cal
Q}_{\acute{a}}^{lm\bf T}{\cal P}_{\acute{b}mij}^{k}
+\frac{1}{15\sqrt 6}{\cal Q}_{\acute{a}}^{kl\bf T}{\cal
P}_{\acute{b}ij}\right]{\mathsf
H}^{ij}_{kl}\right.\nonumber\\
\left.+\left[\frac{1}{6\sqrt {5}}\epsilon_{ijklm}{\cal
Q}_{\acute{a}}^{ip\bf T}{\cal P}_{\acute{b}jn}^{p}
+\frac{1}{6\sqrt {3}}\epsilon_{ijklm}{\cal
Q}_{(S)\acute{a}}^{ip\bf T}{\cal
P}_{\acute{b}p}^{jn}+\frac{1}{60}\epsilon_{ijklm}{\cal
Q}_{\acute{a}}^{ij\bf
T}{\cal P}_{\acute{b}}^{n}\right.\right.\nonumber\\
\left.\left.-\frac{1}{60}\epsilon_{ijklm}{\cal
Q}_{\acute{a}}^{in\bf T}{\cal P}_{\acute{b}}^{j}-\frac{1}{12\sqrt
{15}}\epsilon_{ijklm}{\cal Q}_{(S)\acute{a}}^{in\bf T}{\cal
P}_{\acute{b}}^{j}-\frac{1}{3\sqrt {6}}{\cal Q}_{\acute{a}p}^{n\bf
T}{\cal P}_{\acute{b}klm}^{p}+\frac{1}{3\sqrt {5}}{\cal
Q}_{\acute{a}k}^{n\bf T}{\cal P}_{\acute{b}lm}\right]{\mathsf H}^{klm}_n\right.\nonumber\\
\left.+\left[\frac{1}{6\sqrt {5}}\epsilon^{ijklm}{\cal
Q}_{\acute{a}in}^{p\bf T}{\cal P}_{\acute{b}jp} +\frac{1}{6\sqrt
{3}}\epsilon^{ijklm}{\cal Q}_{\acute{a}in}^{p\bf T}{\cal
P}_{\acute{b}jp}^{(S)}-\frac{1}{60}\epsilon^{ijklm}{\cal
Q}_{\acute{a}n}^{\bf
T}{\cal P}_{\acute{b}ij}\right.\right.\nonumber\\
\left.\left.-\frac{1}{60}\epsilon^{ijklm}{\cal
Q}_{\acute{a}i}^{\bf T}{\cal P}_{\acute{b}jn}-\frac{1}{12\sqrt
{15}}\epsilon^{ijklm}{\cal Q}_{\acute{a}i}^{\bf T}{\cal
P}_{\acute{b}jn}^{(S)}+\frac{1}{3\sqrt {6}}{\cal
Q}_{\acute{a}p}^{klm\bf T}{\cal
P}_{\acute{b}n}^{p}-\frac{1}{3\sqrt {5}}{\cal
Q}_{\acute{a}}^{lm\bf T}{\cal P}_{\acute{b}n}^k\right]{\mathsf
H}_{klm}^n\right\}\nonumber\\
\end{eqnarray}

\subsection{The $\bf{(\overline{144}\times\overline{144}\times 10)}$ couplings}
The $(\overline{144}\times\overline{144}\times 10)$ couplings in
the superpotential are given by

\begin{equation}
{\mathsf
W}^{(10)}=h^{^{(10)}}_{\acute{a}\acute{b}}<\Upsilon^{*}_{(+)\acute{a}\mu}|B\Gamma_{\nu}|\Upsilon_{(+)\acute{b}\mu}>
\Phi_{\nu}
\end{equation}
where $\Phi_{\nu}$ represents the 10-plet of Higgs field. The
superpotential in its $SU(5)\times U(1)$ decomposed form is

\begin{eqnarray} {\mathsf
W}^{(10)}=ih^{^{{(10)(+)}}}_{\acute{a}\acute{b}}\left\{
\left[\frac{1}{\sqrt {15}}\epsilon^{ijlmn}{\cal
P}_{\acute{a}lmn}^{k\bf T}{\cal
P}_{\acute{b}ik}+\frac{1}{3}\epsilon^{ijlmn}{\cal
P}_{\acute{a}lmn}^{k\bf T}{\cal P}_{\acute{b}ik}^{(S)}-\frac{\sqrt
{2}}{5}\epsilon^{ijklm}{\cal P}_{\acute{a}lm}^{\bf T}{\cal
P}_{\acute{b}ik}\right.\right.\nonumber\\
\left.\left.+2\sqrt {2}{\cal P}_{\acute{a}k}^{ij\bf T}{\cal
P}_{\acute{b}i}^{k}-\sqrt{\frac{2}{5}}{\cal P}_{\acute{a}}^{i\bf
T}{\cal P}_{\acute{b}i}^{j}\right]{\mathsf H}_j\right.\nonumber\\
\left.+\left[\frac{2\sqrt 3}{5}{\cal P}_{\acute{a}}^{j\bf T}{\cal
P}_{\acute{b}jk}-\frac{4}{\sqrt 5}{\cal P}_{\acute{a}}^{j\bf
T}{\cal P}_{\acute{b}jk}^{(S)}-2\sqrt{2}{\cal P}_{\acute{a}j}^{\bf
T}{\cal P}_{\acute{b}k}^{j}+\sqrt {2}{\cal P}_{\acute{a}ijk}^{l\bf
T}{\cal P}_{\acute{b}l}^{ij}\right.\right.\nonumber\\
\left.\left.-\frac{2}{\sqrt {15}}{\cal P}_{\acute{a}ij}^{\bf
T}{\cal P}_{\acute{b}k}^{ij}\right]{\mathsf H}^k\right\}
\end{eqnarray}
where we have defined
\begin{equation}
h^{^{{(10)(+)}}}_{\acute{a}\acute{b}}=\frac{1}{2}\left(h^{^{(10)}}_{\acute{a}\acute{b}}+
h^{^{(10)}}_{\acute{b}\acute{a}}\right)
\end{equation}

\subsection{The $\bf{(\overline{144}\times \overline{144}\times 120)}$ coupling}
The $(\overline{144}\times\overline{144}\times 120)$ couplings in
the superpotential are given by

\begin{equation}
{\mathsf
W}^{(120)}=\frac{1}{3!}h^{^{(120)}}_{\acute{a}\acute{b}}<\Upsilon^{*}_{(+)\acute{a}\mu}|B\Gamma_{[\nu}
\Gamma_{\rho}\Gamma_{\lambda ]}|\Upsilon_{(+)\acute{b}\mu}>
\Phi_{\nu\rho\lambda}
\end{equation}
where $\Phi_{\nu\rho\lambda}$ represents the 120-plet of Higgs
field.  The superpotential in its $SU(5)\times U(1)$ decomposed
form is

\begin{eqnarray}
{\mathsf W}^{(120)}=ih^{^{{(120)(-)}}}_{\acute{a}\acute{b}}\left\{
\left[-\frac{1}{3\sqrt {10}}\epsilon^{ijlmn}{\cal
P}_{\acute{a}lmn}^{k\bf T}{\cal P}_{\acute{b}ik}-\frac{1}{3\sqrt 6
}\epsilon^{ijlmn}{\cal P}_{\acute{a}lmn}^{k\bf T}{\cal
P}_{\acute{b}ik}^{(S)}+\frac{1}{5\sqrt 3}\epsilon^{ijklm}{\cal
P}_{\acute{a}lm}^{\bf T}{\cal
P}_{\acute{b}ik}\right.\right.\nonumber\\
\left.\left.-\frac{2}{\sqrt {3}}{\cal P}_{\acute{a}k}^{ij\bf
T}{\cal P}_{\acute{b}i}^{k}+\frac{1}{\sqrt {15}}{\cal
P}_{\acute{a}}^{i\bf
T}{\cal P}_{\acute{b}i}^{j}\right]{\mathsf H}_j\right.\nonumber\\
\left.+\left[\frac{4}{\sqrt {15}}{\cal P}_{\acute{a}}^{j\bf
T}{\cal P}_{\acute{b}jk}-2\sqrt{\frac{2}{3}}{\cal
P}_{\acute{a}}^{j\bf T}{\cal P}_{\acute{b}jk}^{(S)}-\frac{4}{\sqrt
3}{\cal P}_{\acute{a}j}^{\bf T}{\cal
P}_{\acute{b}k}^j\right]{\mathsf H}^k
\right.\nonumber\\
 \left.+\left[-\frac{1}{3\sqrt 3}\epsilon^{ijlmn}{\cal
P}_{\acute{a}k}^{\bf T}{\cal
P}_{\acute{b}lmn}^k+\frac{1}{3}\sqrt{\frac{2}{5}}
\epsilon^{ijklm}{\cal P}_{\acute{a}k}^{\bf T}{\cal
P}_{\acute{b}lm}-\frac{4}{\sqrt {15}}{\cal P}_{\acute{a}}^{k\bf
T}{\cal P}_{\acute{b}k}^{ij}-2\sqrt{\frac{2}{5}}{\cal
P}_{\acute{a}}^{i\bf T}{\cal P}_{\acute{b}}^{j}\right]{\mathsf
H}_{ij}
\right.\nonumber\\
\left.+\left[\frac{4\sqrt 2}{5}{\cal P}_{\acute{a}i}^{k\bf T}{\cal
P}_{\acute{b}jk}+4\sqrt{\frac{2}{15}}{\cal P}_{\acute{a}i}^{k\bf
T}{\cal P}_{\acute{b}jk}^{(S)}\right]{\mathsf H}^{ij}\right.\nonumber\\
\left.+\left[-\frac{1}{3\sqrt {10}}\epsilon^{ijmnp}{\cal
P}_{\acute{a}mnp}^{l\bf T}{\cal P}_{\acute{b}kl}-\frac{1}{3\sqrt
6}\epsilon^{ijmnp}{\cal P}_{\acute{a}mnp}^{l\bf T}{\cal
P}_{\acute{b}kl}^{(S)}+\frac{1}{5\sqrt 3}\epsilon^{ijlmn}{\cal
P}_{\acute{a}mn}^{\bf T}{\cal P}_{\acute{b}kl}\right.\right.\nonumber\\
\left.\left.+\frac{1}{3\sqrt 5}\epsilon^{ijlmn}{\cal
P}_{\acute{a}mn}^{\bf T}{\cal
P}_{\acute{b}kl}^{(S)}-\frac{2}{\sqrt 3}{\cal
P}_{\acute{a}l}^{ij\bf T}{\cal P}_{\acute{b}k}^l-\frac{2}{\sqrt
{15}}{\cal P}_{\acute{a}}^{i\bf T}{\cal
P}_{\acute{b}k}^j\right]{\mathsf
H}^k_{ij}\right.\nonumber\\
\left.+\left[-\frac{2}{\sqrt {3}}{\cal P}_{\acute{a}klm}^{j\bf
T}{\cal P}_{\acute{b}j}^{mn}-\frac{1}{\sqrt {15}}{\cal
P}_{\acute{a}}^{j\bf T}{\cal
P}_{\acute{b}jkl}^n-\frac{4}{3}\sqrt{\frac{2}{5}}{\cal
P}_{\acute{a}km}^{\bf T}{\cal P}_{\acute{b}l}^{mn} +\frac{2\sqrt
2}{15}{\cal P}_{\acute{a}kl}^{\bf T}{\cal
P}_{\acute{b}}^n\right]{\mathsf H}^{kl}_n\right\}\nonumber\\
\end{eqnarray}
where we have defined
\begin{equation}
h^{^{{(120)(-)}}}_{\acute{a}\acute{b}}=\frac{1}{2}\left(h^{^{(120)}}_{\acute{a}\acute{b}}-
h^{^{(120)}}_{\acute{b}\acute{a}}\right)
\end{equation}

\subsection{The $\bf{(\overline{144}\times\overline{144}\times\overline{126})}$ couplings}

The $(\overline{144}\times\overline{144}\times\overline{126})$
coupling in the superpotential is

\begin{equation}
{\mathsf W}^{(\overline {126} )}=\frac{1}{5!}h^{^{(\overline
{126})}}_{\acute{a}\acute{b}}<\Upsilon^{*}_{(+)\acute{a}\mu}|B\Gamma_{[\nu}\Gamma_{\rho}\Gamma_{\sigma}
\Gamma_{\lambda}\Gamma_{\theta ]
}|\Upsilon_{(+)\acute{b}\mu}>\overline{\Phi}_{\nu\rho\sigma\lambda\theta}
\end{equation}
where $\overline{\Phi}_{\nu\rho\sigma\lambda\theta}$ represents
the $\overline{126}$-plet of Higgs field. The superpotential in
its $SU(5)\times U(1)$ decomposed form is
\begin{eqnarray}
{\mathsf W}^{(\overline {126} )}=ih^{^{(\overline
{126})(+)}}_{\acute{a}\acute{b}}\left\{\left[-\frac{8}{5\sqrt
3}{\cal P}_{\acute{a}}^{i\bf T}{\cal
P}_{\acute{b}i}\right]{\mathsf
H}\right.\nonumber\\
\left.+\left[\frac{\sqrt 2}{5}{\cal P}_{\acute{a}}^{j\bf T}{\cal
P}_{\acute{b}jk}-\frac{1}{\sqrt 5}{\cal P}_{\acute{a}}^{j\bf
T}{\cal P}_{\acute{b}jk}^{(S)}-\sqrt{\frac{2}{5}}{\cal
P}_{\acute{a}j}^{\bf T}{\cal P}_{\acute{b}k}^j\right.\right.\nonumber\\
\left.\left.-\frac{1}{3\sqrt {10}}{\cal P}_{\acute{a}ijk}^{l\bf
T}{\cal P}_{\acute{b}l}^{ij} +\frac{1}{15\sqrt {3}}{\cal
P}_{\acute{a}ij}^{\bf T}{\cal P}_{\acute{b}k}^{ij}+\frac{1}{5\sqrt
{15}}{\cal P}_{\acute{a}ik}^{\bf T}{\cal
P}_{\acute{b}}^i\right]{\mathsf
H}^k\right.\nonumber\\
\left.+\left[\frac{1}{3\sqrt {30}}\epsilon^{ijlmn}{\cal
P}_{\acute{a}k}^{\bf T}{\cal P}_{\acute{b}lmn}^k-\frac{1}{15}
\epsilon^{ijklm}{\cal P}_{\acute{a}k}^{\bf T}{\cal
P}_{\acute{b}lm}\right.\right.\nonumber\\
\left.\left.+\frac{2}{5}\sqrt{\frac{2}{3}}{\cal
P}_{\acute{a}}^{k\bf T}{\cal P}_{\acute{b}k}^{ij}+
\frac{2}{5}\sqrt{\frac{2}{15}}{\cal P}_{\acute{a}}^{i\bf T}{\cal
P}_{\acute{b}}^{j}\right]{\mathsf H}_{ij}
\right.\nonumber\\
\left.+\left[-\frac{2}{5}{\cal P}_{\acute{a}i}^{k\bf T}{\cal
P}_{\acute{b}jk}-\frac{2}{\sqrt{15}}{\cal P}_{\acute{a}i}^{k\bf
T}{\cal P}_{\acute{b}jk}^{(S)}\right]{\mathsf H}_{(S)}^{ij}\right.\nonumber\\
\left.+\left[\frac{1}{30}\epsilon^{ijmnp}{\cal
P}_{\acute{a}mnp}^{l\bf T}{\cal P}_{\acute{b}kl}+\frac{1}{6\sqrt
{15}}\epsilon^{ijmnp}{\cal P}_{\acute{a}mnp}^{l\bf T}{\cal
P}_{\acute{b}kl}^{(S)}-\frac{1}{5\sqrt {30}}\epsilon^{ijlmn}{\cal
P}_{\acute{a}mn}^{\bf T}{\cal P}_{\acute{b}kl}\right.\right.\nonumber\\
\left.\left.-\frac{1}{15\sqrt 2}\epsilon^{ijlmn}{\cal
P}_{\acute{a}mn}^{\bf T}{\cal
P}_{\acute{b}kl}^{(S)}+\sqrt{\frac{2}{15}}{\cal
P}_{\acute{a}l}^{ij\bf T}{\cal
P}_{\acute{b}k}^l+\frac{1}{5}\sqrt{\frac{2}{3}}{\cal
P}_{\acute{a}}^{i\bf T}{\cal P}_{\acute{b}k}^j\right]{\mathsf
H}^k_{ij}\right.\nonumber\\
\left.+\left[-\frac{1}{3\sqrt {15}}{\cal P}_{\acute{a}ijk}^{n\bf
T}{\cal P}_{\acute{b}n}^{rs}-\frac{1}{15\sqrt {3}}{\cal
P}_{\acute{a}ij}^{\bf T}{\cal P}_{\acute{b}k}^{rs}+\frac{\sqrt 2
}{15}{\cal P}_{\acute{a}}^{ij\bf T}{\cal
P}_{\acute{b}k}^{rs}\right]{\mathsf H}^{ijk}_{rs}\right\}
\end{eqnarray}
where we have defined
\begin{equation}
h^{^{{(\overline
{126})(+)}}}_{\acute{a}\acute{b}}=\frac{1}{2}\left(h^{^{(\overline
{126})}}_{\acute{a}\acute{b}}+ h^{^{(\overline
{126})}}_{\acute{b}\acute{a}}\right)
\end{equation}

\section{Higgs Sector Quartic Couplings}
We discuss now the quartic couplings  involving four
vector-spinors. We will discuss specifically the quartic couplings
that arise from the cubic couplings discussed in Sec.(4) by
elimination of the 1, 45 and 210 fields assuming they are heavy in
the $144\times \overline{144}$ couplings and by elimination of 10,
120 and $126+\overline{126}$  assuming they are
 heavy for the $144\times 144$ couplings.
We first  discuss the quartic couplings  that arise from the elimination of
 1, 10, 45. In this case we start with  the superpotenial

\begin{eqnarray}
{\mathsf
W}^{(1,45,210)}=h^{^{(1)}}_{\acute{a}\acute{b}}<\Upsilon^{*}_{(-)\acute{a}\mu}|B|\Upsilon_{(+)\acute{b}\mu}>k_{_{{\cal
X}}}^{^{(1)}}\Phi_{\cal X}+\frac{1}{2}\Phi_{\cal X}{\cal
M}^{^{(1)}}_{{\cal X}{\cal X}'}\Phi_{{\cal X}'}\nonumber\\
+\frac{1}{2!}h^{^{(45)}}_{\acute{a}\acute{b}}<\Upsilon^{*}_{(-)\acute{a}\mu}|B\Sigma_{\rho\sigma}
|\Upsilon_{(+)\acute{b}\mu}>k_{_{{\cal
Y}}}^{^{(45)}}\Phi_{\rho\sigma\cal
Y}+\frac{1}{2}\Phi_{\rho\sigma\cal Y}{\cal M}^{^{(45)}}_{{\cal
Y}{\cal Y}'}\Phi_{\rho\sigma{\cal Y}'}\nonumber\\
+\frac{1}{4!}h^{^{(210)}}_{\acute{a}\acute{b}}<\Upsilon^{*}_{(-)\acute{a}\mu}|B\Gamma_{[\nu}\Gamma_{\rho}\Gamma_{\sigma}
\Gamma_{\lambda]}|\Upsilon_{(+)\acute{b}\mu}>k_{_{{\cal
Z}}}^{^{(210)}} \Phi_{\nu\rho\sigma\lambda\cal
Z}\nonumber\\
+\frac{1}{2}\Phi_{\nu\rho\sigma\lambda {\cal Z}}{\cal
M}^{^{(210)}}_{{\cal Z}{\cal Z}'} \Phi_{\nu\rho\sigma\lambda {\cal
Z}'}
\end{eqnarray}
We then eliminate $\Phi_{\cal X}$, $\Phi_{\rho\sigma\cal Y}$,
$\Phi_{\nu\rho\sigma\lambda\cal Z}$  assuming they are superheavy using the
 F-flatness conditions
\begin{equation}
\frac{\partial {\mathsf W}^{(1,45,210)}}{\partial \Phi_{\cal
X}}=0,~~\frac{\partial {\mathsf W}^{(1,45,210)}}{\partial
\Phi_{\rho\sigma\cal Y}}=0,~~\frac{\partial {\mathsf
W}^{(1,45,210)}}{\partial \Phi_{\nu\rho\sigma\lambda\cal Z}}=0
\end{equation}
We discuss now  the individual contribution arising from the
elimination of 1, 45 and 210 separately.

\subsection{The $\bf{{\left(144\times \overline{144}\right)_1 \left(144\times \overline{144}\right)_1}}$ couplings}

The ${\left(144\times \overline{144}\right)_1 \left(144\times \overline{144}\right)_1}$ couplings gotten
by the singlet mediation are  given by

\begin{equation}
{\mathsf
W}_{dim-5}^{(1)}=2\lambda_{\acute{a}\acute{b},\acute{c}\acute{d}}^{^{(1)}}
<\Upsilon^{*}_{(-)\acute{a}\mu}|B|\Upsilon_{(+)\acute{b}\mu}><\Upsilon^{*}_{(-)\acute{c}\lambda}|B|
\Upsilon_{(+)\acute{d}\lambda}>
\end{equation}
where
\begin{eqnarray}
<\Upsilon^{*}_{(-)\acute{a}\mu}|B|\Upsilon_{(+)\acute{b}\mu}>=
i\left\{\frac{3}{5}{\cal Q}_{\acute{a}i}^{\bf T}{\cal
P}^i_{\acute{b}}+{\cal Q}_{\acute{a}}^{i\bf T}{\cal
P}_{\acute{b}i}+\frac{1}{10}{\cal Q}_{\acute{a}}^{ij\bf T}{\cal
P}_{\acute{b}ij}+\frac{1}{2}{\cal Q}_{(S)\acute{a}}^{ij\bf
T}{\cal P}^{(S)}_{\acute{b}ij}\right.\nonumber\\
\left.+{\cal Q}_{\acute{a}j}^{i\bf T}{\cal
P}_{\acute{b}i}^j-\frac{1}{6}{\cal Q}_{\acute{a}l}^{ijk\bf T}{\cal
P}_{\acute{b}ijk}^l-\frac{1}{2}{\cal Q}_{\acute{a}ij}^{k\bf
T}{\cal P}_{\acute{b}k}^{ij}\right\}
\end{eqnarray}
Explicit evaluation of the above quantities gives
\begin{eqnarray} {\mathsf
W}_{dim-5}^{(1)}=\lambda_{\acute{a}\acute{b},\acute{c}\acute{d}}^{^{(1)}}
\left\{-\frac{18}{25}{\cal Q}_{\acute{a}i}^{\bf T}{\cal
P}^i_{\acute{b}}{\cal Q}_{\acute{c}j}^{\bf T}{\cal
P}^j_{\acute{d}}-\frac{12}{5}{\cal Q}_{\acute{a}i}^{\bf T}{\cal
P}^i_{\acute{b}}{\cal Q}_{\acute{c}}^{j\bf T}{\cal
P}_{\acute{d}j}-\frac{6}{25}{\cal Q}_{\acute{a}i}^{\bf T}{\cal
P}^i_{\acute{b}}{\cal Q}_{\acute{c}}^{jk\bf T}{\cal
P}_{\acute{d}jk}\right.\nonumber\\
\left. -\frac{6}{5}{\cal Q}_{\acute{a}i}^{\bf T}{\cal
P}^i_{\acute{b}}{\cal Q}_{(S)\acute{c}}^{jk\bf T}{\cal
P}_{\acute{d}jk}^{(S)}-\frac{12}{5}{\cal Q}_{\acute{a}i}^{\bf
T}{\cal P}^i_{\acute{b}}{\cal Q}_{\acute{c}k}^{j\bf T}{\cal
P}_{\acute{d}j}^k+\frac{2}{5}{\cal Q}_{\acute{a}i}^{\bf T}{\cal
P}^i_{\acute{b}}{\cal Q}_{\acute{c}m}^{jkl\bf T}{\cal
P}_{\acute{d}jkl}^m\right.\nonumber\\
\left.+\frac{6}{5}{\cal Q}_{\acute{a}i}^{\bf T}{\cal
P}^i_{\acute{b}}{\cal Q}_{\acute{c}kl}^{j\bf T}{\cal
P}_{\acute{d}j}^{kl}-2{\cal Q}_{\acute{a}}^{i\bf T}{\cal
P}_{\acute{b}i}{\cal Q}_{\acute{c}}^{j\bf T}{\cal P}_{\acute{d}j}
-\frac{2}{5}{\cal Q}_{\acute{a}}^{i\bf T}{\cal
P}_{\acute{b}i}{\cal Q}_{\acute{c}}^{jk\bf T}{\cal
P}_{\acute{d}jk}\right.\nonumber\\
\left. -2{\cal Q}_{\acute{a}}^{i\bf T}{\cal P}_{\acute{b}i}{\cal
Q}_{(S)\acute{c}}^{jk\bf T}{\cal P}_{\acute{d}jk}^{(S)}-4{\cal
Q}_{\acute{a}}^{i\bf T}{\cal P}_{\acute{b}i}{\cal
Q}_{\acute{c}k}^{j\bf T}{\cal P}_{\acute{d}j}^k+\frac{2}{3} {\cal
Q}_{\acute{a}}^{i\bf T}{\cal P}_{\acute{b}i}{\cal
Q}_{\acute{c}j}^{klm\bf T}{\cal P}_{\acute{d}klm}^j\right.\nonumber\\
\left.+2{\cal Q}_{\acute{a}}^{i\bf T}{\cal P}_{\acute{b}i}{\cal
Q}_{\acute{c}kl}^{j\bf T}{\cal
P}_{\acute{d}j}^{kl}-\frac{1}{50}{\cal Q}_{\acute{a}}^{ij\bf
T}{\cal P}_{\acute{b}ij}{\cal Q}_{\acute{c}}^{kl\bf T}{\cal
P}_{\acute{d}kl}-\frac{1}{5}{\cal Q}_{\acute{a}}^{ij\bf T}{\cal
P}_{\acute{b}ij}{\cal Q}_{(S)\acute{c}}^{kl\bf T}{\cal
P}_{\acute{d}kl}^{(S)}\right.\nonumber\\
\left.-\frac{2}{5}{\cal Q}_{\acute{a}}^{ij\bf T}{\cal
P}_{\acute{b}ij}{\cal Q}_{\acute{c}l}^{k\bf T}{\cal
P}_{\acute{d}k}^l +\frac{1}{15}{\cal Q}_{\acute{a}}^{ij\bf T}{\cal
P}_{\acute{b}ij}{\cal Q}_{\acute{c}n}^{klm\bf T}{\cal
P}_{\acute{d}klm}^n+\frac{1}{5}{\cal Q}_{\acute{a}}^{ij\bf T}{\cal
P}_{\acute{b}ij}{\cal Q}_{\acute{c}lm}^{k\bf T}{\cal
P}_{\acute{d}k}^{lm}\right.\nonumber\\
\left.-\frac{1}{2}{\cal Q}_{(S)\acute{a}}^{ij\bf T}{\cal
P}_{\acute{b}ij}^{(S)} {\cal Q}_{(S)\acute{c}}^{kl\bf T}{\cal
P}_{\acute{d}kl}^{(S)}-2{\cal Q}_{(S)\acute{a}}^{ij\bf T}{\cal
P}_{\acute{d}ij}^{(S)}{\cal Q}_{\acute{c}l}^{k\bf T}{\cal
P}_{\acute{d}k}^{l}+\frac{1}{3}{\cal Q}_{(S)\acute{a}}^{ij\bf
T}{\cal P}_{\acute{d}ij}^{(S)}{\cal Q}_{\acute{c}n}^{klm\bf
T}{\cal P}_{\acute{d}klm}^{n}\right.\nonumber\\
\left.+{\cal Q}_{(S)\acute{a}}^{ij\bf T}{\cal
P}_{\acute{d}ij}^{(S)}{\cal Q}_{\acute{c}lm}^{k\bf T}{\cal
P}_{\acute{d}k}^{lm}-2{\cal Q}_{\acute{a}j}^{i\bf T}{\cal
P}_{\acute{b}i}^j{\cal Q}_{\acute{c}l}^{k\bf T}{\cal
P}_{\acute{d}k}^{l} +\frac{2}{3} {\cal Q}_{\acute{a}j}^{i\bf
T}{\cal P}_{\acute{b}i}^j{\cal Q}_{\acute{c}k}^{lmn\bf T}{\cal
P}_{\acute{d}lmn}^{k} \right.\nonumber\\
\left.+2{\cal Q}_{\acute{a}j}^{i\bf T}{\cal P}_{\acute{b}i}^j{\cal
Q}_{\acute{c}lm}^{k\bf T}{\cal
P}_{\acute{d}k}^{lm}-\frac{1}{18}{\cal Q}_{\acute{a}l}^{ijk\bf
T}{\cal P}_{\acute{b}ijk}^l{\cal Q}_{\acute{c}p}^{mno\bf T}{\cal
P}_{\acute{d}mno}^{p}-\frac{1}{3}{\cal Q}_{\acute{a}l}^{ijk\bf
T}{\cal P}_{\acute{b}ijk}^l{\cal Q}_{\acute{c}no}^{m\bf T}{\cal
P}_{\acute{d}m}^{no}\right.\nonumber\\
\left.-\frac{1}{2}{\cal Q}_{\acute{a}ij}^{k\bf T}{\cal
P}_{\acute{b}k}^{ij}{\cal Q}_{\acute{c}mn}^{l\bf T}{\cal
P}_{\acute{d}mn}^{l}\right\}
\end{eqnarray}

where we have defined
\begin{eqnarray}
\lambda_{\acute{a}\acute{b},\acute{c}\acute{d}}^{^{(1)}}=
h_{\acute{a}\acute{b}}^{^{(1)}}h_{\acute{c}\acute{d}}^{^{(1)}}k_{_{{\cal
X}}}^{^{(1)}}\left[\widetilde{{\cal M}}^{^{(1)}}\left\{{\cal
M}^{^{(1)}}\widetilde{{\cal M}}^{^{(1)}} -\bf{1}\right\}
\right]_{{\cal X}{\cal X}'}k_{_{{\cal X}'}}^{^{(1)}}\nonumber\\
\widetilde{{\cal M}}^{^{(1)}}=\left[{\cal M}^{^{(1)}}+\left({\cal
M}^{^{(1)}}\right)^{\bf {T}}\right]^{-1}
\end{eqnarray}
\\
\subsection{The $\bf{{\left(144\times\overline{144}\right)_{45} \left(144\times
\overline{144}\right)_{45}}}$ couplings}

The ${\left(144\times\overline{144}\right)_{45} \left(144\times\overline{144}\right)_{45}}$ couplings
gotten by the 45 plet mediation are given by
\begin{eqnarray}
 {\mathsf
W}_{dim-5}^{(45)}=\lambda_{\acute{a}\acute{b},\acute{c}\acute{d}}^{^{(45)}}
\left[-4<\Upsilon^{*}_{(-)\acute{a}\mu}|Bb_ib_j|\Upsilon_{(+)\acute{b}\mu}>
<\Upsilon^{*}_{(-)\acute{c}\lambda}|Bb_i^{\dagger}b_j^{\dagger}|\Upsilon_{(+)\acute{d}\lambda}>
~~~\right.\nonumber\\
\left.+4<\Upsilon^{*}_{(-)\acute{a}\mu}|Bb_i^{\dagger}b_j|\Upsilon_{(+)\acute{b}\mu}>
<\Upsilon^{*}_{(-)\acute{c}\lambda}|Bb_j^{\dagger}b_i|\Upsilon_{(+)\acute{d}\lambda}>
~~~\right.\nonumber\\
\left.-4<\Upsilon^{*}_{(-)\acute{a}\mu}|Bb_n^{\dagger}b_n|\Upsilon_{(+)\acute{b}\mu}>
<\Upsilon^{*}_{(-)\acute{c}\lambda}|B|\Upsilon_{(+)\acute{d}\lambda}>~~~\right.\nonumber\\
\left.+5<\Upsilon^{*}_{(-)\acute{a}\mu}|B|\Upsilon_{(+)\acute{b}\mu}><\Upsilon^{*}_{(-)\acute{c}\lambda}|B|
\Upsilon_{(+)\acute{d}\lambda}>\right]~~~
 \end{eqnarray}
 where the explicit evaluation in $SU(5)\times U(1)$ decomposition gives
\begin{eqnarray}
<\Upsilon^{*}_{(-)\acute{a}\mu}|Bb_ib_j|\Upsilon_{(+)\acute{b}\mu}>=
i\left\{-{\cal Q}_{\acute{a}}^{k\bf T}{\cal
P}_{\acute{b}k}^{ij}-\frac{1}{2\sqrt {5}}{\cal
Q}_{\acute{a}}^{i\bf T}{\cal P}^j_{\acute{b}}+\frac{1}{2\sqrt
{5}}{\cal Q}_{\acute{a}}^{j\bf T}{\cal P}^i_{\acute{b}}
 + \sqrt{\frac{2}{15}}{\cal Q}_{\acute{a}}^{ik\bf T}{\cal
P}_{\acute{b}k}^j\right.\nonumber\\
 \left.-\sqrt{\frac{2}{15}}{\cal
Q}_{\acute{a}}^{jk\bf T}{\cal P}_{\acute{b}k}^i +{\cal
Q}_{\acute{a}l}^{ijk\bf T}{\cal P}_{\acute{b}k}^l
+\frac{7}{10\sqrt {6}}\epsilon^{ijklm}{\cal Q}_{\acute{a}k}^{\bf
T}{\cal P}_{\acute{b}lm}\right.\nonumber\\
 \left.-\frac{1}{3\sqrt
{5}}\epsilon^{ijklm}{\cal Q}_{\acute{a}n}^{\bf T}{\cal
P}_{\acute{b}klm}^{n}
+\frac{1}{2}\sqrt{\frac{3}{10}}\epsilon^{ijklm}{\cal
Q}_{\acute{a}kl}^{n\bf T}{\cal P}_{\acute{b}mn}\right.\nonumber\\
\left.+\frac{1}{2\sqrt{2}}\epsilon^{ijklm}{\cal
Q}_{\acute{a}kl}^{n\bf T}{\cal P}^{(S)}_{\acute{b}mn}\right\}
\end{eqnarray}

\begin{eqnarray}
<\Upsilon^{*}_{(-)\acute{a}\mu}|Bb_i^{\dagger}b_j^{\dagger}|\Upsilon_{(+)\acute{b}\mu}>=
i\left\{-{\cal Q}_{\acute{a}ij}^{k\bf T}{\cal
P}_{\acute{b}k}-\frac{1}{2\sqrt {5}}{\cal Q}_{\acute{a}j}^{\bf
T}{\cal P}_{i\acute{b}}+\frac{1}{2\sqrt {5}}{\cal
Q}_{\acute{a}i}^{\bf T}{\cal P}_{\acute{b}j}
 + \sqrt{\frac{2}{15}}{\cal Q}_{\acute{a}i}^{k\bf T}{\cal
P}_{\acute{b}kj}\right.\nonumber\\
\left. -\sqrt{\frac{2}{15}}{\cal
 Q}_{\acute{a}j}^{k\bf T}{\cal P}_{\acute{b}ki} +{\cal
Q}_{\acute{a}l}^{k\bf T}{\cal P}_{\acute{b}ijk}^l
+\frac{7}{10\sqrt {6}}\epsilon_{ijklm}{\cal Q}_{\acute{a}}^{kl\bf
T}{\cal P}_{\acute{b}}^m\right.\nonumber\\
 \left.-\frac{1}{3\sqrt
{5}}\epsilon_{ijklm}{\cal Q}_{\acute{a}n}^{klm\bf T}{\cal
P}_{\acute{b}}^{n}
+\frac{1}{2}\sqrt{\frac{3}{10}}\epsilon_{ijklm}{\cal
Q}_{\acute{a}}^{kn\bf T}{\cal P}_{\acute{b}n}^{lm}\right.\nonumber\\
\left.+\frac{1}{2\sqrt{2}}\epsilon_{ijklm}{\cal
Q}_{\acute{a}(S)}^{kn\bf T}{\cal P}^{lm}_{\acute{b}n}\right\}
\end{eqnarray}

\begin{eqnarray}
<\Upsilon^{*}_{(-)\acute{a}\mu}|Bb_i^{\dagger}b_j
|\Upsilon_{(+)\acute{b}\mu}>= i\left\{{\cal Q}_{\acute{a}ik}^{l\bf
T}{\cal P}_{\acute{b}l}^{kj}-\frac{1}{2\sqrt {5}}{\cal
Q}_{\acute{a}ik}^{j\bf T}{\cal P}_{\acute{b}}^k+\frac{1}{2\sqrt
{5}}{\cal Q}_{\acute{a}k}^{\bf T}{\cal
P}_{\acute{b}i}^{kj}\right.\nonumber\\
\left.-\frac{3}{20}{\cal Q}_{\acute{a}i}^{\bf T}{\cal
P}_{\acute{b}}^j -\frac{1}{20}\delta^j_i{\cal Q}_{\acute{a}k}^{\bf
T}{\cal P}_{\acute{b}}^k +\frac{1}{2}{\cal Q}_{\acute{a}m}^{jkl\bf
T}{\cal
P}_{\acute{b}ikl}^m\right.\nonumber\\
\left.-\frac{1}{6}\delta^j_i{\cal Q}_{\acute{a}n}^{klm\bf T}{\cal
P}_{\acute{b}klm}^n
 -\frac{1}{\sqrt {30}}{\cal
Q}_{\acute{a}i}^{jkl\bf T}{\cal P}_{kl\acute{b}}
-\frac{1}{\sqrt{30}}{\cal Q}_{\acute{a}}^{kl\bf T}{\cal
P}_{\acute{b}ikl}^j\right.\nonumber\\
 \left.+\frac{1}{30}{\cal
Q}_{\acute{a}}^{jk\bf T}{\cal
P}_{\acute{b}ki}+\frac{1}{6}\delta^j_i{\cal Q}_{\acute{a}}^{kl\bf
T}{\cal P}_{\acute{b}kl} -\frac{1}{2}\sqrt{\frac{3}{5}}{\cal
Q}_{\acute{a}}^{jk\bf T}{\cal
P}_{\acute{b}ki}^{(S)}\right.\nonumber\\
\left.+\frac{1}{2}\sqrt{\frac{3}{5}}{\cal Q}_{(S)\acute{a}}^{jk\bf
T}{\cal P}_{\acute{b}ki}-\frac{1}{2}{\cal Q}_{(S)\acute{a}}^{jk\bf
T}{\cal P}_{\acute{b}ki}^{(S)} +\frac{1}{2}\delta^j_i{\cal
Q}_{(S)\acute{a}}^{kl\bf T}{\cal
P}_{\acute{b}kl}^{(S)}\right.\nonumber\\
\left.-{\cal Q}_{\acute{a}k}^{j\bf T}{\cal
P}_{\acute{b}i}^{k}+\delta^j_i{\cal Q}_{\acute{a}k}^{l\bf T}{\cal
P}_{\acute{b}l}^{k}\right\}
\end{eqnarray}

\begin{eqnarray}
<\Upsilon^{*}_{(-)\acute{a}\mu}|Bb_n^{\dagger}b_n|\Upsilon_{(+)\acute{b}\mu}>=
i\left\{4{\cal Q}_{\acute{a}j}^{i\bf T}{\cal
P}_{\acute{b}i}^j+\frac{4}{5}{\cal Q}_{\acute{a}}^{ij\bf T}{\cal
P}_{\acute{b}ij}+2{\cal Q}_{(S)\acute{a}}^{ij\bf T}{\cal
P}^{(S)}_{\acute{b}ij}-{\cal Q}_{\acute{a}ij}^{k\bf T}{\cal
P}_{\acute{b}k}^{ij}\right.\nonumber\\
\left.-\frac{2}{5}{\cal Q}_{\acute{a}i}^{\bf T}{\cal
P}^i_{\acute{b}}-\frac{1}{3}{\cal Q}_{\acute{a}l}^{ijk\bf T}{\cal
P}_{\acute{b}ijk}^l\right\}
\end{eqnarray}
where
\begin{eqnarray}
\lambda_{\acute{a}\acute{b},\acute{c}\acute{d}}^{^{(45)}}=
h_{\acute{a}\acute{b}}^{^{(45)}}h_{\acute{c}\acute{d}}^{^{(45)}}k_{_{{\cal
Y}}}^{^{(45)}}\left[\widetilde{{\cal M}}^{^{(45)}}\left\{{\cal
M}^{^{(45)}}\widetilde{{\cal M}}^{^{(45)}} -\bf{1}\right\}
\right]_{{\cal Y}{\cal Y}'}k_{_{{\cal Y}'}}^{^{(45)}}\nonumber\\
\widetilde{{\cal M}}^{^{(45)}}=\left[{\cal
M}^{^{(45)}}+\left({\cal M}^{^{(45)}}\right)^{\bf {T}}\right]^{-1}
\end{eqnarray}

\subsection{The $\bf{{\left(144\times \overline{144}\right)_{210} \left(144\times
\overline{144}\right)_{210}}}$ couplings}

 The ${\left(144\times \overline{144}\right)_{210} \left(144\times
\overline{144}\right)_{210}}$ couplings are  gotten by 210 mediation and are
given by
\begin{eqnarray}
 {\mathsf
W}_{dim-5}^{(210)}=-\frac{1}{18}\lambda_{ab,cd}^{^{(210)}}
\left[8<\Upsilon^{*}_{(-)\acute{a}\nu}|Bb_i^{\dagger}b_jb_kb_l
|\Upsilon_{(+)\acute{b}\nu}>
<\Upsilon^{*}_{(-)\acute{c}\lambda}|Bb_j^{\dagger}b_k^{\dagger}b_l^{\dagger}b_i
|\Upsilon_{(+)\acute{d}\lambda}>
\right.\nonumber\\
\left.-6<\Upsilon^{*}_{(-)\acute{a}\mu}|Bb_i^{\dagger}b_j^{\dagger}b_kb_l
|\Upsilon_{(+)\acute{b}\mu}>
<\Upsilon^{*}_{(-)\acute{c}\lambda}|Bb_k^{\dagger}b_l^{\dagger}b_ib_j
|\Upsilon_{(+)\acute{d}\lambda}>
\right.\nonumber\\
\left.-2<\Upsilon^{*}_{(-)\acute{a}\mu}|Bb_ib_jb_kb_l
|\Upsilon_{(+)\acute{b}\mu}>
<\Upsilon^{*}_{(-)\acute{c}\lambda}|Bb_i^{\dagger}b_j^{\dagger}b_k^{\dagger}
b_l^{\dagger} |\Upsilon_{(+)\acute{d}\lambda}>
\right.\nonumber\\
\left.+24<\Upsilon^{*}_{(-)\acute{a}\mu}|Bb_i^{\dagger}b_j
|\Upsilon_{(+)\acute{b}\mu}>
<\Upsilon^{*}_{(-)\acute{c}\lambda}|Bb_j^{\dagger}b_n^{\dagger}b_nb_i
|\Upsilon_{(+)\acute{d}\lambda}>
\right.\nonumber\\
\left.-12<\Upsilon^{*}_{(-)\acute{a}\mu}|Bb_i^{\dagger}b_j^{\dagger}
|\Upsilon_{(+)\acute{b}\mu}>
<\Upsilon^{*}_{(-)\acute{c}\lambda}|Bb_n^{\dagger}b_nb_ib_j
|\Upsilon_{(+)\acute{d}\lambda}>
\right.\nonumber\\
\left.-12<\Upsilon^{*}_{(-)\acute{a}\mu}|Bb_ib_j
|\Upsilon_{(+)\acute{b}\mu}>
<\Upsilon^{*}_{(-)\acute{c}\lambda}|Bb_i^{\dagger}b_j^{\dagger}b_n^{\dagger}b_n
|\Upsilon_{(+)\acute{d}\lambda}>
\right.\nonumber\\
\left.-6<\Upsilon^{*}_{(-)\acute{a}\mu}|Bb_m^{\dagger}b_m
|\Upsilon_{(+)\acute{b}\mu}>
<\Upsilon^{*}_{(-)\acute{c}\lambda}|Bb_n^{\dagger}b_n
|\Upsilon_{(+)\acute{d}\lambda}>
\right.\nonumber\\
\left.-6<\Upsilon^{*}_{(-)\acute{a}\mu}|B
|\Upsilon_{(+)\acute{b}\mu}>
<\Upsilon^{*}_{(-)\acute{c}\lambda}|Bb_m^{\dagger}b_n^{\dagger}b_nb_m
|\Upsilon_{(+)\acute{d}\lambda}>
\right.\nonumber\\
\left.+18<\Upsilon^{*}_{(-)\acute{a}\mu}|Bb_ib_j
|\Upsilon_{(+)\acute{b}\mu}>
<\Upsilon^{*}_{(-)\acute{c}\lambda}|Bb_i^{\dagger}b_j^{\dagger}
|\Upsilon_{(+)\acute{d}\lambda}>
\right.\nonumber\\
\left.-18<\Upsilon^{*}_{(-)\acute{a}\mu}|Bb_i^{\dagger}b_j
|\Upsilon_{(+)\acute{b}\mu}>
<\Upsilon^{*}_{(-)\acute{c}\lambda}|Bb_j^{\dagger}b_i
|\Upsilon_{(+)\acute{d}\lambda}>
\right.\nonumber\\
\left.+24<\Upsilon^{*}_{(-)\acute{a}\mu}|B
|\Upsilon_{(+)\acute{b}\mu}>
<\Upsilon^{*}_{(-)\acute{c}\lambda}|Bb_n^{\dagger}b_n
|\Upsilon_{(+)\acute{d}\lambda}>
\right.\nonumber\\
\left.-15<\Upsilon^{*}_{(-)\acute{a}\mu}|B
|\Upsilon_{(+)\acute{b}\mu}> <\Upsilon^{*}_{(-)\acute{c}\lambda}|B
|\Upsilon_{(+)\acute{d}\lambda}>\right]
\end {eqnarray}
We carry out now an $SU(5)\times U(1)$ decomposition of these and get

\begin{eqnarray}
<\Upsilon^{*}_{(-)\acute{a}\mu}|Bb_i^{\dagger}b_jb_kb_l|\Upsilon_{(+)\acute{b}\mu}>
=i\left\{\sqrt{\frac{2}{15}}{\cal Q}_{\acute{a}}^{kl\bf T}{\cal
P}_{\acute{b}i}^j-\sqrt{\frac{2}{15}}\left(\delta^k_i{\cal
Q}_{\acute{a}}^{nl\bf T}-\delta^l_i{\cal Q}_{\acute{a}}^{nk\bf
T}\right){\cal P}_{\acute{b}n}^j \right.\nonumber\\
\left.+ \sqrt{\frac{2}{15}}{\cal Q}_{\acute{a}}^{lj\bf T}{\cal
P}_{\acute{b}i}^k-\sqrt{\frac{2}{15}}\left(\delta^l_i{\cal
Q}_{\acute{a}}^{nj\bf T}-\delta^j_i{\cal Q}_{\acute{a}}^{nl\bf
T}\right){\cal P}_{\acute{b}n}^k \right.\nonumber\\
\left.+\sqrt{\frac{2}{15}}{\cal Q}_{\acute{a}}^{jk\bf T}{\cal
P}_{\acute{b}i}^l-\sqrt{\frac{2}{15}}\left(\delta^j_i{\cal
Q}_{\acute{a}}^{nk\bf T}-\delta^k_i{\cal Q}_{\acute{a}}^{nj\bf
T}\right){\cal P}_{\acute{b}n}^l \right.\nonumber\\
\left.-{\cal Q}_{\acute{a}m}^{jkl\bf T}{\cal
P}_{\acute{b}i}^{m}+\left( \delta^j_i{\cal Q}_{\acute{a}m}^{nkl\bf
T} +\delta^k_i{\cal Q}_{\acute{a}m}^{nlj\bf T}+\delta^l_i{\cal
Q}_{\acute{a}m}^{njk\bf
T}\right){\cal P}_{\acute{b}n}^{m}\right.\nonumber\\
\left.-\sqrt{\frac{3}{10}}\epsilon^{jklmn}{\cal
Q}_{\acute{a}mi}^{p\bf T}{\cal P}_{\acute{b}np}-\frac{1}{\sqrt
{2}}\epsilon^{jklmn}{\cal Q}_{\acute{a}mi}^{p\bf T}{\cal
P}_{\acute{b}np}^{(S)}\right.\nonumber\\
\left.+\frac{1}{10}\sqrt{\frac{3}{2}}\epsilon^{jklmn}{\cal
Q}_{\acute{a}i}^{\bf T}{\cal P}_{\acute{b}mn}
+\frac{1}{10}\sqrt{\frac{3}{2}}\epsilon^{jklmn}{\cal
Q}_{\acute{a}m}^{\bf T}{\cal
P}_{\acute{b}ni}\right.\nonumber\\
\left.+\frac{1}{2\sqrt {10}}\epsilon^{jklmn}{\cal
Q}_{\acute{a}m}^{\bf T}{\cal P}_{\acute{b}ni}^{(S)}\right\}
\end{eqnarray}

\begin{eqnarray}
<\Upsilon^{*}_{(-)\acute{c}\lambda}|Bb_j^{\dagger}b_k^{\dagger}b_l^{\dagger}b_i
|\Upsilon_{(+)\acute{d}\lambda}> =i\left\{\sqrt{\frac{2}{15}}{\cal
Q}_{\acute{c}j}^{i\bf T}{\cal
P}_{\acute{d}kl}-\sqrt{\frac{2}{15}}{\cal Q}_{\acute{c}j}^{m\bf
T}\left(\delta^i_k{\cal P}_{\acute{d}ml}-\delta^i_l{\cal
P}_{\acute{d}mk}\right)\right.\nonumber\\
\left.+\sqrt{\frac{2}{15}}{\cal Q}_{\acute{c}l}^{i\bf T}{\cal
P}_{\acute{d}jk}-\sqrt{\frac{2}{15}}{\cal Q}_{\acute{c}l}^{m\bf
T}\left(\delta^i_j{\cal P}_{\acute{d}mk}-\delta^i_k{\cal
P}_{\acute{d}mj}\right)\right.\nonumber\\
\left.+\sqrt{\frac{2}{15}}{\cal Q}_{\acute{c}k}^{i\bf T}{\cal
P}_{\acute{d}lj}-\sqrt{\frac{2}{15}}{\cal Q}_{\acute{c}k}^{m\bf
T}\left(\delta^i_l{\cal P}_{\acute{d}mj}-\delta^i_j{\cal
P}_{\acute{d}ml}\right)\right.\nonumber\\
\left.-{\cal Q}_{\acute{c}m}^{i\bf T}{\cal
P}_{\acute{d}jkl}^{m}+{\cal Q}_{\acute{c}n}^{m\bf
T}\left(\delta_j^i{\cal P}_{\acute{d}mkl}^{n}+\delta_k^i{\cal
P}_{\acute{d}mlj}^{n}+\delta_l^i{\cal P}_{\acute{d}mjk}^{n}
\right)\right.\nonumber\\
 \left.+\sqrt{\frac{3}{10}}\epsilon_{jklmn}{\cal
Q}_{\acute{c}}^{mp\bf T}{\cal P}_{\acute{d}p}^{ni}+\frac{1}{\sqrt
{2}}\epsilon_{jklmn}{\cal Q}_{(S)\acute{c}}^{mp\bf T}{\cal
P}_{\acute{d}p}^{ni}\right.\nonumber\\
\left.+\frac{1}{10}\sqrt{\frac{3}{2}}\epsilon_{jklmn}{\cal
Q}_{\acute{c}}^{mn\bf T}{\cal P}_{\acute{d}}^i
-\frac{1}{10}\sqrt{\frac{3}{2}}\epsilon_{jklmn}{\cal
Q}_{\acute{c}}^{mi\bf T}{\cal
P}_{\acute{d}}^n\right.\nonumber\\
\left.-\frac{1}{2\sqrt {10}}\epsilon_{jklmn}{\cal
Q}_{(S)\acute{c}}^{mi\bf T}{\cal P}_{\acute{d}}^{n}\right\}
\end{eqnarray}

\begin{eqnarray}
<\Upsilon^{*}_{(-)\acute{a}\mu}|Bb_i^{\dagger}b_j^{\dagger}b_kb_l
|\Upsilon_{(+)\acute{b}\mu}>= i\left\{{\cal Q}_{\acute{a}ij}^{m\bf
T}{\cal P}_{\acute{b}m}^{kl} +\frac{1}{2\sqrt {5}}{\cal
Q}_{\acute{a}ij}^{k\bf T}{\cal P}_{\acute{b}}^{l}-\frac{1}{2\sqrt
{5}}{\cal Q}_{\acute{a}ij}^{l\bf T}{\cal
P}_{\acute{b}}^{k}\right.\nonumber\\
\left.+\frac{1}{2\sqrt {5}}{\cal Q}_{\acute{a}j}^{\bf T}{\cal
P}_{\acute{b}i}^{kl}-\frac{1}{2\sqrt {5}}{\cal
Q}_{\acute{a}i}^{\bf T}{\cal P}_{\acute{b}j}^{kl}
-\frac{1}{20}\left(\delta^k_j {\cal Q}_{\acute{a}i}^{\bf T}
-\delta^k_i {\cal Q}_{\acute{a}j}^{\bf T}\right){\cal P}_{\acute{b}}^{l}\right.\nonumber\\
\left.-\frac{1}{20}\left(\delta^l_i {\cal Q}_{\acute{a}j}^{\bf T}
-\delta^l_j {\cal Q}_{\acute{a}i}^{\bf T}\right){\cal
P}_{\acute{b}}^{k}+{\cal Q}_{\acute{a}n}^{klm\bf T}{\cal
P}_{\acute{b}ijm}^{n}
\right.\nonumber\\
\left.-\frac{1}{2}{\cal Q}_{\acute{a}p}^{kmn\bf T}\left(
\delta^l_j{\cal P}_{\acute{b}imn}^{p}- \delta^l_i{\cal
P}_{\acute{b}jmn}^{p}\right) -\frac{1}{2}{\cal
Q}_{\acute{a}p}^{lmn\bf T}\left( \delta^k_i{\cal
P}_{\acute{b}jmn}^{p}- \delta^k_j{\cal
P}_{\acute{b}imn}^{p}\right)\right.\nonumber\\
\left.+\left(\delta^l_i\delta^k_j-\delta^k_i\delta^l_j
\right)\left(-\frac{1}{6}{\cal Q}_{\acute{a}q}^{mnp\bf T}{\cal
P}_{\acute{b}mnp}^{q}+\frac{7}{30}{\cal Q}_{\acute{a}}^{mn\bf
T}{\cal P}_{\acute{b}mn}+{\cal Q}_{\acute{a}n}^{m\bf T}{\cal
P}_{\acute{b}m}^{n}+\frac{1}{2}{\cal Q}_{(S)\acute{a}}^{mn\bf
T}{\cal P}_{(S)\acute{b}mn}\right)\right.\nonumber\\
\left.+\sqrt{\frac{2}{15}}{\cal Q}_{\acute{a}i}^{klm\bf T}{\cal
P}_{\acute{b}mj}-\sqrt{\frac{2}{15}}{\cal Q}_{\acute{a}j}^{klm\bf
T}{\cal P}_{\acute{b}mi}\right.\nonumber\\
\left.-\frac{1}{\sqrt{30}}\left( \delta^k_j {\cal
Q}_{\acute{a}i}^{lmn\bf T}-\delta^k_i {\cal
Q}_{\acute{a}j}^{lmn\bf T}-\delta^l_j {\cal
Q}_{\acute{a}i}^{kmn\bf T} +\delta^l_i {\cal
Q}_{\acute{a}j}^{kmn\bf T} \right){\cal
P}_{\acute{b}mn}\right.\nonumber\\
\left.-\sqrt{\frac{2}{15}}{\cal Q}_{\acute{a}}^{lm\bf T}{\cal
P}_{\acute{b}mij}^{k}+\sqrt{\frac{2}{15}}{\cal
Q}_{\acute{a}}^{km\bf T}{\cal P}_{\acute{b}mij}^{l}\right.\nonumber\\
\left.-\frac{1}{\sqrt{30}}{\cal Q}_{\acute{a}}^{mn\bf T}\left(
\delta^l_i{\cal P}_{\acute{b}jmn}^{k}-\delta^l_j{\cal
P}_{\acute{b}imn}^{k}-\delta^k_i{\cal P}_{\acute{b}jmn}^{l}
+\delta^k_j{\cal P}_{\acute{b}imn}^{l} \right)\right.\nonumber\\
\left.+\frac{1}{2}\left(\delta^l_i{\cal Q}_{\acute{a}}^{km\bf T}
-\delta^k_i{\cal Q}_{\acute{a}}^{lm\bf
T}\right)\left(\frac{1}{3}{\cal
P}_{\acute{b}mj}-\sqrt{\frac{3}{5}}{\cal P}_{(S)\acute{b}mj}
\right)\right.\nonumber\\
\left.+\frac{1}{2}\left(\delta^k_j{\cal Q}_{\acute{a}}^{lm\bf T}
-\delta^l_j{\cal Q}_{\acute{a}}^{km\bf
T}\right)\left(\frac{1}{3}{\cal
P}_{\acute{b}mi}-\sqrt{\frac{3}{5}}{\cal P}_{(S)\acute{b}mi}
\right)\right.\nonumber\\
\left.+\frac{1}{2}\left(\delta^l_i{\cal Q}_{(S)\acute{a}}^{km\bf
T} -\delta^k_i{\cal Q}_{(S)\acute{a}}^{lm\bf
T}\right)\left(\sqrt{\frac{3}{5}}{\cal P}_{\acute{b}mj}-{\cal
P}_{(S)\acute{b}mj}
\right)\right.\nonumber\\
\left.+\frac{1}{2}\left(\delta^k_j{\cal Q}_{(S)\acute{a}}^{lm\bf
T} -\delta^l_j{\cal Q}_{(S)\acute{a}}^{km\bf
T}\right)\left(\sqrt{\frac{3}{5}}{\cal P}_{\acute{b}mi}-{\cal
P}_{(S)\acute{b}mi}
\right)\right.\nonumber\\
\left.-\left(\delta^k_j{\cal Q}_{\acute{a}m}^{l\bf T}
-\delta^l_j{\cal Q}_{\acute{a}m}^{k\bf T}\right){\cal
P}_{\acute{b}i}^{m} -\left(\delta^l_i{\cal Q}_{\acute{a}m}^{k\bf
T} -\delta^k_i{\cal Q}_{\acute{a}m}^{l\bf T}\right){\cal
P}_{\acute{b}j}^{m}\right.\nonumber\\
\left.+\frac{2}{15}{\cal Q}_{\acute{a}}^{kl\bf T}{\cal
P}_{\acute{b}ij}\right\}
\end{eqnarray}

\begin{eqnarray}
<\Upsilon^{*}_{(-)\acute{a}\mu}|Bb_ib_jb_kb_l
|\Upsilon_{(+)\acute{b}\mu}>=i\left\{-\sqrt{\frac{3}{10}}\epsilon^{ijklm}{\cal
Q}_{\acute{a}}^{n\bf T}{\cal P}_{\acute{b}nm}
+\frac{1}{\sqrt{2}}\epsilon^{ijklm}{\cal Q}_{\acute{a}}^{n\bf
T}{\cal P}_{\acute{b}nm}^{(S)}\right.\nonumber\\
\left. +\frac{2}{\sqrt 5}\epsilon^{ijklm}{\cal
Q}_{\acute{a}n}^{\bf T}{\cal P}_{\acute{b}m}^n\right\}
\end{eqnarray}

\begin{eqnarray}
<\Upsilon^{*}_{(-)\acute{c}\lambda}|Bb_i^{\dagger}b_j^{\dagger}b_k^{\dagger}b_l^{\dagger}
|\Upsilon_{(+)\acute{d}\lambda}>=
i\left\{\sqrt{\frac{3}{10}}\epsilon_{ijklm}{\cal
Q}_{\acute{c}}^{mn\bf T}{\cal P}_{\acute{d}n}
+\frac{1}{\sqrt{2}}\epsilon_{ijklm}{\cal
Q}_{(S)\acute{c}}^{mn\bf T}{\cal P}_{\acute{d}n}\right.\nonumber\\
\left.+\frac{2}{\sqrt 5}\epsilon_{ijklm}{\cal
Q}_{\acute{c}n}^{m\bf T}{\cal P}_{\acute{d}}^n\right\}
\end{eqnarray}

\begin{eqnarray}
<\Upsilon^{*}_{(-)\acute{c}\lambda}|Bb_j^{\dagger}b_n^{\dagger}b_nb_i
|\Upsilon_{(+)\acute{d}\lambda}>=
 i\left\{{\cal Q}_{\acute{c}jk}^{l\bf T}{\cal
P}_{\acute{d}l}^{ki}-\frac{1}{2\sqrt {5}}{\cal
Q}_{\acute{c}jk}^{i\bf T}{\cal P}_{\acute{d}}^k+\frac{1}{2\sqrt
{5}}{\cal Q}_{\acute{c}k}^{\bf T}{\cal
P}_{\acute{d}j}^{ki}\right.\nonumber\\
\left.-\frac{3}{20}{\cal Q}_{\acute{c}j}^{\bf T}{\cal
P}_{\acute{d}}^i -\frac{1}{20}\delta^i_j{\cal Q}_{\acute{c}k}^{\bf
T}{\cal P}_{\acute{d}}^k +\frac{1}{2}{\cal Q}_{\acute{c}m}^{ikl\bf
T}{\cal
P}_{\acute{d}jkl}^m \right.\nonumber\\
\left.-\frac{1}{6} \delta_j^i{\cal Q}_{\acute{c}n}^{klm\bf T}{\cal
P}_{\acute{d}klm}^n -\frac{1}{\sqrt {30}}{\cal
Q}_{\acute{c}j}^{ikl\bf T}{\cal P}_{\acute{d}}^{kl}
-\frac{1}{\sqrt{30}}{\cal Q}_{\acute{c}}^{kl\bf T}{\cal
P}_{\acute{d}jkl}^i\right.\nonumber\\
 \left.+\frac{19}{30}{\cal
Q}_{\acute{c}}^{ik\bf T}{\cal
P}_{\acute{d}kj}+\frac{23}{30}\delta^i_j{\cal
Q}_{\acute{c}}^{kl\bf T}{\cal P}_{\acute{d}kl}
-\frac{3}{2}\sqrt{\frac{3}{5}}{\cal Q}_{\acute{c}}^{ik\bf T}{\cal
P}_{\acute{d}kj}^{(S)}\right.\nonumber\\
\left.+\frac{3}{2}\sqrt{\frac{3}{5}}{\cal Q}_{(S)\acute{c}}^{ik\bf
T}{\cal P}_{\acute{d}kj}-\frac{3}{2}{\cal Q}_{(S)\acute{c}}^{ik\bf
T}{\cal P}_{\acute{d}kj}^{(S)}+\frac{3}{2}\delta^i_j{\cal
Q}_{(S)\acute{c}}^{kl\bf T}{\cal
P}_{\acute{d}kl}^{(S)}\right.\nonumber\\
\left.-3{\cal Q}_{\acute{c}k}^{i\bf T}{\cal
P}_{\acute{d}j}^{k}+3\delta^i_j{\cal Q}_{\acute{c}k}^{l\bf T}{\cal
P}_{\acute{d}l}^{k}\right\}
\end{eqnarray}

\begin{eqnarray}
<\Upsilon^{*}_{(-)\acute{c}\lambda}|Bb_n^{\dagger}b_nb_ib_j
|\Upsilon_{(+)\acute{d}\lambda}>= i\left\{\sqrt{\frac{2}{15}}{\cal
Q}_{\acute{c}}^{ik\bf T}{\cal P}_{\acute{d}k}^j
 -\sqrt{\frac{2}{15}}{\cal
Q}_{\acute{c}}^{jk\bf T}{\cal P}_{\acute{d}k}^i +{\cal
Q}_{\acute{c}l}^{ijk\bf T}{\cal P}_{\acute{d}k}^l\right.\nonumber\\
\left.+\frac{1}{10}\sqrt{\frac{3}{2}}\epsilon^{ijklm}{\cal
Q}_{\acute{c}k}^{\bf T}{\cal P}_{\acute{d}lm}
+\frac{1}{2}\sqrt{\frac{3}{10}}\epsilon^{ijklm}{\cal
Q}_{\acute{c}kl}^{n\bf T}{\cal P}_{\acute{d}mn}\right.\nonumber\\
\left.+\frac{1}{2\sqrt{2}}\epsilon^{ijklm}{\cal
Q}_{\acute{c}kl}^{n\bf T}{\cal P}^{(S)}_{\acute{d}mn}\right\}
\end{eqnarray}

\begin{eqnarray}
<\Upsilon^{*}_{(-)\acute{c}\lambda}|Bb_i^{\dagger}b_j^{\dagger}b_n^{\dagger}b_n
|\Upsilon_{(+)\acute{d}\lambda}>= i\left\{\sqrt{\frac{2}{15}}{\cal
Q}_{\acute{c}i}^{k\bf T}{\cal P}_{\acute{d}kj}
 -\sqrt{\frac{2}{15}}{\cal
Q}_{\acute{c}j}^{k\bf T}{\cal P}_{\acute{d}ki} +{\cal
Q}_{\acute{c}l}^{k\bf T}{\cal P}_{\acute{d}ijk}^l
\right.\nonumber\\
\left.+\frac{1}{10}\sqrt{\frac{3}{2}}\epsilon_{ijklm}{\cal
Q}_{\acute{c}}^{kl\bf T}{\cal P}_{\acute{d}}^m
+\frac{1}{2}\sqrt{\frac{3}{10}}\epsilon_{ijklm}{\cal
Q}_{\acute{c}}^{kn\bf T}{\cal P}_{\acute{d}n}^{lm}\right.\nonumber\\
\left.+\frac{1}{2\sqrt{2}}\epsilon_{ijklm}{\cal
Q}_{\acute{d}(S)}^{kn\bf T}{\cal P}^{lm}_{\acute{b}n}\right\}
\end{eqnarray}

\begin{eqnarray}
<\Upsilon^{*}_{(-)\acute{c}\lambda}|Bb_m^{\dagger}b_n^{\dagger}b_nb_m
|\Upsilon_{(+)\acute{d}\lambda}>= i\left\{12{\cal
Q}_{\acute{c}j}^{i\bf T}{\cal P}_{\acute{d}i}^j+\frac{16}{5}{\cal
Q}_{\acute{c}}^{ij\bf T}{\cal P}_{\acute{d}ij}+6{\cal
Q}_{(S)\acute{c}}^{ij\bf T}{\cal
P}^{(S)}_{\acute{d}ij}\right.\nonumber\\
\left.-{\cal Q}_{\acute{c}ij}^{k\bf T}{\cal P}_{\acute{d}k}^{ij}
-\frac{2}{5}{\cal Q}_{\acute{c}i}^{\bf T}{\cal
P}^i_{\acute{d}}-\frac{1}{3}{\cal Q}_{\acute{c}l}^{ijk\bf T}{\cal
P}_{\acute{d}ijk}^l\right\}
\end{eqnarray}
where
\begin{eqnarray}
\lambda_{\acute{a}\acute{b},\acute{c}\acute{d}}^{^{(210)}}=
h_{\acute{a}\acute{b}}^{^{(210)}}h_{\acute{c}\acute{d}}^{^{(210)}}k_{_{{\cal
Z}}}^{^{(210)}}\left[\widetilde{{\cal M}}^{^{(210)}}\left\{{\cal
M}^{^{(210)}}\widetilde{{\cal M}}^{^{(210)}} -\bf{1}\right\}
\right]_{{\cal Z}{\cal Z}'}k_{_{{\cal Z}'}}^{^{(210)}}\nonumber\\
\widetilde{{\cal M}}^{^{(210)}}=\left[{\cal
M}^{^{(210)}}+\left({\cal M}^{^{(210)}}\right)^{\bf
{T}}\right]^{-1}
\end{eqnarray}

\subsection{The $\bf{{\left(\overline{144}\times
\overline{144}\right)_{10} \left(\overline{144}\times
\overline{144}\right)_{10}}}$ couplings}

Here we consider the quartic interactions that  arise from mediation by
the  10 plet of Higgs. We begin by considering the superpotential

\begin{eqnarray}
{\mathsf W}^{(10)'}=\frac{1}{2}\Phi_{\nu\cal U}{\cal
M}^{^{(10)}}_{{\cal U}{\cal U}'}\Phi_{\nu{\cal
U}'}+h^{^{(10)}}_{\acute{a}\acute{b}}<\Upsilon^{*}_{(+)\acute{a}\mu}|B\Gamma_{\nu}
|\Upsilon_{(+)\acute{b}\mu}>k_{_{{\cal U}}}^{^{(10)}}\Phi_{\nu\cal
U}\nonumber\\
+\bar{h}^{^{(10)}}_{\acute{a}\acute{b}}<\Upsilon^{*}_{(-)\acute{a}\mu}|B\Gamma_{\nu}
|\Upsilon_{(-)\acute{b}\mu}>\bar{k}_{_{{\cal
U}}}^{^{(10)}}\Phi_{\nu\cal U}
\end{eqnarray}
 Elimination of the $\Phi_{\nu\cal U}$  as a superheavy field
using the F-flatness condition
\begin{equation}
\frac{\partial {\mathsf W}^{(10)'}}{\partial \Phi_{\nu\cal U}}=0
\end{equation}
leads to the quartic interaction generated by 10 mediation.

\begin{eqnarray}
{\mathsf W}^{^{^{{(\overline{144}\times
\overline{144})}_{10}{(\overline{144}\times
\overline{144})}_{10}}}}=
2\lambda_{\acute{a}\acute{b},\acute{c}\acute{d}}^{^{(10)}}
<\Upsilon^{*}_{(+)\acute{a}\mu}|B\Gamma_{\rho}|\Upsilon_{(+)\acute{b}\mu}>
<\Upsilon^{*}_{(+)\acute{c}\nu}|B\Gamma_{\rho}|\Upsilon_{(+)\acute{d}\nu}>\nonumber\\
=8\lambda_{\acute{a}\acute{b},\acute{c}\acute{d}}^{^{(10)}}
<\Upsilon^{*}_{(+)\acute{a}\mu}|Bb_i|\Upsilon_{(+)\acute{b}\mu}>
<\Upsilon^{*}_{(+)\acute{c}\nu}|Bb_i^{\dagger}|\Upsilon_{(+)\acute{d}\nu}>\nonumber\\
=4\lambda_{\acute{a}\acute{b},\acute{c}\acute{d}}^{^{(10)(+)}}
\left(8{\bf P}_{\acute{a}i\mu}^{\bf T}{\bf
P}_{\acute{b}\mu}^{ij}{\bf P}_{\acute{c}j\nu}^{\bf T}{\bf
P}_{\acute{d}\nu}-\epsilon_{jklmn}{\bf P}_{\acute{a}i\mu}^{\bf
T}{\bf P}_{\acute{b}\mu}^{ij}{\bf P}_{\acute{c}\nu}^{kl\bf T}{\bf
P}_{\acute{d}\nu}^{mn} \right)
\end{eqnarray}
where
\begin{equation}
\lambda_{\acute{a}\acute{b},\acute{c}\acute{d}}^{^{(10)(+)}}=
h_{\acute{a}\acute{b}}^{^{(10)(+)}}h_{\acute{c}\acute{d}}^{^{(10)(+)}}k_{_{{\cal
U}}}^{^{(10)}}\left[\widetilde{{\cal M}}^{^{(10)}}\left\{{\cal
M}^{^{(10)}}\widetilde{{\cal M}}^{^{(10)}} -\bf{1}\right\}
\right]_{{\cal U}{\cal U}'}k_{_{{\cal U}'}}^{^{(10)}}
\end{equation}
and
\begin{equation}
\widetilde{{\cal M}}^{^{(10)}}=\left[{\cal
M}^{^{(10)}}+\left({\cal M}^{^{(10)}}\right)^{\bf {T}}\right]^{-1}
\end{equation}

\subsection{The $\bf{{\left({144}\times {144}\right)_{10}
\left({144}\times {144}\right)_{10}}}$ couplings} An analysis
similar to the above gives in this case the following

\begin{eqnarray}
{\mathsf
W}^{^{^{{(\overline{144}\times\overline{144})}_{10}{(\overline{144}\times
\overline{144})}_{10}}}}=
2\bar{\lambda}_{\acute{a}\acute{b},\acute{c}\acute{d}}^{^{(10)}}
<\Upsilon^{*}_{(-)\acute{a}\mu}|B\Gamma_{\rho}|\Upsilon_{(-)\acute{b}\mu}>
<\Upsilon^{*}_{(-)\acute{c}\nu}|B\Gamma_{\rho}|\Upsilon_{(-)\acute{d}\nu}>\nonumber\\
=8\bar{\lambda}_{\acute{a}\acute{b},\acute{c}\acute{d}}^{^{(10)}}
<\Upsilon^{*}_{(-)\acute{a}\mu}|Bb_i|\Upsilon_{(-)\acute{b}\mu}>
<\Upsilon^{*}_{(-)\acute{c}\nu}|Bb_i^{\dagger}|\Upsilon_{(-)\acute{d}\nu}>\nonumber\\
=4\bar{\lambda}_{\acute{a}\acute{b},\acute{c}\acute{d}}^{^{(10)(+)}}
\left(-8{\bf Q}_{\acute{a}\mu}^{i\bf T}{\bf
Q}_{\acute{b}ij\mu}{\bf Q}_{\acute{c}\nu}^{j\bf T}{\bf
Q}_{\acute{d}\nu}+\epsilon^{jklmn}{\bf Q}_{\acute{a}\mu}^{i\bf
T}{\bf Q}_{\acute{b}ij\mu}{\bf Q}_{\acute{c}kl\nu}^{\bf T}{\bf
Q}_{\acute{d}mn\nu}\right)
\end{eqnarray}
where
\begin{equation}
\bar{\lambda}_{\acute{a}\acute{b},\acute{c}\acute{d}}^{^{(10)(+)}}=
\bar{h}_{\acute{a}\acute{b}}^{^{(10)(+)}}\bar{h}_{\acute{c}\acute{d}}^{^{(10)(+)}}\bar{k}_{_{{\cal
U}}}^{^{(10)}}\left[\widetilde{{\cal M}}^{^{(10)}}\left\{{\cal
M}^{^{(10)}}\widetilde{{\cal M}}^{^{(10)}} -\bf{1}\right\}
\right]_{{\cal U}{\cal U}'}\bar{k}_{_{{\cal U}'}}^{^{(10)}}
\end{equation}

\subsection{The $\bf{{\left(\overline{144}\times
\overline{144}\right)_{10} \left({144}\times {144}\right)_{10}}}$
couplings} An analysis similar to above  gives

\begin{eqnarray}
{\mathsf W}^{^{^{{(\overline{144}\times
\overline{144})}_{10}{({144}\times
{144})}_{10}}}}=-2\theta_{\acute{a}\acute{b},\acute{c}\acute{d}}^{^{(10)}}
<\Upsilon^{*}_{(+)\acute{a}\mu}|B\Gamma_{\rho}|\Upsilon_{(+)\acute{b}\mu}><\Upsilon^{*}_{(-)\acute{c}\nu}|B\Gamma_{\rho}
|\Upsilon_{(-)\acute{d}\nu}>\nonumber\\
=-4\theta_{\acute{a}\acute{b},\acute{c}\acute{d}}^{^{(10)}}
\left[<\Upsilon^{*}_{(+)\acute{a}\mu}|Bb_i|\Upsilon_{(+)\acute{b}\mu}><\Upsilon^{*}_{(-)\acute{c}\nu}|Bb_i^{\dagger}
|\Upsilon_{(-)\acute{d}\nu}>\right.\nonumber\\
\left.+<\Upsilon^{*}_{(+)\acute{a}\mu}|Bb_i^{\dagger}|\Upsilon_{(+)\acute{b}\mu}><\Upsilon^{*}_{(-)\acute{c}\nu}|Bb_i
|\Upsilon_{(-)\acute{d}\nu}>\right]\nonumber\\
=2\theta_{\acute{a}\acute{b},\acute{c}\acute{d}}^{^{(10)(+)}}
\left(8{\bf P}_{\acute{a}i\nu}^{\bf T}{\bf
P}_{\acute{b}\nu}^{ij}{\bf Q}_{\acute{c}\mu}^{k\bf T}{\bf
Q}_{\acute{d}kj\mu} -8{\bf P}_{\acute{a}i\nu}^{\bf T}{\bf
P}_{\acute{b}\nu}{\bf
Q}_{\acute{c}\mu}^{i\bf T}{\bf Q}_{\acute{d}\mu}\right.\nonumber\\
 \left.-{\bf P}_{\acute{a}\nu}^{ij\bf T}{\bf P}_{\acute{b}\nu}^{kl}{\bf
Q}_{\acute{c}kl\mu}^{\bf T}{\bf Q}_{\acute{d}ij\mu} +{\bf
P}_{\acute{a}\nu}^{ij\bf T}{\bf P}_{\acute{b}\nu}^{kl}{\bf
Q}_{\acute{c}ik\mu}^{\bf T}{\bf Q}_{\acute{d}jl\mu}\right.\nonumber\\
\left.-{\bf P}_{\acute{a}\nu}^{ij\bf T}{\bf
P}_{\acute{b}\nu}^{kl}{\bf Q}_{\acute{c}il\mu}^{\bf T}{\bf
Q}_{\acute{d}jk\mu}+ \epsilon^{ijklm}{\bf P}_{\acute{a}i\nu}^{\bf
T}{\bf P}_{\acute{b}\nu}{\bf Q}_{\acute{c}jk\mu}^{\bf T}{\bf
Q}_{\acute{d}lm\mu}\right.\nonumber\\
\left.+\epsilon_{ijklm}{\bf P}_{\acute{a}\nu}^{ij\bf T}{\bf
P}_{\acute{b}\nu}^{kl}{\bf Q}_{\acute{c}\mu}^{m\bf T}{\bf
Q}_{\acute{d}\mu}\right)
\end{eqnarray}
where
\begin{equation}
\theta_{\acute{a}\acute{b},\acute{c}\acute{d}}^{^{(10)(+)}}=
h_{\acute{a}\acute{b}}^{^{(10)(+)}}\bar{h}_{\acute{c}\acute{d}}^{^{(10)(+)}}k_{_{{\cal
U}}}^{^{(10)}}\widetilde{{\cal M}}^{^{(10)}}_{{\cal U}{\cal
U}'}\bar{k}_{_{{\cal U}'}}^{^{(10)}}
\end{equation}
Further details of  the decomposition of the  couplings generated by 10 mediation are given
in Appendix A.

\subsection{The $\bf{{\left(\overline{144}\times
\overline{144}\right)_{120} \left(\overline{144}\times
\overline{144}\right)_{120}}}$ couplings}

 We begin by considering
the  superpotenial
\begin{eqnarray}
{\mathsf W}^{(120)'}=\frac{1}{2}\Phi_{\nu\rho\lambda V}{\cal
M}^{^{(120)}}_{V V'}\Phi_{\rho\nu\lambda V'}
+\frac{1}{3!}h^{^{(120)}}_{\acute{a}\acute{b}}<\Upsilon^{*}_{(+)\acute{a}\mu}|B\Gamma_{[\nu}
\Gamma_{\rho}\Gamma_{\lambda
]}|\Upsilon_{(+)\acute{b}\mu}>k_{_{V}}^{^{(120)}}
\Phi_{\nu\rho\lambda V}\nonumber\\
+\frac{1}{3!}{\bar
h}^{^{(120)}}_{\acute{a}\acute{b}}<\Upsilon^{*}_{(-)\acute{a}\mu}|B\Gamma_{[\nu}
\Gamma_{\rho}\Gamma_{\lambda ]}|\Upsilon_{(-)\acute{b}\mu}>{\bar
k}_{_{V}}^{^{(120)}} \Phi_{\nu\rho\lambda V}
\end{eqnarray}

Eliminating $\Phi_{\nu\rho\lambda V}$ using the F flatness condition
\begin{equation}
\frac{\partial {\mathsf W}^{(120)'}}{\partial \Phi_{\nu\rho\lambda
V}}=0
\end{equation}
we obtain

\begin{eqnarray}
{\mathsf W}^{^{^{{(\overline{144}\times
\overline{144})}_{120}{(\overline{144}\times
\overline{144})}_{120}}}}=
\frac{1}{18}\lambda_{\acute{a}\acute{b},\acute{c}\acute{d}}^{^{(120)}}
<\Upsilon^{*}_{(+)\acute{a}\mu}|B\Gamma_{[\nu}
\Gamma_{\rho}\Gamma_{\lambda
]}|\Upsilon_{(+)\acute{b}\mu}><\Upsilon^{*}_{(+)\acute{c}\nu}|B\Gamma_{[\nu}
\Gamma_{\rho}\Gamma_{\lambda
]}|\Upsilon_{(+)\acute{d}\nu}>\nonumber\\
=\frac{1}{18}\lambda_{\acute{a}\acute{b},\acute{c}\acute{d}}^{^{(120)}}
\left[<\Upsilon^{*}_{(+)\acute{a}\mu}|B\Gamma_{\nu}
\Gamma_{\rho}\Gamma_{\lambda
}|\Upsilon_{(+)\acute{b}\mu}><\Upsilon^{*}_{(+)\acute{c}\nu}|B\Gamma_{\nu}
\Gamma_{\rho}\Gamma_{\lambda
}|\Upsilon_{(+)\acute{d}\nu}>\right.~~\nonumber\\
\left.-28<\Upsilon^{*}_{(+)\acute{a}\mu}|B\Gamma_{\nu}
|\Upsilon_{(+)\acute{b}\mu}><\Upsilon^{*}_{(+)\acute{c}\nu}|B\Gamma_{\nu}
|\Upsilon_{(+)\acute{d}\nu}>\right]~~~~~~~~\nonumber\\
=\frac{8}{9}\lambda_{\acute{a}\acute{b},\acute{c}\acute{d}}^{^{(120)}}
\left[<\Upsilon^{*}_{(+)\acute{a}\mu}|Bb_ib_jb_k|\Upsilon_{(+)\acute{b}\mu}><\Upsilon^{*}_{(+)\acute{c}\nu}|B
b_i^{\dagger}b_j^{\dagger}b_k^{\dagger}|\Upsilon_{(+)\acute{d}\nu}>\right.~~~~~~~~\nonumber\\
\left.-6<\Upsilon^{*}_{(+)\acute{a}\mu}|Bb_i|\Upsilon_{(+)\acute{b}\mu}><\Upsilon^{*}_{(+)\acute{c}\nu}|B
b_i^{\dagger}|\Upsilon_{(+)\acute{d}\nu}>\right.~~~~~~~~\nonumber\\
\left.+3<\Upsilon^{*}_{(+)\acute{a}\mu}|Bb_i^{\dagger}|\Upsilon_{(+)\acute{b}\mu}><\Upsilon^{*}_{(+)\acute{c}\nu}|B
b_n^{\dagger}b_nb_i|\Upsilon_{(+)\acute{d}\nu}>\right.~~~~~~~~\nonumber\\
\left.+3<\Upsilon^{*}_{(+)\acute{a}\mu}|Bb_i|\Upsilon_{(+)\acute{b}\mu}><\Upsilon^{*}_{(+)\acute{c}\nu}|B
b_i^{\dagger}b_n^{\dagger}b_n|\Upsilon_{(+)\acute{d}\nu}>\right.~~~~~~~~\nonumber\\
\left.+3<\Upsilon^{*}_{(+)\acute{a}\mu}|Bb_i^{\dagger}b_j^{\dagger}b_k|\Upsilon_{(+)\acute{b}\mu}><\Upsilon^{*}_{(+)\acute{c}\nu}|B
b_k^{\dagger}b_ib_j|\Upsilon_{(+)\acute{d}\nu}>\right]~~~~~~~~\nonumber\\
=\frac{8}{3}\lambda_{\acute{a}\acute{b},\acute{c}\acute{d}}^{^{(120)(-)}}
\left[-4{\bf P}_{\acute{a}i\mu}^{\bf T}{\bf P}_{\acute{b}j\mu}{\bf
P}_{\acute{c}\nu}^{ij\bf T}{\bf P}_{\acute{d}\nu} +4{\bf
P}_{\acute{a}\mu}^{\bf T}{\bf P}_{\acute{b}i\mu}{\bf
P}_{\acute{c}\nu}^{ij\bf T}{\bf P}_{\acute{d}j\nu}\right.~~~~~~~~~~~~~~~~~~~\nonumber\\
\left.+\epsilon_{ijklm}{\bf P}_{\acute{a}\mu}^{ij\bf T}{\bf
P}_{\acute{b}\mu}^{kn}{\bf P}_{\acute{c}n\nu}^{\bf T}{\bf
P}_{\acute{d}\nu}^{lm}\right]~~~~~~~~
\end{eqnarray}
where
\begin{equation}
\lambda_{\acute{a}\acute{b},\acute{c}\acute{d}}^{^{(120)(-)}}=
h_{\acute{a}\acute{b}}^{^{(120)(-)}}h_{\acute{c}\acute{d}}^{^{(120)(-)}}k_{_{V}}^{^{(120)}}\left[\widetilde{{\cal
M}} ^{^{(120)}}\left\{{\cal M}^{^{(120)}}\widetilde{{\cal
M}}^{^{(120)}} -\bf{1}\right\} \right]_{VV'}k_{_{V'}}^{^{(120)}}
\end{equation}
and
\begin{equation}
\widetilde{{\cal M}}^{^{(120)}}=\left[{\cal
M}^{^{(120)}}+\left({\cal M}^{^{(120)}}\right)^{\bf
{T}}\right]^{-1}
\end{equation}

\subsection{The $\bf{{\left({144}\times {144}\right)_{120}
\left({144}\times {144}\right)_{120}}}$ couplings} An analysis
similar to the above gives

\begin{eqnarray}
{\mathsf W}^{^{^{{({144}\times {144})}_{120}{({144}\times
{144})}_{120}}}}=
\frac{1}{18}\bar{\lambda}_{\acute{a}\acute{b},\acute{c}\acute{d}}^{^{(120)}}
<\Upsilon^{*}_{(-)\acute{a}\mu}|B\Gamma_{[\nu}
\Gamma_{\rho}\Gamma_{\lambda
]}|\Upsilon_{(-)\acute{b}\mu}><\Upsilon^{*}_{(-)\acute{c}\nu}|B\Gamma_{[\nu}
\Gamma_{\rho}\Gamma_{\lambda
]}|\Upsilon_{(-)\acute{d}\nu}>\nonumber\\
=\frac{1}{18}\bar{\lambda}_{\acute{a}\acute{b},\acute{c}\acute{d}}^{^{(120)}}
\left[<\Upsilon^{*}_{(-)\acute{a}\mu}|B\Gamma_{\nu}
\Gamma_{\rho}\Gamma_{\lambda
}|\Upsilon_{(-)\acute{b}\mu}><\Upsilon^{*}_{(-)\acute{c}\nu}|B\Gamma_{\nu}
\Gamma_{\rho}\Gamma_{\lambda
}|\Upsilon_{(-)\acute{d}\nu}>\right.~~\nonumber\\
\left.-28<\Upsilon^{*}_{(-)\acute{a}\mu}|B\Gamma_{\nu}
|\Upsilon_{(-)\acute{b}\mu}><\Upsilon^{*}_{(-)\acute{c}\nu}|B\Gamma_{\nu}
|\Upsilon_{(-)\acute{d}\nu}>\right]~~~~~~~~\nonumber\\
=\frac{8}{9}\bar{\lambda}_{\acute{a}\acute{b},\acute{c}\acute{d}}^{^{(120)}}
\left[<\Upsilon^{*}_{(-)\acute{a}\mu}|Bb_ib_jb_k|\Upsilon_{(-)\acute{b}\mu}><\Upsilon^{*}_{(-)\acute{c}\nu}|B
b_i^{\dagger}b_j^{\dagger}b_k^{\dagger}|\Upsilon_{(-)\acute{d}\nu}>\right.~~~~~~~~\nonumber\\
\left.-6<\Upsilon^{*}_{(-)\acute{a}\mu}|Bb_i|\Upsilon_{(-)\acute{b}\mu}><\Upsilon^{*}_{(-)\acute{c}\nu}|B
b_i^{\dagger}|\Upsilon_{(-)\acute{d}\nu}>\right.~~~~~~~~\nonumber\\
\left.+3<\Upsilon^{*}_{(-)\acute{a}\mu}|Bb_i^{\dagger}|\Upsilon_{(-)\acute{b}\mu}><\Upsilon^{*}_{(-)\acute{c}\nu}|B
b_n^{\dagger}b_nb_i|\Upsilon_{(-)\acute{d}\nu}>\right.~~~~~~~~\nonumber\\
\left.+3<\Upsilon^{*}_{(-)\acute{a}\mu}|Bb_i|\Upsilon_{(-)\acute{b}\mu}><\Upsilon^{*}_{(-)\acute{c}\nu}|B
b_i^{\dagger}b_n^{\dagger}b_n|\Upsilon_{(-)\acute{d}\nu}>\right.~~~~~~~~\nonumber\\
\left.+3<\Upsilon^{*}_{(-)\acute{a}\mu}|Bb_i^{\dagger}b_j^{\dagger}b_k|\Upsilon_{(-)\acute{b}\mu}><\Upsilon^{*}_{(-)\acute{c}\nu}|B
b_k^{\dagger}b_ib_j|\Upsilon_{(-)\acute{d}\nu}>\right]~~~~~~~~\nonumber\\
=\frac{8}{3}\bar{\lambda}_{\acute{a}\acute{b},\acute{c}\acute{d}}^{^{(120)(-)}}
\left[4{\bf Q}_{\acute{a}i\mu}^{\bf T}{\bf Q}_{\acute{b}j\mu}{\bf
Q}_{\acute{c}\nu}^{ij\bf T}{\bf Q}_{\acute{d}\nu} -4{\bf
Q}_{\acute{a}\mu}^{\bf T}{\bf Q}_{\acute{b}i\mu}{\bf
Q}_{\acute{c}\nu}^{ij\bf T}{\bf Q}_{\acute{d}j\nu}\right.~~~~~~~~~~~~~~~~~~~\nonumber\\
\left.-\epsilon_{ijklm}{\bf Q}_{\acute{a}\mu}^{ij\bf T}{\bf
Q}_{\acute{b}\mu}^{kn}{\bf Q}_{\acute{c}n\nu}^{\bf T}{\bf
Q}_{\acute{d}\nu}^{lm}\right]~~~~~~~~
\end{eqnarray}
where
\begin{equation}
\bar{\lambda}_{\acute{a}\acute{b},\acute{c}\acute{d}}^{^{(120)(-)}}=
\bar{h}_{\acute{a}\acute{b}}^{^{(120)(-)}}\bar{h}_{\acute{c}\acute{d}}^{^{(120)(-)}}\bar{k}_{_{V}}^{^{(120)}}
\left[\widetilde{{\cal M}}^{^{(120)}}\left\{{\cal
M}^{^{(120)}}\widetilde{{\cal M}}^{^{(120)}} -\bf{1}\right\}
\right]_{VV'}\bar{k}_{_{V'}}^{^{(120)}}
\end{equation}

\subsection{The $\bf{{\left(\overline{144}\times
\overline{144}\right)_{120} \left({144}\times
{144}\right)_{120}}}$ couplings}

Starting with cubic  couplings involving the 120-plet of fields
and following the same procedure as above one gets the following

\begin{eqnarray}
{\mathsf W}^{^{^{{(\overline{144}\times
\overline{144})}_{120}{(144\times144)}_{120}}}}=
-\frac{1}{18}\theta_{\acute{a}\acute{b},\acute{c}\acute{d}}^{^{(120)}}
<\Upsilon^{*}_{(+)\acute{a}\mu}|B\Gamma_{[\nu}
\Gamma_{\rho}\Gamma_{\lambda
]}|\Upsilon_{(+)\acute{b}\mu}><\Upsilon^{*}_{(-)\acute{c}\nu}|B\Gamma_{[\nu}
\Gamma_{\rho}\Gamma_{\lambda
]}|\Upsilon_{(-)\acute{d}\nu}>\nonumber\\
=-\frac{1}{18}\theta_{\acute{a}\acute{b},\acute{c}\acute{d}}^{^{(120)}}
\left[<\Upsilon^{*}_{(+)\acute{a}\mu}|B\Gamma_{\nu}
\Gamma_{\rho}\Gamma_{\lambda
}|\Upsilon_{(+)\acute{b}\mu}><\Upsilon^{*}_{(-)\acute{c}\nu}|B\Gamma_{\nu}
\Gamma_{\rho}\Gamma_{\lambda
}|\Upsilon_{(-)\acute{d}\nu}>\right.~~\nonumber\\
\left.-28<\Upsilon^{*}_{(+)\acute{a}\mu}|B\Gamma_{\nu}
|\Upsilon_{(+)\acute{b}\mu}><\Upsilon^{*}_{(-)\acute{c}\nu}|B\Gamma_{\nu}
|\Upsilon_{(-)\acute{d}\nu}>\right]~~~~~~~~\nonumber\\
=-\frac{4}{9}\theta_{\acute{a}\acute{b},\acute{c}\acute{d}}^{^{(120)}}
\left[<\Upsilon^{*}_{(+)\acute{a}\mu}|Bb_ib_jb_k|\Upsilon_{(+)\acute{b}\mu}><\Upsilon^{*}_{(-)\acute{c}\nu}|B
b_i^{\dagger}b_j^{\dagger}b_k^{\dagger}|\Upsilon_{(-)\acute{d}\nu}>\right.~~~~~~~~\nonumber\\
\left.-6<\Upsilon^{*}_{(+)\acute{a}\mu}|Bb_i|\Upsilon_{(+)\acute{b}\mu}><\Upsilon^{*}_{(-)\acute{c}\nu}|B
b_i^{\dagger}|\Upsilon_{(-)\acute{d}\nu}>\right.~~~~~~~~\nonumber\\
\left.+3<\Upsilon^{*}_{(+)\acute{a}\mu}|Bb_i^{\dagger}|\Upsilon_{(+)\acute{b}\mu}><\Upsilon^{*}_{(-)\acute{c}\nu}|B
b_n^{\dagger}b_nb_i|\Upsilon_{(-)\acute{d}\nu}>\right.~~~~~~~~\nonumber\\
\left.+3<\Upsilon^{*}_{(+)\acute{a}\mu}|Bb_i|\Upsilon_{(+)\acute{b}\mu}><\Upsilon^{*}_{(-)\acute{c}\nu}|B
b_i^{\dagger}b_n^{\dagger}b_n|\Upsilon_{(-)\acute{d}\nu}>\right.~~~~~~~~\nonumber\\
\left.+3<\Upsilon^{*}_{(+)\acute{a}\mu}|Bb_i^{\dagger}b_j^{\dagger}b_k|\Upsilon_{(+)\acute{b}\mu}>
<\Upsilon^{*}_{(-)\acute{c}\nu}|B
b_k^{\dagger}b_ib_j|\Upsilon_{(-)\acute{d}\nu}>\right.~~~~~~~~\nonumber\\
+\left.<\Upsilon^{*}_{(-)\acute{a}\nu}|Bb_ib_jb_k|\Upsilon_{(-)\acute{b}\nu}><\Upsilon^{*}_{(+)\acute{c}\mu}|B
b_i^{\dagger}b_j^{\dagger}b_k^{\dagger}|\Upsilon_{(+)\acute{d}\mu}>\right.~~~~~~~~\nonumber\\
\left.-6<\Upsilon^{*}_{(-)\acute{a}\nu}|Bb_i|\Upsilon_{(-)\acute{b}\nu}><\Upsilon^{*}_{(+)\acute{c}\mu}|B
b_i^{\dagger}|\Upsilon_{(+)\acute{d}\mu}>\right.~~~~~~~~\nonumber\\
\left.+3<\Upsilon^{*}_{(-)\acute{a}\nu}|Bb_i^{\dagger}|\Upsilon_{(-)\acute{b}\nu}><\Upsilon^{*}_{(+)\acute{c}\mu}|B
b_n^{\dagger}b_nb_i|\Upsilon_{(+)\acute{d}\mu}>\right.~~~~~~~~\nonumber\\
\left.+3<\Upsilon^{*}_{(-)\acute{a}\nu}|Bb_i|\Upsilon_{(-)\acute{b}\nu}><\Upsilon^{*}_{(+)\acute{c}\mu}|B
b_i^{\dagger}b_n^{\dagger}b_n|\Upsilon_{(+)\acute{d}\mu}>\right.~~~~~~~~\nonumber\\
\left.+3<\Upsilon^{*}_{(-)\acute{a}\nu}|Bb_i^{\dagger}b_j^{\dagger}b_k|\Upsilon_{(-)\acute{b}\nu}><\Upsilon^{*}_{(+)\acute{c}\mu}|B
b_k^{\dagger}b_ib_j|\Upsilon_{(+)\acute{d}\mu}>\right]~~~~~~~~\nonumber\\
=\frac{4}{3}\theta_{\acute{a}\acute{b},\acute{c}\acute{d}}^{^{(120)(-)}}
\left[4{\bf P}_{\acute{a}\nu}^{ij\bf T}{\bf P}_{\acute{b}\nu}{\bf
Q}_{\acute{c}ij\mu}^{\bf T}{\bf Q}_{\acute{d}\mu} -4{\bf
P}_{\acute{a}i\nu}^{\bf T}{\bf P}_{\acute{b}j\nu}{\bf
Q}_{\acute{c}\mu}^{i\bf T}{\bf Q}_{\acute{d}\mu}^j\right.~~~~~~~~~~~~~~~~~~~~~~~~\nonumber\\
\left.-4{\bf P}_{\acute{a}\nu}^{ij\bf T}{\bf
P}_{\acute{b}\nu}^{kl}{\bf Q}_{\acute{c}ij\mu}^{\bf T}{\bf
Q}_{\acute{d}kl\mu}-8{\bf P}_{\acute{a}\nu}^{ij\bf T}{\bf
P}_{\acute{b}\nu}^{kl}{\bf
Q}_{\acute{c}jk\mu}^{\bf T}{\bf Q}_{\acute{d}il\mu}\right.~~~~~~~~\nonumber\\
\left.+\epsilon_{ijklm}{\bf P}_{\acute{a}\nu}^{ij\bf T}{\bf
P}_{\acute{b}\nu}^{kl}{\bf Q}_{\acute{c}\mu}^{\bf T}{\bf
Q}_{\acute{d}\mu}^m-4\epsilon^{ijklm}{\bf P}_{\acute{a}\nu}^{\bf
T}{\bf P}_{\acute{b}i\nu}{\bf Q}_{\acute{c}jk\mu}^{\bf T}{\bf
Q}_{\acute{d}lm\mu}\right.~~~~~~~~\nonumber\\
\left.+8{\bf P}_{\acute{a}\nu}^{\bf T}{\bf P}_{\acute{b}i\nu}{\bf
Q}_{\acute{c}\mu}^{\bf T}{\bf Q}_{\acute{d}\mu}^i +4{\bf
P}_{\acute{a}i\nu}^{\bf T}{\bf P}_{\acute{b}\nu}^{jk}{\bf
Q}_{\acute{c}\mu}^{i\bf T}{\bf Q}_{\acute{d}jk\mu}\right.~~~~~~~~\nonumber\\
\left.-4{\bf P}_{\acute{a}i\nu}^{\bf T}{\bf
P}_{\acute{b}\nu}^{ij}{\bf Q}_{\acute{c}\mu}^{k\bf T}{\bf
Q}_{\acute{d}kj\mu}\right]~~~~~~~~
\end{eqnarray}
where
\begin{equation}
\theta_{\acute{a}\acute{b},\acute{c}\acute{d}}^{^{(120)(-)}}=
h_{\acute{a}\acute{b}}^{^{(120)(-)}}\bar{h}_{\acute{c}\acute{d}}^{^{(120)(-)}}k_{_{V}}^{^{(120)}}\widetilde{{\cal
M}}^{^{(120)}}_{VV'}\bar{k}_{_{{\cal U}'}}^{^{(120)}}
\end{equation}
The $SU(5)\times U(1)$ decomposition of the quartic couplings can be carried out using
the results  given in Appendix B.

\subsection{The $\bf{{\left(\overline{144}\times\overline{144}\right)_{\overline{126}}
\left(144\times144\right)_{126}}}$ couplings}

Here we begin by considering the superpotential
\begin{eqnarray}
{\mathsf
W}^{(126,\overline{126})'}=\frac{1}{2}\Phi_{\nu\rho\sigma\lambda\vartheta
\cal{W}}{\cal M}^{^{(126,\overline{126})}}_{\cal{W}
\cal{W}'}\overline{\Phi}_{\rho\nu\sigma\lambda\vartheta
\cal{W}'}\nonumber\\
+\frac{1}{5!}h^{^{(\overline{126})}}_{\acute{a}\acute{b}}<\Upsilon^{*}_{(+)\acute{a}\mu}|B\Gamma_{[\nu}
\Gamma_{\rho}\Gamma_{\sigma}\Gamma_{\lambda}\Gamma_{\vartheta]}|\Upsilon_{(+)\acute{b}\mu}>k_{_{\cal{W}}}
^{^{(\overline{126})}}
\overline{\Phi}_{\nu\rho\sigma\lambda\vartheta
\cal{W}}\nonumber\\
+\frac{1}{5!}{\bar
h}^{^{(126)}}_{\acute{a}\acute{b}}<\Upsilon^{*}_{(-)\acute{a}\mu}|B\Gamma_{[\nu}
\Gamma_{\rho}\Gamma_{\sigma}\Gamma_{\lambda}\Gamma_{\vartheta]}|\Upsilon_{(-)\acute{b}\mu}>{\bar
k}_{_{\cal{W}}}^{^{(126)}}
\Phi_{\nu\rho\sigma\lambda\vartheta\cal{W}}\nonumber\\
+\frac{1}{5!}{\bar
h}^{^{(126)}}_{\acute{a}\acute{b}}<\Upsilon^{*}_{(-)\acute{a}\mu}|B\Gamma_{[\nu}
\Gamma_{\rho}\Gamma_{\sigma}\Gamma_{\lambda}\Gamma_{\vartheta]}|\Upsilon_{(-)\acute{b}\mu}>{\bar
k}_{_{\cal{W}}}^{^{(126)}}
\Phi_{\nu\rho\sigma\lambda\vartheta\cal{W}}
\end{eqnarray}

Eliminating $\Phi_{\nu\rho\sigma\lambda\vartheta\cal{W}}$,
$\overline{\Phi}_{\nu\rho\sigma\lambda\vartheta\cal{W}}$
  through the F flatness conditions
\begin{equation}
\frac{\partial {\mathsf W}^{(126,\overline{126})'}}{\partial
\Phi_{\nu\rho\sigma\lambda\vartheta\cal{W}}}=0,~~~~~\frac{\partial
{\mathsf W}^{(126,\overline{126})'}}{\partial
\overline{\Phi}_{\nu\rho\sigma\lambda\vartheta\cal{W}}}=0
\end{equation}
gives the quartic interaction below

\begin{eqnarray}
{\mathsf
W}^{^{^{{(\overline{144}\times\overline{144})}_{\overline{126}}{(144\times144)}_{126}}}}=
\frac{1}{7200}\kappa_{\acute{a}\acute{b},\acute{c}\acute{d}}^{^{(126,\overline{126})}}
<\Upsilon^{*}_{(+)\acute{a}\mu}|B\Gamma_{[\nu}
\Gamma_{\rho}\Gamma_{\sigma}\Gamma_{\lambda}\Gamma_{\vartheta]}|\Upsilon_{(+)\acute{b}\mu}>\nonumber\\
\times<\Upsilon^{*}_{(-)\acute{c}\tau}|B\Gamma_{[\nu}
\Gamma_{\rho}\Gamma_{\sigma}\Gamma_{\lambda}\Gamma_{\vartheta]}|\Upsilon_{(-)\acute{d}\tau}>\nonumber\\
=
\frac{2}{15}\kappa_{\acute{a}\acute{b},\acute{c}\acute{d}}^{^{(126,\overline{126})(+)}}
\left[2{\bf P}_{\acute{a}\mu}^{\bf T}{\bf P}_{\acute{b}\mu}{\bf
Q}_{\acute{c}\nu}^{\bf T}{\bf Q}_{\acute{d}\nu}-2{\bf
P}_{\acute{a}\mu}^{\bf T}{\bf P}_{\acute{b}\mu}^{ij}{\bf
Q}_{\acute{c}\nu}^{\bf T}{\bf Q}_{\acute{d}ij\nu}\right.\nonumber\\
\left.+2{\bf P}_{\acute{a}i\mu}^{\bf T}{\bf P}_{\acute{b}j\mu}{\bf
Q}_{\acute{c}\nu}^{i\bf T}{\bf Q}_{\acute{d}\nu}^{j}+48{\bf
P}_{\acute{a}\mu}^{\bf T}{\bf P}_{\acute{b}k\mu}{\bf
Q}_{\acute{c}\nu}^{\bf
T}{\bf Q}_{\acute{d}\nu}^{k}\right.\nonumber\\
\left.-2{\bf P}_{\acute{a}\mu}^{ij\bf T}{\bf
P}_{\acute{b}k\mu}{\bf Q}_{\acute{c}ij\nu}^{\bf T}{\bf
Q}_{\acute{d}\nu}^{k}+{\bf P}_{\acute{a}\mu}^{ij\bf T}{\bf
P}_{\acute{b}j\mu}{\bf
Q}_{\acute{c}ik\nu}^{\bf T}{\bf Q}_{\acute{d}\nu}^{k}\right.\nonumber\\
\left.+6{\bf P}_{\acute{a}\mu}^{ij\bf T}{\bf
P}_{\acute{b}\mu}^{kl}{\bf Q}_{\acute{c}ij\nu}^{\bf T}{\bf
Q}_{\acute{d}kl\nu} -30{\bf P}_{\acute{a}\mu}^{ij\bf T}{\bf
P}_{\acute{b}\mu}^{kl}{\bf
Q}_{\acute{c}ik\nu}^{\bf T}{\bf Q}_{\acute{d}jl\nu}\right.\nonumber\\
\left.+9{\bf P}_{\acute{a}\mu}^{ij\bf T}{\bf
P}_{\acute{b}\mu}^{kl}{\bf Q}_{\acute{c}il\nu}^{\bf T}{\bf
Q}_{\acute{d}jk\nu}+2\epsilon^{ijklm} {\bf P}_{\acute{a}\mu}^{\bf
T}{\bf P}_{\acute{b}i\mu}{\bf
Q}_{\acute{c}jk\nu}^{\bf T}{\bf Q}_{\acute{d}lm\nu}\right.\nonumber\\
\left.+2\epsilon_{ijklm} {\bf P}_{\acute{a}\mu}^{ij\bf T}{\bf
P}_{\acute{b}\mu}^{kl}{\bf Q}_{\acute{c}\nu}^{\bf T}{\bf
Q}_{\acute{d}\nu}^m\right]
\end{eqnarray}
where
\begin{equation}
\kappa_{\acute{a}\acute{b},\acute{c}\acute{d}}^{^{(126,\overline{126})(+)}}=
h^{^{(\overline{126})(+)}}_{\acute{a}\acute{b}}\bar{h}^{^{(126)(+)}}_{\acute{c}\acute{d}}
\bar{k}_{_{\cal W}}^{^{(126)}} \widehat{{\cal
M}}^{^{(126,\overline{126})}} _{\cal {W}\cal{W}'}{k}_{_{\cal
W'}}^{^{(\overline{126})}}
\end{equation}
and
\begin{equation}
\widehat{{\cal M}}^{^{(126,\overline{126})}}=\left({\cal
M}^{^{(126,\overline{126})}}\right)^{-1}\left[\left({\cal
M}^{^{(126,\overline{126})}}\right)^{\bf T}\left({\cal
M}^{^{(126,\overline{126})}}\right)^{-1} -2\cdot\bf{1}\right]
\end{equation}

\subsection{The $\bf{{\left(\overline{144}\times\overline{144}\right)_{\overline{126}}
\left(\overline{16}\times \overline{16}\right)_{126}}}$ couplings}
The analysis here follows a very similar approach as above and one
gets

\begin{eqnarray}
{\mathsf
W}^{^{^{{(\overline{144}\times\overline{144})}_{\overline{126}}{(\overline{16}\times\overline{16})}_{126}}}}=
\frac{1}{7200}\varsigma_{\acute{a}\acute{b},\acute{c}\acute{d}}^{^{(126,\overline{126})}}
<\Upsilon^{*}_{(+)\acute{a}\mu}|B\Gamma_{[\nu}
\Gamma_{\rho}\Gamma_{\sigma}\Gamma_{\lambda}\Gamma_{\vartheta]}|\Upsilon_{(+)\acute{b}\mu}>\nonumber\\
\times<\Psi^{*}_{(-)\acute{c}}|B\Gamma_{[\nu}
\Gamma_{\rho}\Gamma_{\sigma}\Gamma_{\lambda}\Gamma_{\vartheta]}|\Psi_{(-)\acute{d}}>\nonumber\\
=
\frac{2}{15}\varsigma_{\acute{a}\acute{b},\acute{c}\acute{d}}^{^{(126,\overline{126})(+)}}
\left[2{\bf P}_{\acute{a}\mu}^{\bf T}{\bf P}_{\acute{b}\mu}{\bf
N}_{\acute{c}}^{\bf T}{\bf N}_{\acute{d}}-2{\bf
P}_{\acute{a}\mu}^{\bf T}{\bf P}_{\acute{b}\mu}^{ij}{\bf
N}_{\acute{c}}^{\bf T}{\bf N}_{\acute{d}ij}\right.\nonumber\\
\left.+2{\bf P}_{\acute{a}i\mu}^{\bf T}{\bf P}_{\acute{b}j\mu}{\bf
N}_{\acute{c}}^{i\bf T}{\bf N}_{\acute{d}}^{j}+48{\bf
P}_{\acute{a}\mu}^{\bf T}{\bf P}_{\acute{b}k\mu}{\bf
N}_{\acute{c}}^{\bf
T}{\bf N}_{\acute{d}}^{k}\right.\nonumber\\
\left.-2{\bf P}_{\acute{a}\mu}^{ij\bf T}{\bf
P}_{\acute{b}k\mu}{\bf N}_{\acute{c}ij}^{\bf T}{\bf
N}_{\acute{d}}^{k}+{\bf P}_{\acute{a}\mu}^{ij\bf T}{\bf
P}_{\acute{b}j\mu}{\bf
N}_{\acute{c}ik}^{\bf T}{\bf N}_{\acute{d}}^{k}\right.\nonumber\\
\left.+6{\bf P}_{\acute{a}\mu}^{ij\bf T}{\bf
P}_{\acute{b}\mu}^{kl}{\bf N}_{\acute{c}ij}^{\bf T}{\bf
N}_{\acute{d}kl} -30{\bf P}_{\acute{a}\mu}^{ij\bf T}{\bf
P}_{\acute{b}\mu}^{kl}{\bf
N}_{\acute{c}ik}^{\bf T}{\bf N}_{\acute{d}jl}\right.\nonumber\\
\left.+9{\bf P}_{\acute{a}\mu}^{ij\bf T}{\bf
P}_{\acute{b}\mu}^{kl}{\bf N}_{\acute{c}il}^{\bf T}{\bf
N}_{\acute{d}jk}+2\epsilon^{ijklm} {\bf P}_{\acute{a}\mu}^{\bf
T}{\bf P}_{\acute{b}i\mu}{\bf
N}_{\acute{c}jk}^{\bf T}{\bf N}_{\acute{d}lm}\right.\nonumber\\
\left.+2\epsilon_{ijklm} {\bf P}_{\acute{a}\mu}^{ij\bf T}{\bf
P}_{\acute{b}\mu}^{kl}{\bf N}_{\acute{c}}^{\bf T}{\bf
N}_{\acute{d}}^m\right]
\end{eqnarray}
where
\begin{eqnarray}
\varsigma_{\acute{a}\acute{b},\acute{c}\acute{d}}^{^{(126,\overline{126})(+)}}=
h^{^{(\overline{126})(+)}}_{\acute{a}\acute{b}}\bar{f}^{^{(126)(+)}}_{\acute{c}\acute{d}}
\bar{l}_{_{\cal W}}^{^{(126)}} \widehat{{\cal
M}}^{^{(126,\overline{126})}} _{\cal {W}\cal{W}'}{k}_{_{\cal
W'}}^{^{(\overline{126})}}
\end{eqnarray}
Further decomposition in the $SU(5)\times U(1)$ basis can be
carried out using the results in Appendix C.

\section{Matter-Higgs Couplings }
In this section we evaluate the quartic couplings involving two
semi-spinors of matter fields, i.e., the 16 plets of matter fields
and two vector-spinor fields. We will utilize the analysis of
Sec.(4) to compute these quartic couplings. Thus the couplings
would arise by mediation from 10, 120 and $126+\overline{126}$
between the matter sector and the Higgs sector. We discuss now
these computations in detail below.

\subsection{The $\bf{{\left(16\times16\right)_{10}
\left(\overline{144}\times\overline{144}\right)_{10}}}$ couplings}

For the computation of the ${\left(16\times16\right)_{10}
\left(\overline{144}\times\overline{144}\right)_{10}}$ couplings arising from
the 10 mediation we consider the superpotential
\begin{eqnarray}
{\mathsf W}^{(10)''}=\frac{1}{2}\Phi_{\nu\cal U}{\cal
M}^{^{(10)}}_{{\cal U}{\cal U}'}\Phi_{\nu{\cal
U}'}+h^{^{(10)}}_{\acute{a}\acute{b}}<\Upsilon^{*}_{(+)\acute{a}\mu}|B\Gamma_{\nu}
|\Upsilon_{(+)\acute{b}\mu}>k_{_{{\cal U}}}^{^{(10)}}\Phi_{\nu\cal
U}\nonumber\\
+f^{^{(10)}}_{\acute{a}\acute{b}}<\Psi^{*}_{(+)\acute{a}}|B\Gamma_{\nu}
|\Psi_{(+)\acute{b}}>l_{_{{\cal U}}}^{^{(10)}}\Phi_{\nu\cal U}
+\bar{h}^{^{(10)}}_{\acute{a}\acute{b}}<\Upsilon^{*}_{(-)\acute{a}\mu}|B\Gamma_{\nu}
|\Upsilon_{(-)\acute{b}\mu}>\bar{k}_{_{{\cal
U}}}^{^{(10)}}\Phi_{\nu\cal U}
\end{eqnarray}
Eliminating $\Phi_{\nu\cal U}$ using F flatness condition we get

\begin{eqnarray}
{\mathsf
W}^{^{^{{(16\times16)}_{10}{(\overline{144}\times\overline{144})}_{10}}}}
=
-2\xi_{\acute{a}\acute{b},\acute{c}\acute{d}}^{^{(10)}}
<\Psi^{*}_{(+)\acute{a}}|B\Gamma_{\rho}|\Psi_{(+)\acute{b}}><\Upsilon^{*}_{(+)\acute{c}\nu}|\Gamma_{\rho}
|\Upsilon_{(+)\acute{d}\nu}>\nonumber\\
=-4\xi_{\acute{a}\acute{b},\acute{c}\acute{d}}^{^{(10)}}
\left[<\Psi^{*}_{(+)\acute{a}}|Bb_i|\Psi_{(+)\acute{b}}><\Upsilon^{*}_{(+)\acute{c}\nu}|Bb_i^{\dagger}
|\Upsilon_{(+)\acute{d}\nu}>\right.\nonumber\\
\left.+<\Psi^{*}_{(+)\acute{a}}|Bb_i^{\dagger}|\Psi_{(+)\acute{b}}><\Upsilon^{*}_{(+)\acute{c}\nu}|Bb_i
|\Upsilon_{(+)\acute{d}\nu}>\right]\nonumber\\
=2\xi_{\acute{a}\acute{b},\acute{c}\acute{d}}^{^{(10)(+)}}
\left(\epsilon_{jklmn}{\bf M}_{\acute{a}i}^{\bf T}{\bf
M}_{\acute{b}}^{ij}{\bf P}_{\acute{c}\mu}^{kl\bf T}{\bf
P}_{\acute{d}\mu}^{mn}-8{\bf M}_{\acute{a}i}^{\bf T}{\bf
M}_{\acute{b}}^{ij}{\bf P}_{\acute{c}j\mu}^{\bf T}{\bf
P}_{\acute{d}\mu}\right.\nonumber\\
\left.+\epsilon_{jklmn}{\bf M}_{\acute{a}}^{kl\bf T}{\bf
M}_{\acute{b}}^{mn}{\bf P}_{\acute{c}i\mu}^{\bf T}{\bf
P}_{\acute{d}\mu}^{ij}-8{\bf M}_{\acute{a}j}^{\bf T}{\bf
M}_{\acute{b}}{\bf P}_{\acute{c}i\mu}^{\bf T}{\bf
P}_{\acute{d}\mu}^{ij}\right)
\end{eqnarray}
where
\begin{equation}
\xi_{\acute{a}\acute{b},\acute{c}\acute{d}}^{^{(10)(+)}}=
f_{\acute{a}\acute{b}}^{^{(10)(+)}}h_{\acute{c}\acute{d}}^{^{(10)(+)}}l_{_{{\cal
U}}}^{^{(10)}}\widetilde{{\cal M}}^{^{(10)}}_{{\cal U}{\cal
U}'}k_{_{{\cal U}'}}^{^{(10)}}
\end{equation}
and
\begin{equation}
\widetilde{{\cal M}}^{^{(10)}}=\left[{\cal
M}^{^{(10)}}+\left({\cal M}^{^{(10)}}\right)^{\bf {T}}\right]^{-1}
\end{equation}

\subsection{The $\bf{{\left(16\times16\right)_{10}
\left(144\times144\right)_{10}}}$ couplings}

An analysis similar to the above gives
\begin{eqnarray}
{\mathsf W}^{^{^{{(16\times16)}_{10}{(144\times144)}_{10}}}}=
-2\zeta_{\acute{a}\acute{b},\acute{c}\acute{d}}^{^{(10)}}
<\Psi^{*}_{(+)\acute{a}}|B\Gamma_{\rho}|\Psi_{(+)\acute{b}}><\Upsilon^{*}_{(-)\acute{c}\nu}|B\Gamma_{\rho}
|\Upsilon_{(-)\acute{d}\nu}>\nonumber\\
=-4\zeta_{\acute{a}\acute{b},\acute{c}\acute{d}}^{^{(10)}}
\left[<\Psi^{*}_{(+)\acute{a}}|Bb_i|\Psi_{(+)\acute{b}}><\Upsilon^{*}_{(-)\acute{c}\nu}|Bb_i^{\dagger}
|\Upsilon_{(-)\acute{d}\nu}>\right.\nonumber\\
\left.+<\Psi^{*}_{(+)\acute{a}}|Bb_i^{\dagger}|\Psi_{(+)\acute{b}}><\Upsilon^{*}_{(-)\acute{c}\nu}|Bb_i
|\Upsilon_{(-)\acute{d}\nu}>\right]\nonumber\\
=2\zeta_{\acute{a}\acute{b},\acute{c}\acute{d}}^{^{(10)(+)}}
\left(8{\bf M}_{\acute{a}i}^{\bf T}{\bf M}_{\acute{b}}^{ij}{\bf
Q}_{\acute{c}\mu}^{k\bf T}{\bf Q}_{\acute{d}kj\mu} -8{\bf
M}_{\acute{a}i}^{\bf T}{\bf M}_{\acute{b}}{\bf
Q}_{\acute{c}\mu}^{i\bf T}{\bf Q}_{\acute{d}\mu}\right.\nonumber\\
 \left.-{\bf M}_{\acute{a}}^{ij\bf T}{\bf M}_{\acute{b}}^{kl}{\bf
Q}_{\acute{c}kl\mu}^{\bf T}{\bf Q}_{\acute{d}ij\mu} +{\bf
M}_{\acute{a}}^{ij\bf T}{\bf M}_{\acute{b}}^{kl}{\bf
Q}_{\acute{c}ik\mu}^{\bf T}{\bf Q}_{\acute{d}jl\mu}\right.\nonumber\\
\left.-{\bf M}_{\acute{a}}^{ij\bf T}{\bf M}_{\acute{b}}^{kl}{\bf
Q}_{\acute{c}il\mu}^{\bf T}{\bf Q}_{\acute{d}jk\mu}+
\epsilon^{ijklm}{\bf M}_{\acute{a}i}^{\bf T}{\bf
M}_{\acute{b}}{\bf Q}_{\acute{c}jk\mu}^{\bf T}{\bf
Q}_{\acute{d}lm\mu}\right.\nonumber\\
\left.+\epsilon_{ijklm}{\bf M}_{\acute{a}}^{ij\bf T}{\bf
M}_{\acute{b}}^{kl}{\bf Q}_{\acute{c}\mu}^{m\bf T}{\bf
Q}_{\acute{d}\mu}\right)
\end{eqnarray}
where
\begin{equation}
\zeta_{\acute{a}\acute{b},\acute{c}\acute{d}}^{^{(10)(+)}}=
f_{\acute{a}\acute{b}}^{^{(10)(+)}}\bar{h}_{\acute{c}\acute{d}}^{^{(10)(+)}}l_{_{{\cal
U}}}^{^{(10)}}\widetilde{{\cal M}}^{^{(10)}}_{{\cal U}{\cal
U}'}\bar{k}_{_{{\cal U}'}}^{^{(10)}}
\end{equation}
To obtain the reduction of the results of Secs. (6.1) and (6.2) in the $SU(5)\times U(1)$ basis
we use the results of Appendix.A.

\subsection{The $\bf{{\left(16\times16\right)_{120}
\left(\overline{144}\times\overline{144}\right)_{120}}}$
couplings} Here we begin by considering the superpotential

\begin{eqnarray}
{\mathsf W}^{(120)''}=\frac{1}{2}\Phi_{\nu\rho\lambda V}{\cal
M}^{^{(120)}}_{V V'}\Phi_{\rho\nu\lambda V'}
+\frac{1}{3!}h^{^{(120)}}_{\acute{a}\acute{b}}<\Upsilon^{*}_{(+)\acute{a}\mu}|B\Gamma_{[\nu}
\Gamma_{\rho}\Gamma_{\lambda
]}|\Upsilon_{(+)\acute{b}\mu}>k_{_{V}}^{^{(120)}}
\Phi_{\nu\rho\lambda V}\nonumber\\
+\frac{1}{3!}f^{^{(120)}}_{\acute{a}\acute{b}}<\Psi^{*}_{(+)\acute{a}}|B\Gamma_{[\nu}
\Gamma_{\rho}\Gamma_{\lambda
]}|\Psi_{(+)\acute{b}}>l_{_{V}}^{^{(120)}}
\Phi_{\nu\rho\lambda V}\nonumber\\
+\frac{1}{3!}{\bar
h}^{^{(120)}}_{\acute{a}\acute{b}}<\Upsilon^{*}_{(-)\acute{a}\mu}|B\Gamma_{[\nu}
\Gamma_{\rho}\Gamma_{\lambda ]}|\Upsilon_{(-)\acute{b}\mu}>{\bar
k}_{_{V}}^{^{(120)}} \Phi_{\nu\rho\lambda V}
\end{eqnarray}
Eliminating $\Phi_{\nu\rho\lambda V}$ by the F flatness condition
we get

\begin{eqnarray}
{\mathsf
W}^{^{^{{(16\times16)}_{120}{(\overline{144}\times\overline{144})}_{120}}}}=
-\frac{1}{18}\xi_{\acute{a}\acute{b},\acute{c}\acute{d}}^{^{(120)}}
<\Psi^{*}_{(+)\acute{a}}|B\Gamma_{[\nu}
\Gamma_{\rho}\Gamma_{\lambda
]}|\Psi_{(+)\acute{b}}><\Upsilon^{*}_{(+)\acute{c}\nu}|B\Gamma_{[\nu}
\Gamma_{\rho}\Gamma_{\lambda
]}|\Upsilon_{(+)\acute{d}\nu}>\nonumber\\
=-\frac{1}{18}\xi_{\acute{a}\acute{b},\acute{c}\acute{d}}^{^{(120)}}
\left[<\Psi^{*}_{(+)\acute{a}}|B\Gamma_{\nu}
\Gamma_{\rho}\Gamma_{\lambda
}|\Psi_{(+)\acute{b}}><\Upsilon^{*}_{(+)\acute{c}\nu}|B\Gamma_{\nu}
\Gamma_{\rho}\Gamma_{\lambda
}|\Upsilon_{(+)\acute{d}\nu}>\right.~~\nonumber\\
\left.-28<\Psi^{*}_{(+)\acute{a}}|B\Gamma_{\nu}
|\Psi_{(+)\acute{b}}><\Upsilon^{*}_{(+)\acute{c}\nu}|B\Gamma_{\nu}
|\Upsilon_{(+)\acute{d}\nu}>\right]~~~~~~~~\nonumber\\
=-\frac{4}{9}\xi_{\acute{a}\acute{b},\acute{c}\acute{d}}^{^{(120)}}
\left[<\Psi^{*}_{(+)\acute{a}}|Bb_ib_jb_k|\Psi_{(+)\acute{b}}><\Upsilon^{*}_{(+)\acute{c}\nu}|B
b_i^{\dagger}b_j^{\dagger}b_k^{\dagger}|\Upsilon_{(+)\acute{d}\nu}>\right.~~~~~~~~\nonumber\\
\left.-6<\Psi^{*}_{(+)\acute{a}}|Bb_i|\Psi_{(+)\acute{b}}><\Upsilon^{*}_{(+)\acute{c}\nu}|B
b_i^{\dagger}|\Upsilon_{(+)\acute{d}\nu}>\right.~~~~~~~~\nonumber\\
\left.+3<\Psi^{*}_{(+)\acute{a}}|Bb_i^{\dagger}|\Psi_{(+)\acute{b}}><\Upsilon^{*}_{(+)\acute{c}\nu}|B
b_n^{\dagger}b_nb_i|\Upsilon_{(+)\acute{d}\nu}>\right.~~~~~~~~\nonumber\\
\left.+3<\Psi^{*}_{(+)\acute{a}}|Bb_i|\Psi_{(+)\acute{b}}><\Upsilon^{*}_{(+)\acute{c}\nu}|B
b_i^{\dagger}b_n^{\dagger}b_n|\Upsilon_{(+)\acute{d}\nu}>\right.~~~~~~~~\nonumber\\
\left.+3<\Psi^{*}_{(+)\acute{a}}|Bb_i^{\dagger}b_j^{\dagger}b_k|\Psi_{(+)\acute{b}}>
<\Upsilon^{*}_{(+)\acute{c}\nu}|B
b_k^{\dagger}b_ib_j|\Upsilon_{(+)\acute{d}\nu}>\right.~~~~~~~~\nonumber\\
+\left.<\Upsilon^{*}_{(+)\acute{a}\nu}|Bb_ib_jb_k|\Upsilon_{(+)\acute{b}\nu}><\Psi^{*}_{(+)\acute{c}}|B
b_i^{\dagger}b_j^{\dagger}b_k^{\dagger}|\Psi_{(+)\acute{d}}>\right.~~~~~~~~\nonumber\\
\left.-6<\Upsilon^{*}_{(+)\acute{a}\nu}|Bb_i|\Upsilon_{(+)\acute{b}\nu}><\Psi^{*}_{(+)\acute{c}}|B
b_i^{\dagger}|\Psi_{(+)\acute{d}}>\right.~~~~~~~~\nonumber\\
\left.+3<\Upsilon^{*}_{(+)\acute{a}\nu}|Bb_i^{\dagger}|\Upsilon_{(+)\acute{b}\nu}><\Psi^{*}_{(+)\acute{c}}|B
b_n^{\dagger}b_nb_i|\Psi_{(+)\acute{d}}>\right.~~~~~~~~\nonumber\\
\left.+3<\Upsilon^{*}_{(+)\acute{a}\nu}|Bb_i|\Upsilon_{(+)\acute{b}\nu}><\Psi^{*}_{(+)\acute{c}}|B
b_i^{\dagger}b_n^{\dagger}b_n|\Psi_{(+)\acute{d}}>\right.~~~~~~~~\nonumber\\
\left.+3<\Upsilon^{*}_{(+)\acute{a}\nu}|Bb_i^{\dagger}b_j^{\dagger}b_k|\Upsilon_{(+)\acute{b}\nu}><\Psi^{*}_{(+)\acute{c}}|B
b_k^{\dagger}b_ib_j|\Psi_{(+)\acute{d}}>\right]~~~~~~~~\nonumber\\
=\frac{4}{3}\xi_{\acute{a}\acute{b},\acute{c}\acute{d}}^{^{(120)(-)}}
\left[4{\bf M}_{\acute{a}i}^{\bf T}{\bf M}_{\acute{b}j}{\bf
P}_{\acute{c}\nu}^{ij\bf T}{\bf P}_{\acute{d}\nu} +4{\bf
M}_{\acute{a}}^{ij\bf T}{\bf M}_{\acute{b}}{\bf
P}_{\acute{c}i\nu}^{\bf T}{\bf P}_{\acute{d}j\nu}\right.~~~~~~~~~~~~~~~~~~~~~\nonumber\\
\left.-4{\bf M}_{\acute{a}}^{\bf T}{\bf M}_{\acute{b}i}{\bf
P}_{\acute{c}\nu}^{ij\bf T}{\bf P}_{\acute{d}j\nu} -4{\bf
M}_{\acute{a}}^{ij\bf T}{\bf M}_{\acute{b}j}{\bf
P}_{\acute{c}\nu}^{\bf T}{\bf P}_{\acute{d}i\nu}\right.~~~~~~~~\nonumber\\
\left.-\epsilon_{ijklm}{\bf M}_{\acute{a}}^{ij\bf T}{\bf
M}_{\acute{b}}^{kn}{\bf P}_{\acute{c}n\nu}^{\bf T}{\bf
P}_{\acute{d}\nu}^{lm} -\epsilon_{ijklm}{\bf M}_{\acute{a}n}^{\bf
T}{\bf M}_{\acute{b}}^{lm}{\bf P}_{\acute{c}\nu}^{ij\bf T}{\bf
P}_{\acute{d}\nu}^{kn}\right]~~~~~~~~
\end{eqnarray}
where
\begin{equation}
\xi_{\acute{a}\acute{b},\acute{c}\acute{d}}^{^{(120)(-)}}=
f_{\acute{a}\acute{b}}^{^{(120)(-)}}h_{\acute{c}\acute{d}}^{^{(120)(-)}}l_{_{V}}^{^{(120)}}\widetilde{{\cal
M}}^{^{(120)}}_{VV'}k_{_{V'}}^{^{(120)}}
\end{equation}
and
\begin{equation}
\widetilde{{\cal M}}^{^{(120)}}=\left[{\cal
M}^{^{(120)}}+\left({\cal M}^{^{(120)}}\right)^{\bf
{T}}\right]^{-1}
\end{equation}

\subsection{The $\bf{{\left(16\times16\right)_{120}
\left(144\times144\right)_{120}}}$ couplings}

An analysis similar to the above  gives

\begin{eqnarray}
{\mathsf W}^{^{^{{(16\times16)}_{120}{(144\times144)}_{120}}}} =
-\frac{1}{18}\zeta_{\acute{a}\acute{b},\acute{c}\acute{d}}^{^{(120)}}
<\Psi^{*}_{(+)\acute{a}}|B\Gamma_{[\nu}
\Gamma_{\rho}\Gamma_{\lambda
]}|\Psi_{(+)\acute{b}}><\Upsilon^{*}_{(-)\acute{c}\nu}|B\Gamma_{[\nu}
\Gamma_{\rho}\Gamma_{\lambda
]}|\Upsilon_{(-)\acute{d}\nu}>\nonumber\\
=-\frac{1}{18}\zeta_{\acute{a}\acute{b},\acute{c}\acute{d}}^{^{(120)}}
\left[<\Psi^{*}_{(+)\acute{a}}|B\Gamma_{\nu}
\Gamma_{\rho}\Gamma_{\lambda
}|\Psi_{(+)\acute{b}}><\Upsilon^{*}_{(-)\acute{c}\nu}|B\Gamma_{\nu}
\Gamma_{\rho}\Gamma_{\lambda
}|\Upsilon_{(-)\acute{d}\nu}>\right.~~\nonumber\\
\left.-28<\Psi^{*}_{(+)\acute{a}}|B\Gamma_{\nu}
|\Psi_{(+)\acute{b}}><\Upsilon^{*}_{(-)\acute{c}\nu}|B\Gamma_{\nu}
|\Upsilon_{(-)\acute{d}\nu}>\right]~~~~~~~~\nonumber\\
=-\frac{4}{9}\zeta_{\acute{a}\acute{b},\acute{c}\acute{d}}^{^{(120)}}
\left[<\Psi^{*}_{(+)\acute{a}}|Bb_ib_jb_k|\Psi_{(+)\acute{b}}><\Upsilon^{*}_{(-)\acute{c}\nu}|B
b_i^{\dagger}b_j^{\dagger}b_k^{\dagger}|\Upsilon_{(-)\acute{d}\nu}>\right.~~~~~~~~\nonumber\\
\left.-6<\Psi^{*}_{(+)\acute{a}}|Bb_i|\Psi_{(+)\acute{b}}><\Upsilon^{*}_{(-)\acute{c}\nu}|B
b_i^{\dagger}|\Upsilon_{(-)\acute{d}\nu}>\right.~~~~~~~~\nonumber\\
\left.+3<\Psi^{*}_{(+)\acute{a}}|Bb_i^{\dagger}|\Psi_{(+)\acute{b}}><\Upsilon^{*}_{(-)\acute{c}\nu}|B
b_n^{\dagger}b_nb_i|\Upsilon_{(-)\acute{d}\nu}>\right.~~~~~~~~\nonumber\\
\left.+3<\Psi^{*}_{(+)\acute{a}}|Bb_i|\Psi_{(+)\acute{b}}><\Upsilon^{*}_{(-)\acute{c}\nu}|B
b_i^{\dagger}b_n^{\dagger}b_n|\Upsilon_{(-)\acute{d}\nu}>\right.~~~~~~~~\nonumber\\
\left.+3<\Psi^{*}_{(+)\acute{a}}|Bb_i^{\dagger}b_j^{\dagger}b_k|\Psi_{(+)\acute{b}}>
<\Upsilon^{*}_{(-)\acute{c}\nu}|B
b_k^{\dagger}b_ib_j|\Upsilon_{(-)\acute{d}\nu}>\right.~~~~~~~~\nonumber\\
+\left.<\Upsilon^{*}_{(-)\acute{a}\nu}|Bb_ib_jb_k|\Upsilon_{(-)\acute{b}\nu}><\Psi^{*}_{(+)\acute{c}}|B
b_i^{\dagger}b_j^{\dagger}b_k^{\dagger}|\Psi_{(+)\acute{d}}>\right.~~~~~~~~\nonumber\\
\left.-6<\Upsilon^{*}_{(-)\acute{a}\nu}|Bb_i|\Upsilon_{(-)\acute{b}\nu}><\Psi^{*}_{(+)\acute{c}}|B
b_i^{\dagger}|\Psi_{(+)\acute{d}}>\right.~~~~~~~~\nonumber\\
\left.+3<\Upsilon^{*}_{(-)\acute{a}\nu}|Bb_i^{\dagger}|\Upsilon_{(-)\acute{b}\nu}><\Psi^{*}_{(+)\acute{c}}|B
b_n^{\dagger}b_nb_i|\Psi_{(+)\acute{d}}>\right.~~~~~~~~\nonumber\\
\left.+3<\Upsilon^{*}_{(-)\acute{a}\nu}|Bb_i|\Upsilon_{(-)\acute{b}\nu}><\Psi^{*}_{(+)\acute{c}}|B
b_i^{\dagger}b_n^{\dagger}b_n|\Psi_{(+)\acute{d}}>\right.~~~~~~~~\nonumber\\
\left.+3<\Upsilon^{*}_{(-)\acute{a}\nu}|Bb_i^{\dagger}b_j^{\dagger}b_k|\Upsilon_{(-)\acute{b}\nu}><\Psi^{*}_{(+)\acute{c}}|B
b_k^{\dagger}b_ib_j|\Psi_{(+)\acute{d}}>\right]~~~~~~~~\nonumber\\
=\frac{4}{3}\zeta_{\acute{a}\acute{b},\acute{c}\acute{d}}^{^{(120)(-)}}
\left[4{\bf M}_{\acute{a}}^{ij\bf T}{\bf M}_{\acute{b}}{\bf
Q}_{\acute{c}ij\mu}^{\bf T}{\bf Q}_{\acute{d}\mu} -4{\bf
M}_{\acute{a}i}^{\bf T}{\bf M}_{\acute{b}j}{\bf
Q}_{\acute{c}\mu}^{i\bf T}{\bf Q}_{\acute{d}\mu}^j\right.~~~~~~~~~~~~~~~~~~~~~\nonumber\\
\left.-4{\bf M}_{\acute{a}}^{ij\bf T}{\bf M}_{\acute{b}}^{kl}{\bf
Q}_{\acute{c}ij\mu}^{\bf T}{\bf Q}_{\acute{d}kl\mu}-8{\bf
M}_{\acute{a}}^{ij\bf T}{\bf M}_{\acute{b}}^{kl}{\bf
Q}_{\acute{c}jk\mu}^{\bf T}{\bf Q}_{\acute{d}il\mu}\right.~~~~~~~~\nonumber\\
\left.+\epsilon_{ijklm}{\bf M}_{\acute{a}}^{ij\bf T}{\bf
M}_{\acute{b}}^{kl}{\bf Q}_{\acute{c}\mu}^{\bf T}{\bf
Q}_{\acute{d}\mu}^m-4\epsilon^{ijklm}{\bf M}_{\acute{a}}^{\bf
T}{\bf M}_{\acute{b}i}{\bf Q}_{\acute{c}jk\mu}^{\bf T}{\bf
Q}_{\acute{d}lm\mu}\right.~~~~~~~~\nonumber\\
\left.+8{\bf M}_{\acute{a}}^{\bf T}{\bf M}_{\acute{b}i}{\bf
Q}_{\acute{c}\mu}^{\bf T}{\bf Q}_{\acute{d}\mu}^i +4{\bf
M}_{\acute{a}i}^{\bf T}{\bf M}_{\acute{b}}^{jk}{\bf
Q}_{\acute{c}\mu}^{i\bf T}{\bf Q}_{\acute{d}jk\mu}\right.~~~~~~~~\nonumber\\
\left.-4{\bf M}_{\acute{a}i}^{\bf T}{\bf M}_{\acute{b}}^{ij}{\bf
Q}_{\acute{c}\mu}^{k\bf T}{\bf Q}_{\acute{d}kj\mu}\right]~~~~~~~~
\end{eqnarray}
where
\begin{equation}
\zeta_{\acute{a}\acute{b},\acute{c}\acute{d}}^{^{(120)(-)}}=
f_{\acute{a}\acute{b}}^{^{(120)(-)}}\bar{h}_{\acute{c}\acute{d}}^{^{(120)(-)}}l_{_{{\cal
U}}}^{^{(120)}}\widetilde{{\cal
M}}^{^{(120)}}_{VV'}\bar{k}_{_{V'}}^{^{(120)}}
\end{equation}
Further reduction of the above to the $SU(5)\times U(1)$ basis can be achieved by using
Appendix B.

\subsection{The $\bf{{\left(16\times16\right)_{\overline{126}}
\left({144}\times{144}\right)_{126}}}$ couplings}

Finally we consider the matter-Higgs couplings  via
$126+\overline{126}$ mediation.  Here we begin by considering the
superpotential

\begin{eqnarray}
{\mathsf
W}^{(126,\overline{126})''}=\frac{1}{2}\Phi_{\nu\rho\sigma\lambda\vartheta
\cal{W}}{\cal M}^{^{(126,\overline{126})}}_{\cal{W}
\cal{W}'}\overline{\Phi}_{\rho\nu\sigma\lambda\vartheta
\cal{W}'}\nonumber\\
+\frac{1}{5!}f^{^{(\overline{126})}}_{\acute{a}\acute{b}}<\Psi^{*}_{(+)\acute{a}}|B\Gamma_{[\nu}
\Gamma_{\rho}\Gamma_{\sigma}\Gamma_{\lambda}\Gamma_{\vartheta]}|\Psi_{(+)\acute{b}}>l_{_{\cal{W}}}
^{^{(\overline{126})}}
\overline{\Phi}_{\nu\rho\sigma\lambda\vartheta\cal{W}}\nonumber\\
+\frac{1}{5!}{\bar
h}^{^{(126)}}_{\acute{a}\acute{b}}<\Upsilon^{*}_{(-)\acute{a}\mu}|B\Gamma_{[\nu}
\Gamma_{\rho}\Gamma_{\sigma}\Gamma_{\lambda}\Gamma_{\vartheta]}|\Upsilon_{(-)\acute{b}\mu}>{\bar
k}_{_{\cal{W}}}^{^{(126)}}
\Phi_{\nu\rho\sigma\lambda\vartheta\cal{W}}\nonumber\\
\end{eqnarray}
Eliminating $\Phi_{\nu\rho\sigma\lambda\vartheta\cal{W}}$,
$\overline{\Phi}_{\nu\rho\sigma\lambda\vartheta\cal{W}}$, by use of F flatness gives

\begin{eqnarray}
{\mathsf
W}^{^{^{{(16\times16)}_{\overline{126}}{({144}\times{144})}_{126}}}}=
\frac{1}{7200}\varrho_{\acute{a}\acute{b},\acute{c}\acute{d}}^{^{(126,\overline{126})}}
<\Psi^{*}_{(+)\acute{a}}|B\Gamma_{[\nu}
\Gamma_{\rho}\Gamma_{\sigma}\Gamma_{\lambda}\Gamma_{\vartheta]}|\Psi_{(+)\acute{b}}>\nonumber\\
\times<\Upsilon^{*}_{(-)\acute{c}\mu}|B\Gamma_{[\nu}
\Gamma_{\rho}\Gamma_{\sigma}\Gamma_{\lambda}\Gamma_{\vartheta]}|\Upsilon_{(-)\acute{d}\mu}>\nonumber\\
=\frac{2}{15}\varrho_{\acute{a}\acute{b},\acute{c}\acute{d}}^{^{(126,\overline{126})}}
\left[2{\bf M}_{\acute{a}}^{\bf T}{\bf M}_{\acute{b}}{\bf
Q}_{\acute{c}\mu}^{\bf T}{\bf Q}_{\acute{d}\mu}-2{\bf
M}_{\acute{a}}^{\bf T}{\bf M}_{\acute{b}}^{ij}{\bf
Q}_{\acute{c}\mu}^{\bf T}{\bf Q}_{\acute{d}ij\mu}\right.\nonumber\\
\left.+2{\bf M}_{\acute{a}i}^{\bf T}{\bf M}_{\acute{b}j}{\bf
Q}_{\acute{c}\mu}^{i\bf T}{\bf Q}_{\acute{d}\mu}^{j}+48{\bf
M}_{\acute{a}}^{\bf T}{\bf M}_{\acute{b}k}{\bf
Q}_{\acute{c}\mu}^{\bf
T}{\bf Q}_{\acute{d}\mu}^{k}\right.\nonumber\\
\left.-2{\bf M}_{\acute{a}}^{ij\bf T}{\bf M}_{\acute{b}k}{\bf
Q}_{\acute{c}ij\mu}^{\bf T}{\bf Q}_{\acute{d}\mu}^{k}+{\bf
M}_{\acute{a}}^{ij\bf T}{\bf M}_{\acute{b}j}{\bf
Q}_{\acute{c}ik\mu}^{\bf T}{\bf Q}_{\acute{d}\mu}^{k}\right.\nonumber\\
\left.+6{\bf M}_{\acute{a}}^{ij\bf T}{\bf M}_{\acute{b}}^{kl}{\bf
Q}_{\acute{c}ij\mu}^{\bf T}{\bf Q}_{\acute{d}kl\mu} -30{\bf
M}_{\acute{a}}^{ij\bf T}{\bf M}_{\acute{b}}^{kl}{\bf
Q}_{\acute{c}ik\mu}^{\bf T}{\bf Q}_{\acute{d}jl\mu}\right.\nonumber\\
\left.+9{\bf M}_{\acute{a}}^{ij\bf T}{\bf M}_{\acute{b}}^{kl}{\bf
Q}_{\acute{c}il\mu}^{\bf T}{\bf
Q}_{\acute{d}jk\mu}+2\epsilon^{ijklm} {\bf M}_{\acute{a}}^{\bf
T}{\bf M}_{\acute{b}i}{\bf
Q}_{\acute{c}jk\mu}^{\bf T}{\bf Q}_{\acute{d}lm\mu}\right.\nonumber\\
\left.+2\epsilon_{ijklm} {\bf M}_{\acute{a}}^{ij\bf T}{\bf
M}_{\acute{b}}^{kl}{\bf Q}_{\acute{c}\mu}^{\bf T}{\bf
Q}_{\acute{d}\mu}^m\right]
\end{eqnarray}
where
\begin{equation}
\varrho_{\acute{a}\acute{b},\acute{c}\acute{d}}^{^{(126,\overline{126})}}=
f^{^{(\overline{126})}}_{\acute{a}\acute{b}}\bar{h}^{^{(126)}}_{\acute{c}\acute{d}}
\bar{k}_{_{\cal W}}^{^{(126)}} \widehat{{\cal
M}}^{^{(126,\overline{126})}} _{\cal {W}\cal{W}'}{l}_{_{\cal
W'}}^{^{(\overline{126})}}
\end{equation}
and
\begin{equation}
\widehat{{\cal M}}^{^{(126,\overline{126})}}=\left({\cal
M}^{^{(126,\overline{126})}}\right)^{-1}\left[\left({\cal
M}^{^{(126,\overline{126})}}\right)^{\bf T}\left({\cal
M}^{^{(126,\overline{126})}}\right)^{-1} -2\cdot\bf{1}\right]
\end{equation}
A further reduction of the quartic interactions to the $SU(5)\times U(1)$ basis
can  be achieved  by use of Appendix C.

\section{The Gauge Couplings of Vector-Spinors}

In this section we compute the interactions of the 144 and
$\overline {144}$ with gauge tensors 1 and 45.

\subsection{The $\bf{{\overline{144}^{\dagger}\times\overline{144}\times 1}}$
 couplings}

These couplings are given by
\begin{equation}
{\mathsf
L}^{(1)}_{++}=g^{^{(1)}}_{\acute{a}\acute{b}}<\Upsilon_{(+)\acute{a}\mu}|
\gamma^0\gamma^A|\Upsilon_{(+)\acute{b} \mu}>\Phi_{A\rho\sigma}
\end{equation}
where $\gamma^A (A,B=0-3)$ spans the Clifford algebra associated
with the Lorentz group. An explicit analysis in the $SU(5)\times U(1)$ basis
gives

\begin{eqnarray}
{\mathsf
L}^{(1)}_{++}=g^{^{(1)}}_{\acute{a}\acute{b}}\left\{\left[
\overline{{\cal P}}_{\acute{a}j}^{i}\gamma^A{\cal
P}_{\acute{b}i}^j+ \overline{{\cal
P}}_{\acute{a}}^{ij}\gamma^A{\cal P}_{\acute{b}ij}
+\frac{1}{2}\overline{{\cal P}}_{(S)\acute{a}}^{ij}\gamma^A{\cal
P}^{(S)}_{\acute{b}ij}
+\frac{1}{2}\overline{{\cal P}}_{\acute{a}ij}^{k}\gamma^A{\cal P}_{\acute{b}k}^{ij}\right.\right.\nonumber\\
\left.\left.+\overline{{\cal
P}}_{\acute{a}i}\gamma^A{\cal
P}^i_{\acute{b}}+\frac{1}{6}\overline{{\cal
P}}_{\acute{a}}^{i}\gamma^A{\cal
P}_{\acute{b}i}+\frac{1}{2}\overline{{\cal
P}}_{\acute{a}l}^{ijk}\gamma^A{\cal
P}_{\acute{b}ijk}^l\right]{\mathsf G}_A\right.
\end{eqnarray}

\subsection{The $\bf{{144^{\dagger}\times144\times 1}}$ couplings}
These couplings are given by
\begin{equation}
{\mathsf
L}^{(1)}_{--}=\bar{g}^{^{(1)}}_{\acute{a}\acute{b}}<\Upsilon_{(-)\acute{a}\mu}|
\gamma^0\gamma^A|\Upsilon_{(-)\acute{b} \mu}>\Phi_{A\rho\sigma}
\end{equation}
An explicit analysis in the $SU(5)\times U(1)$ basis gives
\begin{eqnarray}
{\mathsf
L}^{(1)}_{--}=\bar{g}^{^{(1)}}_{\acute{a}\acute{b}}\left\{\left[\overline{{\cal Q}}_{\acute{a}i}^{j}\gamma^A{\cal
Q}_{\acute{b}j}^i+\overline{{\cal
Q}}_{\acute{a}ij}\gamma^A{\cal Q}_{\acute{b}}^{ij}+\frac{1}{2}
\overline{{\cal Q}}_{\acute{a}ij}^{(S)}\gamma^A{\cal
Q}_{(S)\acute{b}}^{ij} +
\frac{1}{2}\overline{{\cal Q}}_{\acute{a}k}^{ij}\gamma^A{\cal Q}_{\acute{b}ij}^{k}\right.\right.\nonumber\\
\left.\left.+\overline{{\cal
Q}}_{\acute{a}}^i\gamma^A{\cal
Q}_{\acute{b}i}+ \frac{1}{6}\overline{{\cal
Q}}_{\acute{a}i}\gamma^A{\cal
Q}_{\acute{b}}^i+\frac{1}{2}\overline{{\cal
Q}}_{\acute{a}ijk}^{l}\gamma^A{\cal
Q}_{\acute{b}l}^{ijk}\right]{\mathsf G}_A\right.
\end{eqnarray}

\subsection{The $\bf{{\overline{144}^{\dagger}\times\overline{144}\times 45}}$
couplings} These couplings are defined by
\begin{equation}
{\mathsf
L}^{(45)}_{++}=\frac{1}{i}\frac{1}{2!}g^{^{(45)}}_{\acute{a}\acute{b}}<\Upsilon_{(+)\acute{a}\mu}|
\gamma^0\gamma^A\Sigma_{\rho\sigma}|\Upsilon_{(+)\acute{b}
\mu}>\Phi_{A\rho\sigma}
\end{equation}
An expansion in the $SU(5)\times U(1)$ basis using the Basic Theorem gives

\begin{eqnarray}
{\mathsf
L}^{(45)}_{++}=g^{^{(45)}}_{\acute{a}\acute{b}}\left\{\left[-\frac{3}{\sqrt
{5}}\overline{{\cal P}}_{\acute{a}j}^{i}\gamma^A{\cal
P}_{\acute{b}i}^j-\frac{7}{10\sqrt {5}}\overline{{\cal
P}}_{\acute{a}}^{ij}\gamma^A{\cal P}_{\acute{b}ij}-\frac{3}{2\sqrt
{5}}\overline{{\cal P}}_{(S)\acute{a}}^{ij}\gamma^A{\cal
P}^{(S)}_{\acute{b}ij}
+\frac{1}{2\sqrt {5}}\overline{{\cal P}}_{\acute{a}ij}^{k}\gamma^A{\cal P}_{\acute{b}k}^{ij}\right.\right.\nonumber\\
\left.\left.+\frac{21}{5\sqrt {5}}\overline{{\cal
P}}_{\acute{a}i}\gamma^A{\cal
P}^i_{\acute{b}}-\sqrt{5}\overline{{\cal
P}}_{\acute{a}}^{i}\gamma^A{\cal
P}_{\acute{b}i}+\frac{1}{6\sqrt{5}}\overline{{\cal
P}}_{\acute{a}l}^{ijk}\gamma^A{\cal
P}_{\acute{b}ijk}^l\right]{\mathsf G}_A\right.\nonumber\\
\left.+\left[\frac{1}{\sqrt 2}\overline{{\cal
P}}_{\acute{a}}^{k}\gamma^A{\cal
P}_{\acute{b}k}^{lm}+\frac{1}{\sqrt {10}}\overline{{\cal
P}}_{\acute{a}}^{l}\gamma^A{\cal P}^m_{\acute{b}} + \frac{2}{\sqrt
{15}}\overline{{\cal P}}_{\acute{a}}^{lk}\gamma^A{\cal
P}_{\acute{b}k}^m+\frac{1}{\sqrt 2}\overline{{\cal
P}}_{\acute{a}n}^{klm}\gamma^A{\cal
P}_{\acute{b}k}^n\right.\right.\nonumber\\
\left.\left.-\frac{1}{20\sqrt {3}}\epsilon^{ijklm}\overline{{\cal
P}}_{\acute{a}i}\gamma^A{\cal P}_{\acute{b}jk} +\frac{1}{3\sqrt
{10}}\epsilon^{ijklm}\overline{{\cal P}}_{\acute{a}n}\gamma^A{\cal
P}_{\acute{b}ijk}^{n}
 -\frac{1}{4}\sqrt{\frac{3}{5}}\epsilon^{ijklm}\overline{{\cal P}}_{\acute{a}ij}^{n}\gamma^A
 {\cal P}_{\acute{b}nk}\right.\right.\nonumber\\
\left.\left.+\frac{1}{4}\epsilon^{ijklm}\overline{{\cal
P}}_{\acute{a}ij}^{n}\gamma^A{\cal
P}^{(S)}_{\acute{b}nk}\right]{\mathsf G}_{Alm}\right.\nonumber\\
\left.+\left[\frac{1}{\sqrt 2}\overline{{\cal
P}}_{\acute{a}lm}^{k}\gamma^A{\cal P}_{\acute{b}k}-\frac{1}{\sqrt
{10}}\overline{{\cal P}}_{\acute{a}l}\gamma^A{\cal P}_{\acute{b}m}
+ \frac{2}{\sqrt {15}}\overline{{\cal
P}}_{\acute{a}l}^{k}\gamma^A{\cal P}_{\acute{b}km}+\frac{1}{\sqrt
2}\overline{{\cal P}}_{\acute{a}n}^{k}\gamma^A{\cal
P}_{\acute{b}klm}^n\right.\right.\nonumber\\
\left.\left.-\frac{1}{20\sqrt {3}}\epsilon_{ijklm}\overline{{\cal
P}}_{\acute{a}}^{ij}\gamma^A{\cal P}^{k}_{\acute{b}}
+\frac{1}{3\sqrt {10}}\epsilon_{ijklm}\overline{{\cal
P}}_{\acute{a}n}^{ijk}\gamma^A{\cal P}_{\acute{b}}^{n}
+\frac{1}{4}\sqrt{\frac{3}{5}}
\epsilon_{ijklm}\overline{{\cal P}}_{\acute{a}}^{in}\gamma^A{\cal P}_{\acute{b}n}^{jk}\right.\right.\nonumber\\
\left.\left.+\frac{1}{4}\epsilon_{ijklm}\overline{{\cal
P}}_{(S)\acute{a}}^{in}\gamma^A{\cal
P}^{jk}_{\acute{b}n}\right]{\mathsf G}^{lm}_A\right.\nonumber\\
\left.+\left[\sqrt 2\overline{{\cal
P}}_{\acute{a}ik}^{l}\gamma^A{\cal P}_{\acute{b}l}^{kj}
-\frac{1}{\sqrt {10}}\overline{{\cal
P}}_{\acute{a}ik}^{j}\gamma^A{\cal P}_{\acute{b}}^k+\frac{1}{\sqrt
{10}}\overline{{\cal P}}_{\acute{a}k}\gamma^A{\cal
P}_{\acute{b}i}^{kj}-\frac{3}{10\sqrt
{2}}\overline{{\cal P}}_{ai}\gamma^A{\cal P}_{\acute{b}}^j\right.\right.\nonumber\\
\left.\left.+\frac{1}{\sqrt {2}}\overline{{\cal
P}}_{\acute{a}m}^{jkl}\gamma^A{\cal P}_{\acute{b}kli}^m
-\frac{1}{\sqrt {15}}\overline{{\cal
P}}_{\acute{a}i}^{jkl}\gamma^A{\cal P}_{\acute{b}kl}
-\frac{1}{\sqrt {15}}\overline{{\cal
P}}_{\acute{a}}^{kl}\gamma^A{\cal P}_{\acute{b}ikl}^j
-\frac{17}{15\sqrt {2}}\overline{{\cal
P}}_{\acute{a}}^{jk}\gamma^A{\cal
P}_{\acute{b}ki}\right.\right.\nonumber\\
\left.\left.+\sqrt{\frac{3}{10}}\overline{{\cal
P}}_{\acute{a}}^{jk}\gamma^A{\cal P}_{\acute{b}ki}^{(S)}
-\sqrt{\frac{3}{10}}\overline{{\cal
P}}_{(S)\acute{a}}^{jk}\gamma^A{\cal
P}_{\acute{b}ki}+\frac{1}{\sqrt 2 }\overline{{\cal
P}}_{(S)\acute{a}}^{jk}\gamma^A{\cal
P}_{\acute{b}ki}^{(S)}+\sqrt{2}\overline{{\cal
P}}_{\acute{a}k}^{j}\gamma^A{\cal
P}_{\acute{b}i}^{k}\right]{\mathsf G}^{i}_{Aj}\right\}\nonumber\\
\end{eqnarray}
The barred matter fields are defined so that $\overline{{\cal
P}}_{\acute{a}}^{jk}={\cal P}_{\acute{a}}^{jk\dagger}\gamma^0$,
etc..

\subsection{The $\bf{{144^{\dagger}\times144\times 45}}$ couplings}
These gauge couplings are defined by

\begin{equation}
{\mathsf
L}^{(45)}_{--}=\frac{1}{i}\frac{1}{2!}\bar{g}^{^{(45)}}_{\acute{a}\acute{b}}<\Upsilon_{(-)\acute{a}\mu}|
\gamma^0\gamma^A\Sigma_{\rho\sigma}|\Upsilon_{(-)\acute{b}
\mu}>\Phi_{A\rho\sigma}
\end{equation}
An analysis in the $SU(5)\times U(1)$ basis using the Basic Theorem gives
\begin{eqnarray}
{\mathsf
L}^{(45)}_{--}=\bar{g}^{^{(45)}}_{\acute{a}\acute{b}}\left\{\left[\frac{3}{\sqrt
{5}}\overline{{\cal Q}}_{\acute{a}i}^{j}\gamma^A{\cal
Q}_{\acute{b}j}^i+\frac{7}{10\sqrt {5}}\overline{{\cal
Q}}_{\acute{a}ij}\gamma^A{\cal Q}_{\acute{b}}^{ij}+\frac{3}{2\sqrt
{5}}\overline{{\cal Q}}_{\acute{a}ij}^{(S)}\gamma^A{\cal
Q}_{(S)\acute{b}}^{ij}-
\frac{1}{2\sqrt {5}}\overline{{\cal Q}}_{\acute{a}k}^{ij}\gamma^A{\cal Q}_{\acute{b}ij}^{k}\right.\right.\nonumber\\
\left.\left.-\frac{21}{5\sqrt {5}}\overline{{\cal
Q}}_{\acute{a}}^i\gamma^A{\cal
Q}_{\acute{b}i}-\sqrt{5}\overline{{\cal
Q}}_{\acute{a}i}\gamma^A{\cal
Q}_{\acute{b}}^i-\frac{1}{6\sqrt{5}}\overline{{\cal
Q}}_{\acute{a}ijk}^{l}\gamma^A{\cal
Q}_{\acute{b}l}^{ijk}\right]{\mathsf G}_A\right.\nonumber\\
\left.+\left[\frac{1}{\sqrt 2}\overline{{\cal
Q}}_{\acute{a}k}^{lm}\gamma^A{\cal
Q}_{\acute{b}}^{k}+\frac{1}{\sqrt {10}}\overline{{\cal
Q}}_{\acute{a}}^{m}\gamma^A{\cal Q}^l_{\acute{b}} + \frac{2}{\sqrt
{15}}\overline{{\cal Q}}_{\acute{a}k}^{m}\gamma^A{\cal
Q}_{\acute{b}}^{lk}+\frac{1}{\sqrt 2}\overline{{\cal
Q}}_{\acute{a}k}^{n}\gamma^A{\cal
Q}_{\acute{b}n}^{klm}\right.\right.\nonumber\\
\left.\left.-\frac{1}{20\sqrt {3}}\epsilon^{ijklm}\overline{{\cal
Q}}_{\acute{a}jk}\gamma^A{\cal Q}_{\acute{b}i} +\frac{1}{3\sqrt
{10}}\epsilon^{ijklm}\overline{{\cal
Q}}_{\acute{a}ijk}^n\gamma^A{\cal Q}_{\acute{b}n}
-\frac{1}{4}\sqrt{\frac{3}{5}}\epsilon^{ijklm}\overline{{\cal
Q}}_{\acute{a}nk}\gamma^A
{\cal Q}_{\acute{b}ij}^n\right.\right.\nonumber\\
\left.\left.+\frac{1}{4}\epsilon^{ijklm}\overline{{\cal
Q}}_{\acute{a}nk}^{(S)}\gamma^A{\cal
Q}_{\acute{b}ij}^n\right]{\mathsf G}_{Alm}\right.\nonumber\\
\left.+\left[\frac{1}{\sqrt 2}\overline{{\cal
Q}}_{\acute{a}k}\gamma^A{\cal Q}_{\acute{b}lm}^k-\frac{1}{\sqrt
{10}}\overline{{\cal Q}}_{\acute{a}m}\gamma^A{\cal Q}_{\acute{b}l}
+ \frac{2}{\sqrt {15}}\overline{{\cal
Q}}_{\acute{a}km}\gamma^A{\cal Q}_{\acute{b}l}^k+\frac{1}{\sqrt
2}\overline{{\cal Q}}_{\acute{a}klm}^{n}\gamma^A{\cal
Q}_{\acute{b}n}^k\right.\right.\nonumber\\
\left.\left.-\frac{1}{20\sqrt {3}}\epsilon_{ijklm}\overline{{\cal
Q}}_{\acute{a}}^{k}\gamma^A{\cal Q}^{ij}_{\acute{b}}
+\frac{1}{3\sqrt {10}}\epsilon_{ijklm}\overline{{\cal
Q}}_{\acute{a}}^{n}\gamma^A{\cal Q}_{\acute{b}n}^{ijk}
+\frac{1}{4}\sqrt{\frac{3}{5}}
\epsilon_{ijklm}\overline{{\cal Q}}_{\acute{a}n}^{jk}\gamma^A{\cal Q}_{\acute{b}}^{in}\right.\right.\nonumber\\
\left.\left.+\frac{1}{4}\epsilon_{ijklm}\overline{{\cal
Q}}_{\acute{a}n}^{jk}\gamma^A{\cal
Q}_{(S)\acute{b}}^{in}\right]{\mathsf G}^{lm}_A\right.\nonumber\\
\left.+\left[-\sqrt 2\overline{{\cal
Q}}_{\acute{a}l}^{kj}\gamma^A{\cal Q}_{\acute{b}ik}^{l}
+\frac{1}{\sqrt {10}}\overline{{\cal
Q}}_{\acute{a}}^{k}\gamma^A{\cal Q}_{\acute{b}ik}^j-\frac{1}{\sqrt
{10}}\overline{{\cal Q}}_{\acute{a}i}^{kj}\gamma^A{\cal
Q}_{\acute{b}k}+\frac{3}{10\sqrt
{2}}\overline{{\cal Q}}_{\acute{a}}^j\gamma^A{\cal Q}_{\acute{b}i}\right.\right.\nonumber\\
\left.\left.-\frac{1}{\sqrt {2}}\overline{{\cal
Q}}_{\acute{a}ikl}^{m}\gamma^A{\cal Q}_{\acute{b}m}^{jkl}
+\frac{1}{\sqrt {15}}\overline{{\cal
Q}}_{\acute{a}kl}\gamma^A{\cal Q}_{\acute{b}i}^{jkl}
+\frac{1}{\sqrt {15}}\overline{{\cal
Q}}_{\acute{a}ikl}^{j}\gamma^A{\cal Q}_{\acute{b}}^{kl}
+\frac{17}{15\sqrt {2}}\overline{{\cal
Q}}_{\acute{a}ki}\gamma^A{\cal
Q}_{\acute{b}}^{jk}\right.\right.\nonumber\\
\left.\left.-\sqrt{\frac{3}{10}}\overline{{\cal
Q}}_{\acute{a}ki}^{(S)}\gamma^A{\cal Q}_{\acute{b}}^{jk}
+\sqrt{\frac{3}{10}}\overline{{\cal Q}}_{\acute{a}ki}\gamma^A{\cal
Q}_{(S)\acute{b}}^{jk}-\frac{1}{\sqrt 2 }\overline{{\cal
Q}}_{\acute{a}ki}^{(S)}\gamma^A{\cal
Q}_{(S)\acute{b}}^{jk}-\sqrt{2}\overline{{\cal
Q}}_{\acute{a}i}^{k}\gamma^A{\cal
Q}_{\acute{b}k}^{j}\right]{\mathsf G}^{i}_{Aj}\right\}\nonumber\\
\end{eqnarray}

The above concludes our analysis of the interactions of the vector-spinors
with Higgs multiplets in  tensor representations, self-couplings of the
vector-spinors, and  of the couplings of the vector-spinors  with the
matter in the spinor 16-plet representations. These couplings are of considerable
value in the analysis of spontaneous breaking of the $SO(10)$gauge symmetry,
in the analysis of proton life time, and in the analysis of quark-lepton
textures and in the study of neutrino masses.   In the next section we
give few illustrative examples of the utility of the couplings for these
analyses.

\section{Use of Vector-Spinor Couplings in Model Building}
In this section we illustrate to the reader the use of
vector-spinor couplings for  further development of $SO(10)$ model
building discussed in Ref.\cite{bgns}.
 In  particular, we discuss
the breaking of $SO(10)$ group down to the Standard Model group,
doublet-triplet splitting, mass growth of quarks and leptons, and
baryon and lepton violating dimension five operators.

In Ref.\cite{bgns} it was shown that breaking of SO(10) to the
Standard Model gauge  group can be  accomplished in one step. In
the following we give a simpler illustration of how this comes
about. This simpler example includes in superpotential a masss
term for $144\times \overline{144}$ and interaction terms
mediated by 45 and 210 and is given by

\begin{eqnarray}
 {\mathsf W}= {\mathsf W}^{(\overline{144}_H\times 144_H)}+{\mathsf W}^{ (\overline{144}_H\times
160_H)_{45}
 (\overline{144}_H\times 144_H)_{45}}\nonumber\\
 +{\mathsf W}^{ (\overline{144}_H\times
144_H)_{210}
 (\overline{144}_H\times 144_H)_{210}}
 \label{simplew}
\end{eqnarray}
Explicit  forms of these couplings are
\begin{eqnarray}
{\mathsf W}^{(\overline{144}_H\times 144_H)}
=M<\Upsilon^{*}_{(-)\mu}|B|\Upsilon_{(+)\mu}>\nonumber\\
 {\mathsf
W}^{ (\overline{144}_H\times 144_H)_{45}
 (\overline{144}_H\times
 144_H)_{45}}=\frac{\Lambda_{45}}{M'}<\Upsilon^{*}_{(-)\mu}|B\Sigma_{\rho\lambda}|\Upsilon_{(+)\mu}>\nonumber\\
 \cdot<\Upsilon^{*}_{(-)\nu}|B\Sigma_{\rho\lambda}|\Upsilon_{(+)\nu}>\nonumber\\
{\mathsf W}^{ (\overline{144}_H\times 144_H)_{210}
 (\overline{144}_H\times 144_H)_{210}}=\frac{\Lambda_{210}}{M'}<\Upsilon^{*}_{(-)\mu}|B\Gamma_{[\rho}
 \Gamma_{\sigma}\Gamma_{\lambda}\Gamma_{\xi
 ]}|\Upsilon_{(+)\mu}>\nonumber\\
\cdot<\Upsilon^{*}_{(-)\nu}|B\Gamma_{[\rho}
 \Gamma_{\sigma}\Gamma_{\lambda}\Gamma_{\xi ]}|\Upsilon_{(+)\nu}>
\end{eqnarray}

\subsection{One step breaking of $\bf{SO(10)}$ GUT symmetry}

The  terms that contribute to one step breaking of GUT symmetry:
$SO(10)\rightarrow SU(3)_C\times SU(2)_L\times U(1)_Y$ are
\begin{eqnarray}
{\mathsf W}_{_{GS}}= M {\cal Q}^i_j{\cal P}^j_i + \alpha_{_{1}}
{\cal Q}^i_j{\cal P}^j_i {\cal Q}^k_l{\cal P}^l_k +\alpha_{_{2}}
{\cal Q}^i_k {\cal P}^k_j {\cal Q}^j_l {\cal P}^l_i
\end{eqnarray}
where
\begin{eqnarray}
 \alpha_{_{1}}=\frac{1}{M'} \left(-\Lambda_{45}+\frac{1}{6}
\Lambda_{210}\right)
\nonumber\\
\alpha_{_{2}}=\frac{1}{M'} \left(-4\Lambda_{45}-
\Lambda_{210}\right)
\end{eqnarray}
For symmetry breaking we invoke the following vacuum expectation
values (VEV's)
\begin{eqnarray}
 <{\cal Q}^i_j>= q ~diag(2,2,2,-3,-3),~~ <{\cal
P}^i_j>= p ~diag(2,2,2,-3,-3)
\end{eqnarray}
and together with the minimization of ${\mathsf W}_{_{GS}}$, we
find
\begin{eqnarray}\label{symmetrybreaking}
\frac{MM'}{qp}=116\lambda_{45}+4\lambda_{210}
\end{eqnarray}
 The D-flatness condition
$<144>=<\overline{144}>$ gives $q=p$. With the above VEV,
 spontaneous breaking of $SO(10)$  occurs down to the Standard Model group.

\subsection{Doublet triplet splitting}
As discussed in Ref.\cite{bgns}in  the scenario with one step
breaking of SO(10) both Higgs doublets and the Higgs triplets will
be  heavy. However, it is possible to get a pair of light Higgs
doublets by fine tuning, a procedure which is justified in the
context of landscape scenarios as discussed in Ref.\cite{bgns}.
Here we illustrate this explicitly for the case of the
superpotential of Eq.(\ref{simplew}). To this end we collect the
relevant  terms using mixed $SO(10)$ and $SU(5)$ indices:
\begin{eqnarray}
{\mathsf W}_{_{DT}}= M\left({\bf Q}_{\mu}{\bf P}_{\mu}-\frac{1}{2}
{\bf Q}_{ij\mu}{\bf
P}_{\mu}^{ij}\right)+\frac{\Lambda_{45}}{M'}\left( 8{\bf
Q}_{\mu}^i{\bf P}_{j\mu}{\bf Q}_{ik\nu}{\bf P}_{\nu}^{kj}+{\bf
Q}_{\mu}^i{\bf P}_{i\mu}{\bf Q}_{jk\nu}{\bf
P}_{\nu}^{jk}\right.\nonumber\\
\left.+6{\bf Q}_{\mu}{\bf P}_{\mu}{\bf Q}_{\nu}^i{\bf
P}_{i\nu}\right)+ \frac{\Lambda_{210}}{M'}\left( -\frac{2}{3}{\bf
Q}_{\mu}^i{\bf P}_{j\mu}{\bf Q}_{ik\nu}{\bf P}_{\nu}^{kj}-
\frac{1}{6}{\bf Q}_{\mu}^i{\bf P}_{i\mu}{\bf Q}_{jk\nu}{\bf
P}_{\nu}^{jk}\right.\nonumber\\
\left.-\frac{1}{3}{\bf Q}_{\mu}{\bf P}_{\mu}{\bf Q}_{\nu}^i{\bf
P}_{i\nu}-\frac{8}{3}{\bf Q}_{\mu}{\bf P}_{i\mu}{\bf
Q}_{\nu}^i{\bf P}_{\nu}\right)
\end{eqnarray}
when expanded in purely $SU(5)$ indices, we get,
\begin{eqnarray}
{\mathsf W}_{_{DT}}=
\left[M+\frac{1}{M'}\left(6\Lambda_{45}-\frac{1}{3}\Lambda_{210}\right)<{\bf
Q}^m_n><{\bf P}^n_m>\right]\left[{\bf Q}_i{\bf P}^i+{\bf Q}^i{\bf
P}_i\right]\nonumber\\
-\frac{8}{3}\frac{\Lambda_{210}}{M'}<{\bf Q}^i_m><{\bf P}^m_j>{\bf
Q}_i{\bf P}^j\nonumber\\
\left[-\frac{1}{2}M+\frac{1}{M'}\left(\Lambda_{45}-\frac{1}{6}\Lambda_{210}\right)<{\bf
Q}^m_n><{\bf P}^n_m>\right]\left[{\bf
Q}_{ij}^k-\frac{1}{4}\left(\delta^k_i{\bf Q}_j-\delta^k_j{\bf
Q}_i\right)\right]\nonumber\\
\times \left[{\bf P}^{ij}_k-\frac{1}{4}\left(\delta^i_k{\bf
P}^j-\delta^j_k{\bf P}^i\right)\right]\nonumber\\
+\left[\frac{1}{M'}\left(\Lambda_{45}-\frac{1}{6}\Lambda_{210}\right)<{\bf
Q}^i_m><{\bf P}^m_j>\right] \left[{\bf
Q}_{il}^k-\frac{1}{4}\left(\delta^k_i{\bf Q}_l-\delta^k_l{\bf
Q}_i\right)\right]\nonumber\\
\times \left[{\bf P}^{lj}_k-\frac{1}{4}\left(\delta^l_k{\bf
P}^j-\delta^j_k{\bf P}^l\right)\right]
\end{eqnarray}

Note that in addition to the pairs of doublets: (${\bf
Q}_{\alpha}$, ${\bf P}^{\alpha}$), (${\bf Q}^{\alpha}$, ${\bf
P}_{\alpha}$) ($\alpha, \beta, \gamma = 4,5$) and pairs of
triplets: (${\bf Q}_{a}$, ${\bf P}^{a}$), (${\bf Q}^{a}$, ${\bf
P}_{a}$) ($a,b,c = 1,2,3$) there are also pairs of $SU(2)$
doublets and $SU(3)_C$ triplets and anti-triplets  that reside in
 ${\bf Q}_{ij}^k$ and ${\bf P}^{ij}_k$. We denote them by
(${\bf {\widetilde Q}}_{\alpha}$, ${\bf {\widetilde
P}}^{\alpha}$), (${\bf {\widetilde Q}}_{a}$, ${\bf {\widetilde
P}}^{a}$), (${\bf {\widetilde Q}}^{a}$, ${\bf {\widetilde
P}}_{a}$). These can be decomposed as follows
\begin{eqnarray}
{\bf Q}_{b\alpha}^{b}=-{\bf Q}_{\beta\alpha}^{\beta}={\bf
{\widetilde Q}}_{\alpha},~~{\bf P}^{b\alpha}_{b}=-{\bf
P}^{\beta\alpha}_{\beta}={\bf {\widetilde
P}}^{\alpha}\nonumber\\
{\bf Q}_{ba}^{b}=-{\bf Q}_{\beta a}^{\beta}={\bf {\widetilde
Q}}_{a},~~{\bf P}^{ba}_{b}=-{\bf P}^{\beta a}_{\beta}={\bf
{\widetilde P}}^{a}
\nonumber\\
{\bf Q}_{b\alpha}^{a}={\bf {\widetilde
Q}}^{a}_{b\alpha}+\frac{1}{3}\delta^a_b{\bf {\widetilde
Q}}_{\alpha},~~{\bf P}^{a\alpha}_{b}={\bf {\widetilde
P}}_{b}^{a\alpha}+\frac{1}{3}\delta^a_b{\bf {\widetilde
P}}^{\alpha},~~{\bf {\widetilde Q}}^{b}_{b\alpha}=0={\bf
{\widetilde P}}_{b}^{b\alpha}\nonumber\\
{\bf Q}_{\beta a}^{\alpha}={\bf {\widetilde Q}}^{\alpha}_{\beta
a}-\frac{1}{2}\delta^{\alpha}_{\beta}{\bf {\widetilde
Q}}_{a},~~{\bf P}^{ab}_{\alpha}={\bf {\widetilde
P}}_{\beta}^{\alpha a}-\frac{1}{2}\delta^{\alpha}_{\beta}{\bf
{\widetilde P}}^{a},~~{\bf {\widetilde Q}}^{\alpha}_{\alpha
b}=0={\bf{\widetilde P}}_{\alpha}^{\alpha b}\nonumber\\
{\bf Q}_{bc}^{a}={\bf {\widetilde
Q}}^{a}_{bc}+\frac{1}{2}\left(\delta^{a}_{b}{\bf {\widetilde
Q}}_{c}-\delta^{a}_{c}{\bf {\widetilde Q}}_{b}\right),~~{\bf
P}^{ab}_{c}={\bf {\widetilde
P}}^{ab}_{c}+\frac{1}{2}\left(\delta^{a}_{c}{\bf {\widetilde
P}}^{b}-\delta^{b}_{c}{\bf {\widetilde P}}_{a}\right),~~{\bf
{\widetilde Q}}^{a}_{a
b}=0={\bf{\widetilde P}}_{a}^{a b}\nonumber\\
{\bf Q}_{\beta\gamma}^{\alpha}={\bf {\widetilde
Q}}^{\alpha}_{\beta\gamma}+\left(\delta^{\alpha}_{\gamma}{\bf
{\widetilde Q}}_{\beta}-\delta^{\alpha}_{\beta}{\bf {\widetilde
Q}}_{\gamma}\right),~~{\bf P}^{\alpha\beta}_{\gamma}={\bf
{\widetilde
P}}^{\alpha\beta}_{\gamma}+\left(\delta^{\beta}_{\gamma}{\bf
{\widetilde P}}^{\alpha}-\delta^{\alpha}_{\gamma}{\bf {\widetilde
P}}_{\beta}\right),~~{\bf {\widetilde Q}}^{\alpha}_{\alpha
\beta}=0={\bf{\widetilde P}}_{\alpha}^{\alpha \beta}\nonumber\\
{\bf Q}_{\alpha\beta}^{a}=\epsilon_{\alpha\beta}{\bf {\widetilde
Q}}^a,~~{\bf P}^{\alpha\beta}_{a}=\epsilon^{\alpha\beta}{\bf
{\widetilde P}}_a~~~~
\end{eqnarray}
The  kinetic energy of the 45 and ${\overline{45}}$ fields is
given by
\begin{eqnarray}
-\partial_A{\bf Q}_{ij}^{k}\partial^A{\bf Q}_{ij}^{k\dagger}
-\partial_A{\bf P}^{ij}_{k}\partial^A{\bf P}^{ij\dagger}_{k}
=-\partial_A{\cal {\widetilde Q}}_{\alpha}\partial^A{\cal
{\widetilde Q}}_{\alpha}^{\dagger}-\partial_A{\cal {\widetilde
Q}}_{a}\partial^A{\cal {\widetilde
Q}}_{a}^{\dagger}-\partial_A{\cal {\widetilde
Q}}^{a}\partial^A{\cal {\widetilde Q}}^{a\dagger}\nonumber\\
-\partial_A{\cal {\widetilde P}}^{\alpha}\partial^A{\cal
{\widetilde P}}^{\alpha\dagger}-\partial_A{\cal {\widetilde
P}}^{a}\partial^A{\cal {\widetilde P}}^{a\dagger}-\partial_A{\cal
{\widetilde P}}_{a}\partial^A{\cal {\widetilde
P}}_a^{\dagger}-....
\end{eqnarray}
so that the doublet and triplet fields are normalized according to
\begin{eqnarray}
{\bf {\widetilde Q}}_{\alpha}=\frac{1}{2}\sqrt{\frac{3}{2}}{\cal
{\widetilde Q}}_{\alpha},~~{\bf {\widetilde
Q}}_{a}=\sqrt{\frac{2}{3}}{\cal {\widetilde Q}}_{a},~~{\bf
{\widetilde Q}}^{a}=\frac{1}{\sqrt 2}{\cal {\widetilde
Q}}^{a}\nonumber\\
{\bf {\widetilde P}}^{\alpha}=\frac{1}{2}\sqrt{\frac{3}{2}}{\cal
{\widetilde P}}^{\alpha},~~{\bf {\widetilde
P}}^{a}=\sqrt{\frac{2}{3}}{\cal {\widetilde P}}^{a},~~{\bf
{\widetilde P}}_{a}=\frac{1}{\sqrt 2}{\cal {\widetilde P}}_{a}
\end{eqnarray}

The mass matrix of the Higgs doublets is given by

{\scriptsize
 { \beqn
 \left[\matrix{
\frac{3}{5}M+\frac{p^2}{M'}(\frac{666}{5}\Lambda_{45}-\frac{273}{10}\Lambda_{210})
&\frac{1}{4}\sqrt{\frac{15}{2}}\frac{p^2}{M'}(8\Lambda_{45}-\frac{2}{3}\Lambda_{210})
& 0 \cr
 \frac{1}{4}\sqrt{\frac{15}{2}}\frac{p^2}{M'}(8\Lambda_{45}-\frac{2}{3}\Lambda_{210}) &
 -\frac{1}{2}M+\frac{p^2}{M'}(-37\Lambda_{45}+\frac{7}{12}\Lambda_{210})
& 0 \cr
 0 & 0 &
 M+\frac{p^2}{M'}(180\Lambda_{45}-10\Lambda_{210})
 }\right]~~~
\eeqn }}

\noindent where the columns  are labelled by  (${\cal {
Q}}_{\alpha},{\cal {\widetilde Q}}_{\alpha},{\cal { P}}_{\alpha}$)
and rows by (${\cal { P}}^{\alpha},{\cal {\widetilde
P}}^{\alpha},{\cal { Q}}^{\alpha}$). The triplet mass matrix in
the basis  where the columns  are labelled by (${\cal {
Q}}_{a},{\cal {\widetilde Q}}_{a},{\cal { P}}_{a}, {\cal
{\widetilde P}}_{a}$) and rows by (${\cal { P}}^{a},{\cal
{\widetilde P}}^{a},{\cal { Q}}^{a}, {\cal {\widetilde Q}}^{a}$) is
given by
{\scriptsize
 {
 \beqn
 \left[\matrix{
\frac{3}{5}M+\frac{p^2}{M'}(\frac{696}{5}\Lambda_{45}-\frac{257}{15}\Lambda_{210})
&\frac{1}{2}\sqrt{\frac{10}{3}}\frac{p^2}{M'}(8\Lambda_{45}-\frac{2}{3}\Lambda_{210})
& 0 &0\cr
 \frac{1}{2}\sqrt{\frac{10}{3}}\frac{p^2}{M'}(8\Lambda_{45}-\frac{2}{3}\Lambda_{210}) &
 -\frac{2}{3}M+\frac{p^2}{M'}(-16\Lambda_{45}-2\Lambda_{210})
& 0 &0\cr
 0 & 0 &
 M+\frac{p^2}{M'}(180\Lambda_{45}-10\Lambda_{210})&0\cr
 0&0&0&-\frac{1}{2}M+\frac{p^2}{M'}(-42\Lambda_{45}+\Lambda_{210})
 }\right]\nonumber\\
\eeqn }}

It is clear from the above Higgs mass matrices that one needs  to diagonalize
in the Higgs doublet sub-sectors
(${\cal {\widetilde Q}}_{\alpha}, {\cal {\widetilde P}}^{\alpha}$)
and (${\cal { Q}}_{\alpha}, {\cal { P}}^{\alpha}$) and
in the Higgs triplet subsectors
(${\cal {\widetilde Q}}_{a}, {\cal {\widetilde P}}^{a}$) and
(${\cal { Q}}_{a}, {\cal { P}}^{a}$).
After, diagonalization we have the
following  pairs of doublets and triplets:
\begin{eqnarray}
{\mathsf D}_1:~({\cal Q}^{\alpha}, {\cal
P}_{\alpha}),~~~~~~~{\mathsf T}_1:~({\cal
Q}^{a}, {\cal P}_{a})\nonumber\\
{\mathsf D}_2:~({\cal Q}_{\alpha}^{\prime},{\cal
P}^{\prime\alpha}),~~~~~~~{\mathsf T}_2:~({\cal
Q}_{a}^{\prime},{\cal
P}^{\prime a})\nonumber\\
{\mathsf D}_3:~({\cal {\widetilde Q}}_{\alpha}^{\prime}, {\cal
{\widetilde P}}^{\prime \alpha }),~~~~~~~{\mathsf T}_3:~({\cal
{\widetilde Q}}^{\prime }_a, {\cal
{\widetilde P}}^{ \prime a})\nonumber\\
{\mathsf T}_4:~({\cal {\widetilde Q}}^{a}, {\cal {\widetilde
P}}_a)
\end{eqnarray}

The prime fields above are expressed in terms of the original ones
through the following transformation matrices
\begin{eqnarray}
\left[\matrix{ ({\cal { Q}}^{\prime }_a,{\cal { P}}^{\prime a}
)\cr({\cal {\widetilde Q}}^{\prime}_a,{\cal {\widetilde
P}}^{\prime a})}\right] = \left[\matrix{ \cos\vartheta_{\mathsf T}
& \sin\vartheta_{\mathsf T}\cr -\sin\vartheta_{\mathsf T} &
\cos\vartheta_{\mathsf T}}\right]\left[\matrix{({\cal { Q}}_{a},
{\cal
{ P}}^{a})\cr ({\cal {\widetilde Q}}_{ a},{\cal {\widetilde P}}^{ a}) }\right]\\
\nonumber\\
 \left[\matrix{({\cal { Q}}^{\prime
}_{\alpha},{\cal { P}}^{\prime \alpha} )\cr ({\cal {\widetilde
Q}}^{\prime}_{\alpha},{\cal {\widetilde P}}^{\prime \alpha})
}\right] = \left[\matrix{ \cos\vartheta_{\mathsf D} &
\sin\vartheta_{\mathsf D}\cr -\sin\vartheta_{\mathsf D} &
\cos\vartheta_{\mathsf D}}\right]\left[\matrix{({\cal {
Q}}_{\alpha}, {\cal { P}}^{\alpha})\cr ({\cal {\widetilde Q}}_{
\alpha},{\cal {\widetilde P}}^{ \alpha})
 }\right]
\end{eqnarray}
where
\begin{equation}
\tan\vartheta_{\mathsf T}=\frac{1}{{\mathsf t_3}}\left({\mathsf
t_2}+\sqrt{{\mathsf t_2}^2+{\mathsf
t_3}^2}\right),~~~\tan\vartheta_{\mathsf D}=\frac{1}{{\mathsf
d_3}}\left({\mathsf d_2}+\sqrt{{\mathsf d_2}^2+{\mathsf
d_3}^2}\right)
\end{equation}
and that
\begin{eqnarray}
{\mathsf
d}_1=\frac{1}{10}M+\frac{p^2}{M'}\left(\frac{481}{5}\Lambda_{45}
-\frac{1603}{60}\Lambda_{210}\right)\nonumber\\
{\mathsf
d}_2=-\frac{11}{10}M+\frac{p^2}{M'}\left(-\frac{851}{5}\Lambda_{45}
+\frac{1673}{60}\Lambda_{210}\right)\nonumber\\
{\mathsf
d_3}=\frac{1}{2}\sqrt{\frac{15}{2}}\frac{p^2}{M'}\left(8\Lambda_{45}
-\frac{2}{3}\Lambda_{210}\right)\nonumber\\
 {\mathsf
t}_1=-\frac{1}{15}M+\frac{p^2}{M'}\left(\frac{616}{5}\Lambda_{45}
-\frac{287}{15}\Lambda_{210}\right)\nonumber\\
{\mathsf
t}_2=-\frac{19}{15}M+\frac{p^2}{M'}\left(-\frac{776}{5}\Lambda_{45}
+\frac{227}{15}\Lambda_{210}\right)\nonumber\\
{\mathsf t_3}=\sqrt{\frac{10}{3}}\frac{p^2}{M'}\left(8\Lambda_{45}
-\frac{2}{3}\Lambda_{210}\right)
\end{eqnarray}
The mass eigenvalues are found to be
\begin{eqnarray}
M_{{\mathsf D}_2,{\mathsf D}_3}=\frac{1}{2}\left({\mathsf
d_1}\pm\sqrt{{\mathsf d_2}^2+{\mathsf d_3}^2}\right)\nonumber\\
 M_{{\mathsf T}_2,{\mathsf
T}_3}=\frac{1}{2}\left({\mathsf t_1}\pm\sqrt{{\mathsf
t_2}^2+{\mathsf t_3}^2}\right)
\end{eqnarray}
and of course
\begin{eqnarray}
M_{{\mathsf D}_1}=M_{{\mathsf
T}_1}=M+\frac{p^2}{M'}\left(180\Lambda_{45}
-10\Lambda_{210}\right)\nonumber\\
M_{{\mathsf
T}_4}=-\frac{1}{2}M+\frac{p^2}{M'}\left(-42\Lambda_{45}
+\Lambda_{210}\right)
\end{eqnarray}
As an illustration we discuss in further detail the implication of the
 massless-ness condition for the doublet ${\mathsf D}_2$.
 Here  the condition $M_{{\mathsf D}_2}=0 $
 together with the symmetry breaking
condition, Eq. (\ref{symmetrybreaking}) gives a relationship among
the parameters $\Lambda_{45}$ and  $\Lambda_{210}$
\begin{eqnarray}
\left(\frac{539}{5}\Lambda_{45}-\frac{1579}{60}\Lambda_{210}\right)^2
=\left(-\frac{1489}{5}\Lambda_{45}+\frac{1409}{60}\Lambda_{210}\right)^2
+\frac{15}{8}\left(8\Lambda_{45}-\frac{2}{3}\Lambda_{210}\right)^2
\end{eqnarray}
The two  roots to the equations  above are
\begin{equation}
\Lambda_{210}\approx 7.4\Lambda_{45}, ~~\Lambda_{210}\approx
-37.2\Lambda_{45}
\end{equation}
Using the roots above  the full doublet-triplet Higgs mass  spectrum
can now be  computed. The results are summarized in  the Table
below.
 {\small{ \small{
\begin{center} \begin{tabular}{|c|c|c|c|c|c|c|c|}
\multicolumn{8}{c}{ Massless Doublet ${\mathsf D}_2$. $M_D$ and
$M_T$ are in units of
$ {\overline M}\equiv\frac{p^2}{M'}\Lambda_{45}$ }\\
\hline
$\frac{M}{{\overline
M}}$&$\frac{\Lambda_{210}}{\Lambda_{45}}$ &
$M_{{\mathsf D}_1}$& $M_{{\mathsf D}_3}$& $M_{{\mathsf T}_1}$ & $M_{{\mathsf T}_2}$& $M_{{\mathsf T}_3}$ & $M_{{\mathsf T}_4}$\\
\hline\hline
145.6 &7.4& 251.6&-87.0&251.6&99.8&127.9&92.6\\
\hline
-32.8 &-37.2& 519.0&1086&519.0&757.7&78.8&137.2\\
\hline
\end{tabular}
\end{center}
}}}
\subsection{Quark, lepton and neutrino masses}
As pointed out in Ref.\cite{bgns} the quark, lepton and neutrino masses
can arise from the quartic couplings involving two 16-plets of matter  and
two 144-plet of Higgs fields.   Cubic Yukawa couplings arise  when one of the
two 144-plets is replaced by a VEV while  mass terms arise when the remaining
Higgs field in the  cubic interaction develops a VEV.  As an illustration of how
this comes about in a concrete way we will consider the following quartic coupling
for computing the masses of quarks and leptons: $ (16\times
16)_{120}({144}\times{144})_{120},~(16\times 16)_{{120}}
(\overline{144}\times \overline{144})_{120},~
 (16\times
16)_{\overline{126}} (144\times 144)_{126}$. However, this
subsection is to be treated as an independent one. That is we do
not make use of the results of the previous subsections here.

 The
relevant terms in Eqs. (94), (97) and (100) that gives mass growth
to quarks and leptons are

\begin{eqnarray}
{\mathsf
W}^{^{(120)}}_{mass}=\xi_{\acute{a}\acute{b},\acute{c}\acute{d}}^{^{(120)(-)}}\left[-\frac{4}{3}\epsilon_{ijklm}
{\bf M}_{\acute{a}}^{ij\bf T}{\bf
M}_{\acute{b}}^{kn}\frac{4}{3}{\cal P}_{\acute{c}n}^{x\bf T}{\cal
P}_{\acute{d}x}^{lm}-\frac{4}{3\sqrt 5} \epsilon_{ijklm} {\bf
M}_{\acute{a}}^{ij\bf T}{\bf M}_{\acute{b}}^{kn}{\cal
P}_{\acute{c}n}^{l\bf T}{\cal P}_{\acute{d}}^{m}\right.\nonumber\\
\left.-\frac{16}{3} {\bf M}_{\acute{a}}^{ij\bf T}{\bf
M}_{\acute{b}j}{\cal P}_{\acute{c}k}^{\bf T}{\cal
P}_{\acute{d}i}^k-\frac{16}{3}{\bf M}_{\acute{a}}^{\bf T}{\bf
M}_{\acute{b}i}{\cal P}_{\acute{c}k}^{ij\bf T}{\cal
P}_{\acute{d}j}^k\right.\nonumber\\
\left.-\frac{8}{3\sqrt 5}{\bf M}_{\acute{a}}^{\bf T}{\bf
M}_{\acute{b}i}{\cal P}_{\acute{c}}^{j\bf T}{\cal
P}_{\acute{d}j}^i +\frac{32}{15} {\bf M}_{\acute{a}i}^{\bf T}{\bf
M}_{\acute{b}j}{\cal P}_{\acute{c}}^{j\bf T}{\cal
P}_{\acute{d}}^i\right]\nonumber\\
+\zeta_{\acute{a}\acute{b},\acute{c}\acute{d}}^{^{(120)(-)}}\left[\frac{4}{3}\epsilon_{ijklm}
{\bf M}_{\acute{a}}^{ij\bf T}{\bf M}_{\acute{b}}^{kl}{\cal
Q}_{\acute{c}}^{n\bf T}{\cal Q}_{\acute{d}n}^m+\frac{16}{3} {\bf
M}_{\acute{a}i}^{\bf T}{\bf M}_{\acute{b}}^{jk}{\cal
Q}_{\acute{c}x}^{i\bf T}{\cal
Q}_{\acute{d}jk}^x\right.\nonumber\\
\left.-\frac{16}{3} {\bf M}_{\acute{a}i}^{\bf T}{\bf
M}_{\acute{b}}^{ij}{\cal Q}_{\acute{c}x}^{k\bf T}{\cal
Q}_{\acute{d}kj}^x+\frac{16}{3\sqrt 5} {\bf M}_{\acute{a}i}^{\bf
T}{\bf M}_{\acute{b}}^{jk}{\cal Q}_{\acute{c}j}^{i\bf T}{\cal
Q}_{\acute{d}k}\right.\nonumber\\
\left.+\frac{8}{3\sqrt 5} {\bf M}_{\acute{a}i}^{\bf T}{\bf
M}_{\acute{b}}^{ij}{\cal Q}_{\acute{c}j}^{k\bf T}{\cal
Q}_{\acute{d}k} -\frac{16}{3} {\bf M}_{\acute{a}}^{\bf T}{\bf
M}_{\acute{b}i}{\cal Q}_{\acute{c}}^{j\bf T}{\cal
Q}_{\acute{d}j}^i\right]
\end{eqnarray}
and
\begin{eqnarray}
{\mathsf
W}^{^{({\overline{126}},126)}}_{mass}=\varrho_{\acute{a}\acute{b},\acute{c}\acute{d}}^{^{(126,\overline{126})(+)}}
\left[\frac{4}{15}\epsilon_{ijklm} {\bf M}_{\acute{a}}^{ij\bf
T}{\bf M}_{\acute{b}}^{kl}{\cal Q}_{\acute{c}}^{n\bf T}{\cal
Q}_{\acute{d}n}^m\right.\nonumber\\
\left.-\frac{4}{15}{\bf M}_{\acute{a}}^{ij\bf T}{\bf
M}_{\acute{b}k}\left({\cal Q}_{\acute{c}ij}^{l\bf T}{\cal
Q}_{\acute{d}l}^k-\frac{1}{\sqrt 5}{{\cal
Q}}_{\acute{c}j}^{\bf T}{\cal Q}_{\acute{d}i}^k\right)\right.\nonumber\\
\left.+\frac{1}{15}{\bf M}_{\acute{a}}^{ij\bf T}{\bf
M}_{\acute{b}j}\left(2{\cal Q}_{\acute{c}ik}^{l\bf T}{\cal
Q}_{\acute{d}l}^k+\frac{1}{\sqrt 5}{{\cal Q}}_{\acute{c}k}^{\bf
T}{{\cal
Q}}_{\acute{d}i}^k\right)\right.\nonumber\\
\left.+\frac{32}{5}{\bf M}_{\acute{a}}^{\bf T}{\bf
M}_{\acute{b}i}{\cal Q}_{\acute{c}}^{j\bf T}{\cal
Q}_{\acute{d}j}^i+\frac{16}{15\sqrt 5}{\bf M}_{\acute{a}}^{\bf
T}{\bf M}_{\acute{b}}{\cal Q}_{\acute{c}}^{i\bf T}{\cal
Q}_{\acute{d}i}\right]
\end{eqnarray}

For completeness we identify the Standard Model particles as
follows:
\begin{eqnarray}\label{particles}
{\bf M}_{\acute{a}}=\Nu_{L\acute{a}}^{\cal C};~~~~{\bf
M}_{\acute{a}\alpha}={\bf D}_{L\acute{a}\alpha}^{\cal C}; ~~~~{\bf
M}_{\acute{a}}^{\alpha\beta}=\epsilon^{\alpha\beta\gamma}{\bf
U}_{L\acute{a}\gamma}^{\cal C};
~~~~{\bf M}_{\acute{a}4}={\bf E}_{L\acute{a}}^-\nonumber\\
{\bf M}_{\acute{a}}^{4\alpha}={\bf U}_{L\acute{a}}^{\alpha};
~~~~{\bf M}_{\acute{a}5}=\Nu_{L\acute{a}};~~~~{\bf
M}_{\acute{a}}^{5\alpha}={\bf D}_{L\acute{a}}^{\alpha}; ~~~~{\bf
M}_{\acute{a}}^{45}={\bf E}_{L\acute{a}}^+
\end{eqnarray}
where $\alpha, \beta, \gamma=1,2,3$ are color indices and the
superscript $^{\cal C}$ denotes charge conjugation. We adopt the
convention that all particles are left handed($L$).

We now single out the terms that are candidates for Majorana and
Dirac neutrinos, Type II see-saw mechanism, down-type and up-type
quarks and charged leptons.
\\
\\
Candidates for \textsc{Majorana Neutrinos}:
\begin{eqnarray}
{\bf M}_{\acute{a}}{\bf M}_{\acute{b}}\left\{\left[
 \frac{16}{15\sqrt
5}\varrho_{\acute{a}\acute{b},\acute{c}\acute{d}}^{^{(126,\overline{126})(+)}}
\right]{{\cal Q}}_{\acute{c}}^i{{\cal Q}}_{\acute{d}i}\right\}
\end{eqnarray}
\\
Candidates for \textsc{Dirac Neutrinos}:
\begin{eqnarray}
{\bf M}_{\acute{a}i}{\bf
M}_{\acute{b}}\left\{\left[\frac{8}{3\sqrt
5}\xi_{\acute{a}\acute{b},\acute{c}\acute{d}}^{^{(120)(-)}}\right]{{\cal
P}}_{\acute{c}}^j{{\cal P}}_{\acute{d}j}^i+\left[
\frac{16}{3}\xi_{\acute{a}\acute{b},\acute{c}\acute{d}}^{^{(120)(-)}}\right]{{\cal
P}}_{\acute{c}k}^{ij}{{\cal P}}_{\acute{d}j}^k\right.\nonumber\\
\left.+\left[\frac{32}{5}\varrho_{\acute{a}\acute{b},\acute{c}\acute{d}}^{^{(126,\overline{126})(+)}}
+\frac{16}{3}\zeta_{\acute{a}\acute{b},\acute{c}\acute{d}}^{^{(120)(-)}}\right]{{\cal
Q}}_{\acute{c}}^{j}{{\cal Q}}_{\acute{d}j}^i\right \}
\end{eqnarray}
\\
Candidates for \textsc{Type II See-Saw Mechanism}:
\begin{eqnarray}
{\bf M}_{\acute{a}i}{\bf M}_{\acute{b}j}\left[
-\frac{32}{15}\xi_{\acute{a}\acute{b},\acute{c}\acute{d}}^{^{(120)(-)}}
\right]{{\cal P}}_{\acute{c}}^{i}{{\cal P}}_{\acute{d}}^j
\end{eqnarray}
\\
Candidates for \textsc{Down-type Quarks} and \textsc{Charged
Leptons}:
\begin{eqnarray}
{\bf M}_{\acute{a}}^{ij}{\bf M}_{\acute{b}j}\left\{\left[
-\frac{16}{3}\xi_{\acute{a}\acute{b},\acute{c}\acute{d}}^{^{(120)(-)}}\right]{{\cal
P}}_{\acute{c}i}^k{{\cal P}}_{\acute{d}k}\right.\nonumber\\
\left.+\left[\frac{1}{15\sqrt
5}\varrho_{\acute{a}\acute{b},\acute{c}\acute{d}}^{^{(126,\overline{126})(+)}}
+\frac{8}{3\sqrt
5}\zeta_{\acute{a}\acute{b},\acute{c}\acute{d}}^{^{(120)(-)}}\right]{{\cal
Q}}_{\acute{c}k}{{\cal Q}}_{\acute{d}j}^k \right.\nonumber\\
\left.+\left[\frac{2}{15}\varrho_{\acute{a}\acute{b},\acute{c}\acute{d}}^{^{(126,\overline{126})(+)}}
-\frac{16}{3}\zeta_{\acute{a}\acute{b},\acute{c}\acute{d}}^{^{(120)(-)}}\right]{{\cal
Q}}_{\acute{c}ik}^l{{\cal Q}}_{\acute{d}l}^k\right\}\nonumber\\
+{\bf M}_{\acute{a}}^{ij}{\bf M}_{\acute{b}k}\left\{\frac{1}{\sqrt
5}\left[\frac{4}{15\sqrt
5}\varrho_{\acute{a}\acute{b},\acute{c}\acute{d}}^{^{(126,\overline{126})(+)}}
+\frac{16}{3\sqrt
5}\zeta_{\acute{a}\acute{b},\acute{c}\acute{d}}^{^{(120)(-)}}\right]{\cal
Q}_{\acute{c}j}{{\cal Q}}_{\acute{d}i}^k \right.\nonumber\\
\left.+\left[-\frac{4}{15}\varrho_{\acute{a}\acute{b},\acute{c}\acute{d}}^{^{(126,\overline{126})(+)}}
+\frac{16}{3}\zeta_{\acute{a}\acute{b},\acute{c}\acute{d}}^{^{(120)(-)}}\right]{\cal
Q}_{\acute{c}ij}^l{{\cal Q}}_{\acute{d}l}^k \right\}
\end{eqnarray}
\\
Candidates for \textsc{Up-type Quarks}:
\begin{eqnarray}
\epsilon_{ijklm}{\bf M}_{\acute{a}}^{ij}{\bf
M}_{\acute{b}}^{kl}\left\{\left[
\frac{4}{15}\varrho_{\acute{a}\acute{b},\acute{c}\acute{d}}^{^{(126,\overline{126})(+)}}
+\frac{4}{3}\zeta_{\acute{a}\acute{b},\acute{c}\acute{d}}^{^{(120)(-)}}\right]{{\cal
Q}}_{\acute{c}}^n{{\cal Q}}_{\acute{d}n}^m\right\}\nonumber\\
+\epsilon_{ijklm}{\bf M}_{\acute{a}}^{in}{\bf
M}_{\acute{b}}^{jk}\left\{\left[\frac{4}{3}\xi_{\acute{a}\acute{b},\acute{c}\acute{d}}^{^{(120)(-)}}\right]{\cal
P}_{\acute{c}n}^p{{\cal P}}_{\acute{d}p}^{lm}
+\left[\frac{4}{3\sqrt 5}
\xi_{\acute{a}\acute{b},\acute{c}\acute{d}}^{^{(120)(-)}} \right]
{\cal P}_{\acute{c}n}^l{{\cal P}}_{\acute{d}}^{m}\right\}
\end{eqnarray}\\
\\
Next we identify the $SU(3)_C\times U(1)_{em}$  conserving VEV's:
\begin{eqnarray}
{<{\cal Q}_{\acute{c}j}^{i}>\choose {<\cal
P}_{\acute{c}j}^{i}>}={q_{\acute{c }}\choose p_{\acute{c
}}}diag(2,2,2,-3,-3)\nonumber\\
<{\cal
Q}_{\acute{c}j5}^{k}>=\frac{1}{2}\sqrt{\frac{3}{2}}<{\widetilde{\cal
Q}}_{{\acute
{c}}5}>\left(\frac{1}{4}\delta^k_j-\delta^k_4\delta^4_j
     \right)\nonumber\\
     <{\cal
P}_{\acute{c}k}^{j5}>=\frac{1}{2}\sqrt{\frac{3}{2}}<{\widetilde{\cal
P}}_{{\acute {c}}}^{
5}>\left(\frac{1}{4}\delta^j_k-\delta^j_4\delta^4_k
     \right)
\end{eqnarray}
To make further progress, we define the following mass parameters
\[
\begin{array}{cc}
\alpha^{(120)}_{1~\acute {a}\acute
{b}}=13\sqrt{\frac{2}{3}}\zeta_{\acute{a}\acute{b},\acute{c}\acute{d}}^{^{(120)(-)}}<{\widetilde{\cal
Q}}_{{\acute {c}}5}>q_{\acute{d }}, &
 \alpha^{(120)}_{2~\acute
{a}\acute {b}}=-\frac{8}{3\sqrt
5}\zeta_{\acute{a}\acute{b},\acute{c}\acute{d}}^{^{(120)(-)}}<{\cal
{Q}}_{\acute{c}5}>q_{\acute{d }}\\
\alpha^{(120)}_{3~\acute {a}\acute
{b}}=16\xi_{\acute{a}\acute{b},\acute{c}\acute{d}}^{^{(120)(-)}}<{\cal
{P}}_{\acute{c}5}>p_{\acute{d }},& \alpha^{(126)}_{1~\acute
{a}\acute {b}}=-\frac{11}{5\sqrt
5}\varrho_{\acute{a}\acute{b},\acute{c}\acute{d}}^{^{(126,\overline{126})(+)}}<{{\cal
Q}}_{{\acute {c}}5}>q_{\acute{d }}\\
\alpha^{(126)}_{2~\acute {a}\acute {b}}=-\frac{11}{20\sqrt
6}\varrho_{\acute{a}\acute{b},\acute{c}\acute{d}}^{^{(126,\overline{126})(+)}}<{\widetilde{\cal
Q}}_{{\acute {c}}5}>q_{\acute{d }}\\
 \\
 a^{(120)}_{1~\acute {a}\acute {b}}=-16\zeta_{\acute{a}\acute{b},\acute{c}\acute{d}}^{^{(120)(-)}}<{\cal
{Q}}_{\acute{c}}^5>q_{\acute{d }}, & a^{(120)}_{2~\acute {a}\acute
{b}}=-\frac{8}{\sqrt 5}\sqrt
{6}\xi_{\acute{a}\acute{b},\acute{c}\acute{d}}^{^{(120)(-)}}<{{\cal
P}}_{{\acute {c}}}^{ 5}>p_{\acute{d }}\\
a^{(120)}_{3~\acute {a}\acute {b}}=3\sqrt
3\xi_{\acute{a}\acute{b},\acute{c}\acute{d}}^{^{(120)(-)}}<{\widetilde{\cal
P}}_{{\acute {c}}}^{ 5}>p_{\acute{d }}, & a^{(126)}_{~\acute
{a}\acute
{b}}=-\frac{32}{5}\varrho_{\acute{a}\acute{b},\acute{c}\acute{d}}^{^{(126,\overline{126})(+)}}<{{\cal
Q}}_{{\acute {c}}}^{5}>q_{\acute{d }}\\
\end{array}
\]

We now compute the down quark ($M^{down}$), charged lepton
($M^{electron}$), up quark ($M^{up}$), Dirac neutrino
($M^{Dirac~\nu}$), RR type neutrino ($M^{RR}$) and LL type
neutrino ($M^{LL}$) mass matrices in terms of the mass parameters
defined above.

\begin{eqnarray}
M^{down}_{\acute {a}\acute {b}}=\left({ A} +{B}\right)_{\acute
{a}\acute
 {b}}\nonumber\\
M^{electron}_{\acute {a}\acute {b}}=\left({ A} -3{
B}\right)_{\acute {a}\acute {b}}
\end{eqnarray}
where
\begin{eqnarray}
 { A}_{\acute {a}\acute {b}}=\left[
 \frac{59}{52}\alpha^{(120)}_{1}+\frac{7}{2}\alpha^{(120)}_{2}+\alpha^{(120)}_{3}
 +\frac{16}{11}\alpha^{(126)}_{1}+\frac{29}{22}\alpha^{(126)}_{2}\right]_{\acute {a}\acute
 {b}}
 \end{eqnarray}
 \begin{eqnarray}
{ B}_{\acute {a}\acute {b}}=\left[\frac{7}{52}\alpha^{(120)}_{1}
+\frac{5}{2}\alpha^{(120)}_{2}+\frac{5}{11}\alpha^{(126)}_{1}+\frac{7}{22}\alpha^{(126)}_{2}\right]_{\acute
{a}\acute {b}}
\end{eqnarray}
and for the up quark and Dirac neutrino masses one has
\begin{eqnarray}
M^{up}_{\acute {a}\acute
{b}}=\left[a^{(120)}_{1}+a^{(120)}_{2}+a^{(120)}_{3}
+a^{(126)}\right]_{\acute {a}\acute
 {b}}
 \end{eqnarray}
\begin{eqnarray}
M^{Dirac~\nu}_{\acute {a}\acute
{b}}=\left[-a^{(120)}_{1}-a^{(120)}_{2}+\frac{5}{3}a^{(120)}_{3}+\frac{1}{3}a^{(126)}\right]_{\acute
{a}\acute
 {b}}
\end{eqnarray}
 Majorana masses of RR and LL type for the neutrinos
 is given by
\begin{eqnarray}
M^{RR}_{\acute {a}\acute {b}}= -\frac{16}{15\sqrt 5}
\varrho_{\acute{a}\acute{b},\acute{c}\acute{d}}^{^{(126,\overline{126})(+)}}
<{{\cal Q}}_{{\acute {c}}}^{5}><{{\cal Q}}_{{\acute {d}}5}>
\end{eqnarray}
\begin{eqnarray}
 M^{LL}_{\acute {a}\acute {b}}=
\frac{32}{15}\xi_{\acute{a}\acute{b},\acute{c}\acute{d}}^{^{(120)(-)}}
<{{\cal P}}_{{\acute {c}}}^{5}><{{\cal P}}_{{\acute {d}}}^5>
\end{eqnarray}
For real model building one may  now consider one at a time each
of the doublets $({\cal { Q}}_{\acute {a}\alpha}, {\cal {
P}}^{\alpha}_{\acute {a}})$, $({\cal { Q}}^{\alpha}_{\acute {a}},
{\cal { P}}_{\acute {a}\alpha})$, $({\widetilde{\cal Q}}_{{\acute
{a}}\alpha},{\widetilde{\cal P}}_{{\acute {a}}}^{\alpha})$
massless and find the corresponding contribution to quark and
lepton masses.

\subsection{Baryon and lepton number violating dimension five
operators}
In supesymmetric theories with R parity the  dominant proton decay arises
from dimension five  operators\cite{pdecay1,pdecay2,pdecay3}.
 Here we look for baryon and lepton number violating
dimension five operators in\footnote{For a complete analysis see
Ref.\cite{bgns1}.}
 $\left(16\times16\right)_{10}$ $\left(144\times
144\right)_{10}$, $\left(16\times16\right)_{10}
\left({\overline{144}}\times {\overline{144}}\right)_{10}$ and
$(16\times 16)_{\overline{126}} (144\times 144)_{126}$. We first
collect all the terms from the three quartic couplings which
contribute to baryon and lepton violating interactions. These are
\begin{eqnarray}
{\mathsf W}={\bf M}_{\acute{a}}^{ij}{\bf
M}_{\acute{b}j}\left\{\left[
16\xi_{\acute{a}\acute{b},\acute{c}\acute{d}}^{^{(10)(+)}}\right]<{{\cal
P}}_{\acute{c}i}^k>{{\cal P}}_{\acute{d}k}
+\left[\frac{1}{15\sqrt5}\varrho_{\acute{a}\acute{b},\acute{c}\acute{d}}^{^{(126,\overline{126})(+)}}
+\frac{8}{\sqrt
5}\zeta_{\acute{a}\acute{b},\acute{c}\acute{d}}^{^{(10)(+)}}
\right]{{\cal
Q}}_{\acute{c}k}<{{\cal Q}}_{\acute{d}j}^k> \right.\nonumber\\
\left.+\left[\frac{2}{15}\varrho_{\acute{a}\acute{b},\acute{c}\acute{d}}^{^{(126,\overline{126})(+)}}
+16\zeta_{\acute{a}\acute{b},\acute{c}\acute{d}}^{^{(10)(+)}}
\right]{{\cal
Q}}_{\acute{c}ik}^l<{{\cal Q}}_{\acute{d}l}^k>\right\}\nonumber\\
+{\bf M}_{\acute{a}}^{ij}{\bf
M}_{\acute{b}k}\left\{\left[\frac{4}{15\sqrt
5}\varrho_{\acute{a}\acute{b},\acute{c}\acute{d}}^{^{(126,\overline{126})(+)}}
\right]{\cal Q}_{\acute{c}j}<{{\cal Q}}_{\acute{d}i}^k>
+\left[-\frac{4}{15}\varrho_{\acute{a}\acute{b},\acute{c}\acute{d}}^{^{(126,\overline{126})(+)}}
\right]{\cal
Q}_{\acute{c}ij}^l<{{\cal Q}}_{\acute{d}l}^k>\right\} \nonumber\\
+\epsilon_{ijklm}{\bf M}_{\acute{a}}^{ij}{\bf
M}_{\acute{b}}^{kl}\left\{\left[
\frac{4}{15}\varrho_{\acute{a}\acute{b},\acute{c}\acute{d}}^{^{(126,\overline{126})(+)}}
+2\zeta_{\acute{a}\acute{b},\acute{c}\acute{d}}^{^{(10)(+)}}\right]{{\cal
Q}}_{\acute{c}}^n<{{\cal Q}}_{\acute{d}n}^m>
+2\left[\xi_{\acute{a}\acute{b},\acute{c}\acute{d}}^{^{(10)(+)}}\right]<{\cal
P}_{\acute{c}n}^p>{{\cal P}}_{\acute{d}p}^{nm}
\right.\nonumber\\
\left.-\frac{1}{\sqrt
5}\left[\xi_{\acute{a}\acute{b},\acute{c}\acute{d}}^{^{(10)(+)}}\right]<{\cal
P}_{\acute{c}n}^m>{{\cal P}}_{\acute{d}}^{n}\right\}~~~
\end{eqnarray}
Expanding and collecting the relevant terms and inserting the
triplet mass terms responsible for proton decay we find
\begin{eqnarray}\label{bandl}
{\mathsf W}_{_{B\&L}}=J_{(1)a}{\cal {P}}^{ a}+K^a_{(1)}{\cal
{Q}}_{ a} +{\mathrm M}_{({\cal {Q}}_{ a},{\cal {P}}^{ a})}{\cal
{Q}}_{ a}{\cal {P}}^{ a}\nonumber\\
+J_{(2)a}{\cal {\widetilde P}}^{ a}+K^a_{(2)}{\cal {\widetilde
Q}}_{ a} +{\mathrm M}_{({\cal {\widetilde Q}}_{ a},{\cal
{\widetilde P}}^{
a})}{\cal{\widetilde Q}}_{ a}{\cal {\widetilde P}}^{ a}\nonumber\\
+J_{(3)}^a{\cal {P}}_{ a}+K_{(3)a}{\cal {Q}}^a +{\mathrm
M}_{({\cal {Q}}^{ a},{\cal {P}}_{ a})}{\cal
{Q}}^a{\cal {P}}_{ a}\nonumber\\
+K_{(4)a}{\cal {\widetilde Q}}^{ a}+{\mathrm M}_{({\cal
{\widetilde Q}}^{ a},{\cal {\widetilde P}}_{ a})}{\cal {\widetilde
Q}}^{ a}{\cal {\widetilde P}}_{ a}
\end{eqnarray}
where we have defined
\begin{eqnarray}
J_{(1)a}=\left[-\frac{2}{\sqrt
5}p\xi_{\acute{a}\acute{b}}^{^{(10)(+)}}\right]\epsilon_{aijkl}{\bf
M}_{\acute{a}}^{ij}{\bf M}_{\acute{b}}^{kl}\nonumber\\
J_{(2)a}=\left[10\sqrt{\frac{2}{3}}p\xi_{\acute{a}\acute{b}}^{^{(10)(+)}}\right]\epsilon_{aijkl}{\bf
M}_{\acute{a}}^{ij}{\bf M}_{\acute{b}}^{kl}\nonumber\\
J_{(3)}^a=\left[-32p\xi_{\acute{a}\acute{b}}^{^{(10)(+)}}\right]{\bf
M}_{\acute{a}i}{\bf M}_{\acute{b}}^{ia}\nonumber\\
K^a_{(1)}=\left[-\frac{16}{\sqrt
5}q\zeta_{\acute{a}\acute{b}}^{^{(10)(+)}}+\frac{2}{3\sqrt
5}q\varrho_{\acute{a}\acute{b}}^{^{(126,\overline{126})(+)}}\right]{\bf
M}_{\acute{a}\alpha}{\bf M}_{\acute{b}}^{\alpha a}\nonumber\\
+\left[-\frac{16}{\sqrt
5}q\zeta_{\acute{a}\acute{b}}^{^{(10)(+)}}-\frac{2}{3\sqrt
5}q\varrho_{\acute{a}\acute{b}}^{^{(126,\overline{126})(+)}}\right]{\bf
M}_{\acute{a}b}{\bf M}_{\acute{b}}^{b a}\nonumber\\
K^a_{(2)}=\left[80\sqrt{\frac{2}{3}}q\zeta_{\acute{a}\acute{b}}^{^{(10)(+)}}-\frac{2}{15}
 \sqrt{\frac{2}{3}}q\varrho_{\acute{a}\acute{b}}^{^{(126,\overline{126})(+)}}\right]{\bf
M}_{\acute{a}\alpha}{\bf M}_{\acute{b}}^{\alpha a}\nonumber\\
+\left[80\sqrt{\frac{2}{3}}q\zeta_{\acute{a}\acute{b}}^{^{(10)(+)}}+\frac{2}{15}
 \sqrt{\frac{2}{3}}q\varrho_{\acute{a}\acute{b}}^{^{(126,\overline{126})(+)}}\right]{\bf
M}_{\acute{a}b}{\bf M}_{\acute{b}}^{b a}\nonumber\\
K_{(3)a}=\left[4q\zeta_{\acute{a}\acute{b}}^{^{(10)(+)}}+\frac{8}{15}q
\varrho_{\acute{a}\acute{b}}^{^{(126,\overline{126})(+)}}\right]
\epsilon_{aijkl}{\bf
M}_{\acute{a}}^{ij}{\bf M}_{\acute{b}}^{kl}\nonumber\\
K_{(4)a}=\left[-\frac{4\sqrt
2}{15}q\varrho_{\acute{a}\acute{b}}^{^{(126,\overline{126})(+)}}\right]{\bf
M}_{\acute{a}a}{\bf
M}_{\acute{b}}^{\alpha\beta}\epsilon_{\alpha\beta}
\end{eqnarray}
Integrating out the Higgs triplet fields in Eq.(\ref{bandl}) and
expanding the results in Standard Model particle states, we get
\begin{eqnarray}
{\mathsf W}_{_{B\&L}}^{dim-5}=128pq\left\{\left(\frac{1}{5{\mathrm
M}_{({\cal {Q}}_{ a},{\cal {P}}^{ a})}}+\frac{50}{3{\mathrm
M}_{({\cal {\widetilde Q}}_{ a},{\cal {\widetilde P}}^{
a})}}\right)\xi_{\acute{a}\acute{b}}^{^{(10)(+)}}\zeta_{\acute{c}\acute{d}}^{^{(10)(+)}}\right.\nonumber\\
\left.-\frac{4}{{\mathrm M}_{({\cal {Q}}^{ a},{\cal {P}}_{
a})}}\left(\zeta_{\acute{a}\acute{b}}^{^{(10)(+)}}
+\frac{2}{15}\varrho_{\acute{a}\acute{b}}^{^{(126,\overline{126})(+)}}\right)\xi_{\acute{c}\acute{d}}^{^{(10)(+)}}\right\}\nonumber\\
\times \left[\epsilon_{abc}{\bf U}_{L\acute{a}}^a{\bf
D}_{L\acute{b}}^b\left({\bf E}_{L\acute{c}}^{-}{\bf
U}_{L\acute{d}}^c+ \Nu_{L\acute{c}}{\bf
D}_{L\acute{d}}^c\right)+2\epsilon^{abc}{\bf
U}_{L\acute{a}a}^{\cal C}{\bf E}_{L\acute{b}}^{+}{\bf
D}_{L\acute{c}b}^{\cal C}{\bf U}_{L\acute{d}c}^{\cal C}
\right]\nonumber\\
-\frac{16}{3}pq\left\{\left(\frac{1}{5{\mathrm M}_{({\cal {Q}}_{
a},{\cal {P}}^{ a})}}+\frac{2}{3{\mathrm M}_{({\cal {\widetilde
Q}}_{ a},{\cal {\widetilde P}}^{
a})}}\right)\xi_{\acute{a}\acute{b}}^{^{(10)(+)}}\varrho_{\acute{c}\acute{d}}^{^{(126,\overline{126})(+)}}\right\}\nonumber\\
\times \left[\epsilon_{abc}{\bf U}_{L\acute{a}}^a{\bf
D}_{L\acute{b}}^b\left({\bf E}_{L\acute{c}}^{-}{\bf
U}_{L\acute{d}}^c+ \Nu_{L\acute{c}}{\bf
D}_{L\acute{d}}^c\right)-2\epsilon^{abc}{\bf
U}_{L\acute{a}a}^{\cal C}{\bf E}_{L\acute{b}}^{+}{\bf
D}_{L\acute{c}b}^{\cal C}{\bf U}_{L\acute{d}c}^{\cal C} \right]
\end{eqnarray}

\section{Conclusions}
In this paper we have given an analysis of the couplings of the
$144+\overline{144}$ multiplets. This multiplet is interesting
since it allows for the breaking of $SO(10)$ symmetry in a single
step down to the Standard Model gauge group symmetry
$SU(3)_C\times SU(2)_L\times U(1)_Y$. The $144$ multiplet is a
vector-spinor representation of  $SO(10)$ with a constraint. The
constraint is needed  to reduce the components of the
vector-spinor from 160 down to 144. These features make the
analysis of the couplings of $144$ and of $\overline{144}$ more
complex than the couplings of ordinary spinors and tensor
representations of $SO(10)$. In this paper we have utilized the
techniques of the basic theorem to compute a variety of couplings
involving the constrained vector-spinors: cubic couplings
involving vector-spinors and tensors, self-couplings of the
vector-spinors, and  couplings  of the vector-spinors with the 16
and $\overline{16}$  spinor representations of  $SO(10)$. These
couplings all enter in model building involving the spinors. Of
course, the full set of couplings involving the vector-spinors are
even larger, but these can also be computed using the techniques
discussed here.
We have also given illustrative examples of how Yukawa couplings,
quark-lepton masses, and  Dirac and Majorana neutrino masses  arise
from the couplings involving the 144 plet of Higgs. Finally we have
exhibited how the baryon and  lepton number violating interactions
arise from the matter and 144 -plet couplings.  It is hoped that the
techniques and the results presented here  will be helpful in further model building
involving the vector-spinor representations.

\begin{center}
{\bf ACKNOWLEDGEMENTS}
\end{center}
The analysis of this paper  was motivated by the work of Refs. \cite{bgns,bgns1}  with Kaladi S. Babu
and Ilia Gogoladze on the vector-spinor  multiplet.   It is a pleasure to acknowledge
fruitful and illuminating communications with them on many aspects of the
 vector-spinor multiplet and its application  for SO(10) model building.
The work is supported in part by NSF grant  PHY-0546568.

\section{Appendix A: Details of couplings from 10-plet mediation}
In this Appendix we expand the $SO(10)$ coupling structures that
enter in 10-plet mediation in Secs.(4,5,6) in a $SU(5)\times U(1)$
basis. We list these structures below

\begin{eqnarray}
h^{^{{(10)(+)}}}_{\acute{c}\acute{d}}{\bf P}_{\acute{c}j\nu}^{\bf
T}{\bf
P}_{\acute{d}\nu}=h^{^{{(10)(+)}}}_{\acute{c}\acute{d}}\left[\frac{\sqrt{6}
}{5}{\cal P}_{\acute{c}jk}^{\bf T}{\cal P}_{\acute{d}}^k
+\sqrt{\frac{2}{5}}{\cal P}_{\acute{c}jk}^{(S)\bf T}{\cal
P}_{\acute{d}}^k+{\cal P}_{\acute{c}j}^{k\bf T}{\cal
P}_{\acute{d}k}\right]
\end{eqnarray}
\begin{eqnarray}
h^{^{{(10)(+)}}}_{\acute{c}\acute{d}}\epsilon_{jklmn}{\bf
P}_{\acute{c}\nu}^{kl\bf T}{\bf
P}_{\acute{d}\nu}^{mn}=h^{^{{(10)(+)}}}_{\acute{c}\acute{d}}\left[4{\cal
P}_{\acute{c}r}^{pq\bf T}{\cal
P}_{\acute{d}pqj}^r-4\sqrt{\frac{2}{15}}{\cal
P}_{\acute{c}pq}^{\bf T}{\cal P}_{\acute{d}j}^{pq} -\frac{4\sqrt
6}{5}{\cal P}_{\acute{c}}^{p\bf T}{\cal P}_{\acute{d}pj}\right]
\end{eqnarray}
\begin{eqnarray}
h^{^{{(10)(+)}}}_{\acute{a}\acute{b}}{\bf P}_{\acute{a}\nu}^{ij\bf
T}{\bf
P}_{\acute{b}\nu}^{kl}=h^{^{{(10)(+)}}}_{\acute{a}\acute{b}}\left[
\frac{1}{6}\epsilon^{ijpqr}{\cal P}_{\acute{a}pqr}^{s\bf T}{\cal
P}_{\acute{b}s}^{kl}+\frac{1}{6}\epsilon^{klpqr}{\cal
P}_{\acute{a}s}^{ij\bf
T}{\cal P}_{\acute{b}pqr}^{s}\right.\nonumber\\
\left.+\frac{1}{12\sqrt{5}}\epsilon^{ijpqr}{\cal
P}_{\acute{a}pqr}^{k\bf T}{\cal
P}_{\acute{b}}^{l}-\frac{1}{12\sqrt{5}}\epsilon^{klpqr}{\cal
P}_{\acute{a}}^{i\bf
T}{\cal P}_{\acute{b}pqr}^{j}\right.\nonumber\\
\left.-\frac{1}{12\sqrt{5}}\epsilon^{ijpqr}{\cal
P}_{\acute{a}pqr}^{l\bf T}{\cal
P}_{\acute{b}}^{k}+\frac{1}{12\sqrt{5}}\epsilon^{klpqr}{\cal
P}_{\acute{a}}^{j\bf
T}{\cal P}_{\acute{b}pqr}^{i}\right.\nonumber\\
\left.+\frac{1}{10\sqrt{6}}\epsilon^{ijlpq}{\cal
P}_{\acute{a}pq}^{\bf T}{\cal
P}_{\acute{b}}^{k}-\frac{1}{10\sqrt{6}}\epsilon^{klipq}{\cal
P}_{\acute{a}}^{j\bf
T}{\cal P}_{\acute{b}pq}\right.\nonumber\\
\left.-\frac{1}{10\sqrt{6}}\epsilon^{ijkpq}{\cal
P}_{\acute{a}pq}^{\bf T}{\cal
P}_{\acute{b}}^{l}+\frac{1}{10\sqrt{6}}\epsilon^{kljpq}{\cal
P}_{\acute{a}}^{i\bf
T}{\cal P}_{\acute{b}pq}\right.\nonumber\\
\left.-\frac{1}{\sqrt{30}}\epsilon^{ijpqr}{\cal
P}_{\acute{a}pq}^{\bf T}{\cal
P}_{\acute{b}r}^{kl}-\frac{1}{\sqrt{30}}\epsilon^{klpqr}{\cal
P}_{\acute{a}p}^{ij\bf T}{\cal P}_{\acute{b}qr}\right]
\end{eqnarray}
\begin{eqnarray}
\bar{h}^{^{{(10)(+)}}}_{\acute{a}\acute{b}}{\bf
Q}_{\acute{a}\mu}^{i\bf T}{\bf
Q}_{\acute{b}ij\mu}=\bar{h}^{^{{(10)(+)}}}_{\acute{a}\acute{b}}\left[-\frac{1}{2\sqrt
{30}}\epsilon_{jklmn}{\cal Q}_{\acute{a}}^{kp\bf T}{\cal
Q}_{\acute{b}p}^{lmn}-\frac{1}{6\sqrt {2}}\epsilon_{jklmn}{\cal
Q}_{(S)\acute{a}}^{kp\bf T}{\cal
Q}_{\acute{b}p}^{lmn}\right.\nonumber\\
\left.+\frac{1}{10}\epsilon_{jklmn}{\cal Q}_{\acute{a}}^{kl\bf
T}{\cal Q}_{\acute{b}}^{mn}+{\cal Q}_{\acute{a}k}^{l\bf T}{\cal
Q}_{\acute{b}lj}^{k}-\frac{1}{2\sqrt {5}}{\cal
Q}_{\acute{a}j}^{k\bf T}{\cal Q}_{\acute{b}k}\right]
\end{eqnarray}
\begin{eqnarray}
\bar{h}^{^{{(10)(+)}}}_{\acute{c}\acute{d}}{\bf
Q}_{\acute{c}\nu}^{j\bf T}{\bf
Q}_{\acute{d}\nu}=\bar{h}^{^{{(10)(+)}}}_{\acute{c}\acute{d}}\left[
\frac{\sqrt{6}}{5}{\cal Q}_{\acute{c}}^{jk\bf T}{\cal
Q}_{\acute{d}k} +\sqrt{\frac{2}{5}}{\cal Q}_{(S)\acute{c}}^{jk\bf
T}{\cal Q}_{\acute{d}k}+{\cal Q}_{\acute{c}k}^{j\bf T}{\cal
Q}_{\acute{d}}^k\right]
\end{eqnarray}
\begin{eqnarray}
\bar{h}^{^{{(10)(+)}}}_{\acute{c}\acute{d}}\epsilon^{jklmn}{\bf
Q}_{\acute{c}kl\nu}^{\bf T}{\bf
Q}_{\acute{d}mn\nu}=\bar{h}^{^{{(10)(+)}}}_{\acute{c}\acute{d}}\left[4{\cal
Q}_{\acute{c}pq}^{r\bf T}{\cal
Q}_{\acute{d}r}^{pqj}-4\sqrt{\frac{2}{15}}{\cal
Q}_{\acute{c}}^{pq\bf T}{\cal Q}_{\acute{d}pq}^{j} -\frac{4\sqrt
6}{5}{\cal Q}_{\acute{c}p}^{\bf T}{\cal Q}_{\acute{d}}^{pj}\right]
\end{eqnarray}
\begin{eqnarray}
\bar{h}^{^{{(10)(+)}}}_{\acute{a}\acute{b}}{\bf
Q}_{\acute{c}ij\nu}^{\bf T}{\bf
Q}_{\acute{d}kl\nu}=\bar{h}^{^{{(10)(+)}}}_{\acute{a}\acute{b}}\left[
\frac{1}{6}\epsilon_{ijpqr}{\cal Q}_{\acute{a}s}^{pqr\bf T}{\cal
Q}_{\acute{b}kl}^{s}+\frac{1}{6}\epsilon_{klpqr}{\cal
Q}_{\acute{a}ij}^{s\bf
T}{\cal Q}_{\acute{b}s}^{pqr}\right.\nonumber\\
\left.+\frac{1}{12\sqrt{5}}\epsilon_{ijpqr}{\cal
Q}_{\acute{a}k}^{pqr\bf T}{\cal
Q}_{\acute{b}l}-\frac{1}{12\sqrt{5}}\epsilon_{klpqr}{\cal
Q}_{\acute{a}i}^{\bf
T}{\cal Q}_{\acute{b}j}^{pqr}\right.\nonumber\\
\left.-\frac{1}{12\sqrt{5}}\epsilon_{ijpqr}{\cal
Q}_{\acute{a}l}^{pqr\bf T}{\cal
Q}_{\acute{b}k}+\frac{1}{12\sqrt{5}}\epsilon_{klpqr}{\cal
Q}_{\acute{a}j}^{\bf
T}{\cal Q}_{\acute{b}i}^{pqr}\right.\nonumber\\
\left.+\frac{1}{10\sqrt{6}}\epsilon_{ijlpq}{\cal
Q}_{\acute{a}}^{pq\bf T}{\cal
Q}_{\acute{b}k}-\frac{1}{10\sqrt{6}}\epsilon_{klipq}{\cal
Q}_{\acute{a}j}^{\bf
T}{\cal Q}_{\acute{b}}^{pq}\right.\nonumber\\
\left.-\frac{1}{10\sqrt{6}}\epsilon_{ijkpq}{\cal
Q}_{\acute{a}}^{pq\bf T}{\cal
Q}_{\acute{b}l}+\frac{1}{10\sqrt{6}}\epsilon_{kljpq}{\cal
Q}_{\acute{a}i}^{\bf
T}{\cal Q}_{\acute{b}}^{pq}\right.\nonumber\\
\left.-\frac{1}{\sqrt{30}}\epsilon_{ijpqr}{\cal
Q}_{\acute{a}}^{pq\bf T}{\cal
Q}_{\acute{b}kl}^{r}-\frac{1}{\sqrt{30}}\epsilon_{klpqr}{\cal
Q}_{\acute{a}ij}^{p\bf T}{\cal Q}_{\acute{b}}^{qr}\right]
\end{eqnarray}

\section{Appendix B: Details of couplings from 120-plet mediation}
In this Appendix we expand the $SO(10)$ coupling structures that
enter in 120-plet mediation in Secs.(4,5,6) in a $SU(5)\times
U(1)$ basis. We list these structures below

\begin{eqnarray}
h_{\acute{c}\acute{d}}^{^{(120)(-)}}{\bf P}_{\acute{c}n\mu}^{\bf
T}{\bf P}_{\acute{d}\mu}^{lm}=h_{\acute{c}\acute{d}}^{^{(120)(-)}}
\left[\frac{1}{2\sqrt {30}}\epsilon^{ijklm}{\cal
P}_{\acute{c}np}^{\bf T}{\cal
P}_{\acute{d}ijk}^{p}-\frac{1}{10}\epsilon^{ijklm}{\cal
P}_{\acute{c}ni}^{\bf T}{\cal P}_{\acute{d}jk}\right.\nonumber\\
\left.+\frac{1}{6\sqrt {2}}\epsilon^{ijklm}{\cal
P}_{\acute{c}np}^{(S)\bf T}{\cal P}_{\acute{d}ijk}^{p}
-\frac{1}{2\sqrt {15}}\epsilon^{ijklm}{\cal
P}_{\acute{c}ni}^{(S)\bf T}{\cal P}_{\acute{d}jk}\right.\nonumber\\
\left.+{\cal P}_{\acute{c}n}^{k\bf T}{\cal P}_{\acute{d}k}^{lm}
+\frac{1}{2\sqrt {5}}{\cal P}_{\acute{c}n}^{l\bf T}{\cal
P}_{\acute{d}}^{m}-\frac{1}{2\sqrt {5}}{\cal P}_{\acute{c}n}^{m\bf
T}{\cal P}_{\acute{d}}^{l}\right]
\end{eqnarray}
\begin{eqnarray}
h_{\acute{a}\acute{b}}^{^{(120)(-)}}{\bf P}_{\acute{a}i\mu}^{\bf
T}{\bf P}_{\acute{b}j\mu}=h_{\acute{a}\acute{b}}^{^{(120)(-)}}
\left[ \sqrt{\frac{3}{10}}{\cal P}_{\acute{a}i}^{k\bf T}{\cal
P}_{\acute{b}jk}-\sqrt{\frac{3}{10}}{\cal P}_{\acute{a}j}^{k\bf
T}{\cal P}_{\acute{b}ik}\right.\nonumber\\
\left.+\sqrt{2}{\cal P}_{\acute{a}i}^{k\bf T}{\cal
P}_{\acute{b}jk}^{(S)}-\sqrt{2}{\cal P}_{\acute{a}j}^{k\bf T}{\cal
P}_{\acute{b}ik}^{(S)}\right]
\end{eqnarray}
\begin{eqnarray}
h_{\acute{c}\acute{d}}^{^{(120)(-)}}{\bf P}_{\acute{c}\mu}^{ij\bf
T}{\bf P}_{\acute{d}\mu}=h_{\acute{c}\acute{d}}^{^{(120)(-)}}
\left[\frac{2}{\sqrt 5}{\cal P}_{\acute{c}k}^{ij\bf T}{\cal
P}_{\acute{d}}^k+\frac{2}{5}{\cal P}_{\acute{c}}^{j\bf T}{\cal
P}_{\acute{d}}^i\right.\nonumber\\
\left.+\frac{1}{6}\epsilon^{ijklm}{\cal P}_{\acute{c}klm}^{n\bf
T}{\cal P}_{\acute{d}n}-\frac{1}{\sqrt{30}}\epsilon^{ijklm}{\cal
P}_{\acute{c}kl}^{\bf T}{\cal P}_{\acute{d}m}\right]
\end{eqnarray}
\begin{eqnarray}
h_{\acute{a}\acute{b}}^{^{(120)(-)}}{\bf P}_{\acute{a}\mu}^{\bf
T}{\bf P}_{\acute{b}i\mu}=h_{\acute{a}\acute{b}}^{^{(120)(-)}}
\left[ \frac{\sqrt 6}{5}{\cal P}_{\acute{a}}^{j\bf T}{\cal
P}_{\acute{b}ij}+\sqrt{\frac{2}{5}}{\cal P}_{\acute{a}}^{j\bf
T}{\cal P}_{\acute{b}ij}^{(S)}+{\cal P}_{\acute{a}j}^{\bf T}{\cal
P}_{\acute{b}i}^{j}\right]
\end{eqnarray}
\begin{eqnarray}
h_{\acute{c}\acute{d}}^{^{(120)(-)}}{\bf P}_{\acute{c}\mu}^{ij\bf
T}{\bf P}_{\acute{d}j\mu}=h_{\acute{c}\acute{d}}^{^{(120)(-)}}
\left[{\cal P}_{\acute{c}l}^{ik\bf T}{\cal P}_{\acute{d}k}^{l}+
\frac{1}{2\sqrt 5}{\cal P}_{\acute{c}}^{j\bf T}{\cal
P}_{\acute{d}j}^{i}\right.\nonumber\\
\left.+\frac{1}{2\sqrt {30}}\epsilon^{iklmn}{\cal
P}_{\acute{c}lmn}^{p\bf T}{\cal P}_{\acute{d}kp}+ \frac{1}{6\sqrt
{2}}\epsilon^{iklmn}{\cal P}_{\acute{c}lmn}^{p\bf T}{\cal
P}_{\acute{d}kp}^{(S)}\right.\nonumber\\
\left.-\frac{1}{10}\epsilon^{iklmn}{\cal P}_{\acute{c}mn}^{\bf
T}{\cal P}_{\acute{d}kl}^{(S)}\right]
\end{eqnarray}
\begin{eqnarray}
\bar{h}_{\acute{a}\acute{b}}^{^{(120)(-)}}\epsilon^{ijklm}{\bf
Q}_{\acute{a}ij\mu}^{\bf T}{\bf
Q}_{\acute{b}kn\mu}=\bar{h}_{\acute{a}\acute{b}}^{^{(120)(-)}}\left[4{\cal
Q}_{\acute{a}p}^{qlm\bf T}{\cal
Q}_{\acute{b}qn}^{p}+\frac{2}{\sqrt 5}{\cal Q}_{\acute{a}p}^{\bf
T}{\cal Q}_{\acute{b}n}^{plm}+4\sqrt{\frac{2}{15}}{\cal
Q}_{\acute{a}}^{pm\bf T}{\cal Q}_{\acute{b}pn}^{l}\right.\nonumber\\
\left.-4\sqrt{\frac{2}{15}}{\cal Q}_{\acute{a}}^{pl\bf T}{\cal
Q}_{\acute{b}pn}^{m}+\frac{4}{5}\sqrt{\frac{2}{3}}{\cal
Q}_{\acute{a}n}^{\bf T}{\cal Q}_{\acute{b}}^{lm}\right.\nonumber\\
\left.+{\cal Q}_{\acute{a}pq}^{r\bf T}\left(\delta^m_n{\cal
Q}_{\acute{b}r}^{pql}-\delta^l_n{\cal Q}_{\acute{b}r}^{pqm}
\right)\right.\nonumber\\
\left.+\sqrt{\frac{2}{15}}\left(\delta^l_n{\cal
Q}_{\acute{a}pq}^{m\bf T}-\delta^m_n{\cal Q}_{\acute{a}pq}^{l\bf T
}\right)
{\cal Q}_{\acute{b}}^{pq}\right.\nonumber\\
\left.+\frac{1}{5}\sqrt{\frac{2}{3}}\left(\delta^l_n{\cal
Q}_{\acute{a}}^{pm\bf T}-\delta^m_n{\cal Q}_{\acute{a}}^{pl\bf T
}\right) {\cal Q}_{\acute{b}p}\right]
\end{eqnarray}
\begin{eqnarray}
\bar{h}_{\acute{c}\acute{d}}^{^{(120)(-)}}{\bf
Q}_{\acute{c}\mu}^{n\bf T}{\bf
Q}_{\acute{d}lm\mu}=\bar{h}_{\acute{c}\acute{d}}^{^{(120)(-)}}
\left[\frac{1}{2\sqrt {30}}\epsilon_{ijklm}{\cal
Q}_{\acute{c}}^{np\bf T}{\cal
Q}_{\acute{d}p}^{ijk}-\frac{1}{10}\epsilon_{ijklm}{\cal
Q}_{\acute{c}}^{ni\bf T}{\cal Q}_{\acute{d}}^{jk}\right.\nonumber\\
\left.+\frac{1}{6\sqrt {2}}\epsilon_{ijklm}{\cal
Q}_{(S)\acute{c}}^{np\bf T}{\cal Q}_{\acute{d}p}^{ijk}
-\frac{1}{2\sqrt {15}}\epsilon_{ijklm}{\cal
Q}_{(S)\acute{c}}^{ni\bf T}{\cal Q}_{\acute{d}}^{jk}\right.\nonumber\\
\left.+{\cal Q}_{\acute{c}k}^{n\bf T}{\cal Q}_{\acute{d}lm}^{k}
+\frac{1}{2\sqrt {5}}{\cal Q}_{\acute{c}l}^{n\bf T}{\cal
Q}_{\acute{d}m}-\frac{1}{2\sqrt {5}}{\cal Q}_{\acute{c}m}^{n\bf
T}{\cal Q}_{\acute{d}l}\right]
\end{eqnarray}
\begin{eqnarray}
\bar{h}_{\acute{a}\acute{b}}^{^{(120)(-)}}{\bf
Q}_{\acute{a}\mu}^{i\bf T}{\bf
Q}_{\acute{b}\mu}^j=\bar{h}_{\acute{a}\acute{b}}^{^{(120)(-)}}
\left[ \sqrt{\frac{3}{10}}{\cal Q}_{\acute{a}k}^{i\bf T}{\cal
Q}_{\acute{b}}^{jk}-\sqrt{\frac{3}{10}}{\cal Q}_{\acute{a}k}^{j\bf
T}{\cal Q}_{\acute{b}}^{ik}\right.\nonumber\\
\left.+\sqrt{2}{\cal Q}_{\acute{a}k}^{i\bf T}{\cal
Q}_{(S)\acute{b}}^{jk}-\sqrt{2}{\cal Q}_{\acute{a}k}^{j\bf T}{\cal
Q}_{(S)\acute{b}}^{ik}\right]
\end{eqnarray}
\begin{eqnarray} \bar{h}_{\acute{c}\acute{d}}^{^{(120)(-)}}{\bf
Q}_{\acute{c}ij\mu}^{\bf T}{\bf
Q}_{\acute{d}\mu}=\bar{h}_{\acute{c}\acute{d}}^{^{(120)(-)}}
\left[\frac{2}{\sqrt 5}{\cal Q}_{\acute{c}ij}^{k\bf T}{\cal
Q}_{\acute{d}k}+\frac{2}{5}{\cal Q}_{\acute{c}j}^{\bf T}{\cal
Q}_{\acute{d}i}\right.\nonumber\\
\left.+\frac{1}{6}\epsilon_{ijklm}{\cal Q}_{\acute{c}n}^{klm\bf
T}{\cal Q}_{\acute{d}}^n-\frac{1}{\sqrt{30}}\epsilon_{ijklm}{\cal
Q}_{\acute{c}}^{kl\bf T}{\cal Q}_{\acute{d}}^m\right]
\end{eqnarray}
\begin{eqnarray}
\bar{h}_{\acute{a}\acute{b}}^{^{(120)(-)}}{\bf
Q}_{\acute{a}\mu}^{\bf T}{\bf
Q}_{\acute{b}\mu}^i=\bar{h}_{\acute{a}\acute{b}}^{^{(120)(-)}}
\left[ \frac{\sqrt 6}{5}{\cal Q}_{\acute{a}j}^{\bf T}{\cal
Q}_{\acute{b}}^{ij}+\sqrt{\frac{2}{5}}{\cal Q}_{\acute{a}j}^{\bf
T}{\cal Q}_{(S)\acute{b}}^{ij}+{\cal Q}_{\acute{a}}^{j\bf T}{\cal
Q}_{\acute{b}j}^{i}\right]
\end{eqnarray}
\begin{eqnarray}
\bar{h}_{\acute{c}\acute{d}}^{^{(120)(-)}}{\bf
Q}_{\acute{c}ij\mu}^{\bf T}{\bf
Q}_{\acute{d}\mu}^{j}=\bar{h}_{\acute{c}\acute{d}}^{^{(120)(-)}}
\left[{\cal Q}_{\acute{c}ik}^{l\bf T}{\cal Q}_{\acute{d}l}^{k}+
\frac{1}{2\sqrt 5}{\cal Q}_{\acute{c}j}^{\bf T}{\cal
Q}_{\acute{d}i}^{j}\right.\nonumber\\
\left.+\frac{1}{2\sqrt {30}}\epsilon_{iklmn}{\cal
Q}_{\acute{c}p}^{lmn\bf T}{\cal Q}_{\acute{d}}^{kp}+
\frac{1}{6\sqrt {2}}\epsilon_{iklmn}{\cal Q}_{\acute{c}p}^{lmn\bf
T}{\cal
Q}_{(S)\acute{d}}^{kp}\right.\nonumber\\
\left.-\frac{1}{10}\epsilon_{iklmn}{\cal Q}_{\acute{c}}^{mn\bf
T}{\cal Q}_{(S)\acute{d}}^{kl}\right]
\end{eqnarray}

\section{Appendix C: Details of couplings from $\bf{126+\overline{126}}$-plet
mediation}

In this Appendix we expand the $SO(10)$ coupling structures that
enter in $126+\overline{126}$-plet mediation in Secs.(4,5,6) in a
$SU(5)\times U(1)$ basis. We list these structures below

\begin{eqnarray}
f_{\acute{a}\acute{b}}^{^{(\overline{126})(+)}}{\bf
P}_{\acute{a}\mu}^{\bf T}{\bf
P}_{\acute{b}\mu}=f_{\acute{a}\acute{b}}^{^{(\overline{126})(+)}}
\left[\frac{4}{\sqrt{5}}{\cal P}_{\acute{a}}^{i\bf T}{\cal
P}_{\acute{b}i}\right]
\end{eqnarray}
\begin{eqnarray}
f_{\acute{a}\acute{b}}^{^{(\overline{126})(+)}}{\bf
P}_{\acute{a}\mu}^{\bf T}{\bf
P}_{\acute{b}\mu}^{ij}=f_{\acute{a}\acute{b}}^{^{(\overline{126})(+)}}
\left[\frac{2}{\sqrt{5}}{\cal P}_{\acute{a}}^{k\bf T}{\cal
P}_{\acute{b}k}^{ij}+\frac{1}{6}\epsilon^{ijklm}{\cal
P}_{\acute{a}n}^{\bf T}{\cal P}_{\acute{b}klm}^{n}\right.\nonumber\\
\left.-\frac{1}{\sqrt{30}}\epsilon^{ijklm}{\cal
P}_{\acute{a}k}^{\bf T}{\cal P}_{\acute{b}lm}\right]
\end{eqnarray}
\begin{eqnarray}
f_{\acute{a}\acute{b}}^{^{(\overline{126})(+)}}{\bf
P}_{\acute{a}i\mu}^{\bf T}{\bf
P}_{\acute{b}j\mu}=f_{\acute{a}\acute{b}}^{^{(\overline{126})(+)}}
\left[\sqrt{\frac{3}{10}}{\cal P}_{\acute{a}i}^{k\bf T}{\cal
P}_{\acute{b}jk}+\sqrt{\frac{3}{10}}{\cal P}_{\acute{a}j}^{k\bf
T}{\cal P}_{\acute{b}ik}+\frac{1}{\sqrt{2}}{\cal
P}_{\acute{a}i}^{k\bf T}{\cal
P}_{\acute{b}jk}^{(S)}\right.\nonumber\\
\left.+\frac{1}{\sqrt{2}}{\cal P}_{\acute{a}j}^{k\bf T}{\cal
P}_{\acute{b}ik}^{(S)}\right]
\end{eqnarray}
\begin{eqnarray}
f^{^{{(\overline{126})(+)}}}_{\acute{a}\acute{b}}{\bf
P}_{\acute{a}\mu}^{\bf T}{\bf
P}_{\acute{b}j\mu}=f^{^{{(\overline{126})(+)}}}_{\acute{a}\acute{b}}\left[\frac{\sqrt{6}
}{5}{\cal P}_{\acute{a}}^{k\bf T}{\cal P}_{\acute{b}jk}
+\sqrt{\frac{2}{5}}{\cal P}_{\acute{a}}^{k\bf T}{\cal
P}_{\acute{b}jk}^{(S)}+{\cal P}_{\acute{a}k}^{\bf T}{\cal
P}_{\acute{b}j}^k\right]
\end{eqnarray}
\begin{eqnarray}
f_{\acute{a}\acute{b}}^{^{(\overline{126})(+)}}{\bf
P}_{\acute{a}\mu}^{ij\bf T}{\bf
P}_{\acute{b}k\mu}=f_{\acute{a}\acute{b}}^{^{(\overline{126})(+)}}
\left[\frac{1}{2\sqrt{30}}\epsilon^{ijpqr}{\cal
P}_{\acute{a}pqr}^{s\bf T}{\cal
P}_{\acute{b}ks}+\frac{1}{6\sqrt{2}}\epsilon^{ijpqr}{\cal
P}_{\acute{a}pqr}^{s\bf T}{\cal P}_{\acute{b}ks}^{(S)}\right.\nonumber\\
\left.-\frac{1}{10}\epsilon^{ijpqr}{\cal P}_{\acute{a}pq}^{\bf
T}{\cal P}_{\acute{b}kr}-\frac{1}{2\sqrt{15}}\epsilon^{ijpqr}{\cal
P}_{\acute{a}pq}^{\bf T}{\cal P}_{\acute{b}kr}^{(S)}\right.\nonumber\\
\left.+{\cal P}_{\acute{a}p}^{ij\bf T}{\cal P}_{\acute{b}k}^{p}
+\frac{1}{2\sqrt{5}}{\cal P}_{\acute{a}}^{j\bf T}{\cal
P}_{\acute{b}k}^{i}-\frac{1}{2\sqrt{5}}{\cal P}_{\acute{a}}^{i\bf
T}{\cal P}_{\acute{b}k}^{j}\right]
\end{eqnarray}
\begin{eqnarray}
f_{\acute{a}\acute{b}}^{^{(\overline{126})(+)}}{\bf
P}_{\acute{a}\mu}^{ij\bf T}{\bf
P}_{\acute{b}\mu}^{kl}=f_{\acute{a}\acute{b}}^{^{(\overline{126})(+)}}\left[
\frac{1}{6}\epsilon^{ijpqr}{\cal P}_{\acute{a}pqr}^{s\bf T}{\cal
P}_{\acute{b}s}^{kl}+\frac{1}{6}\epsilon^{klpqr}{\cal
P}_{\acute{a}s}^{ij\bf
T}{\cal P}_{\acute{b}pqr}^{s}\right.\nonumber\\
\left.+\frac{1}{12\sqrt{5}}\epsilon^{ijpqr}{\cal
P}_{\acute{a}pqr}^{k\bf T}{\cal
P}_{\acute{b}}^{l}-\frac{1}{12\sqrt{5}}\epsilon^{klpqr}{\cal
P}_{\acute{a}}^{i\bf
T}{\cal P}_{\acute{b}pqr}^{j}\right.\nonumber\\
\left.-\frac{1}{12\sqrt{5}}\epsilon^{ijpqr}{\cal
P}_{\acute{a}pqr}^{l\bf T}{\cal
P}_{\acute{b}}^{k}+\frac{1}{12\sqrt{5}}\epsilon^{klpqr}{\cal
P}_{\acute{a}}^{j\bf
T}{\cal P}_{\acute{b}pqr}^{i}\right.\nonumber\\
\left.+\frac{1}{10\sqrt{6}}\epsilon^{ijlpq}{\cal
P}_{\acute{a}pq}^{\bf T}{\cal
P}_{\acute{b}}^{k}-\frac{1}{10\sqrt{6}}\epsilon^{klipq}{\cal
P}_{\acute{a}}^{j\bf
T}{\cal P}_{\acute{b}pq}\right.\nonumber\\
\left.-\frac{1}{10\sqrt{6}}\epsilon^{ijkpq}{\cal
P}_{\acute{a}pq}^{\bf T}{\cal
P}_{\acute{b}}^{l}+\frac{1}{10\sqrt{6}}\epsilon^{kljpq}{\cal
P}_{\acute{a}}^{i\bf
T}{\cal P}_{\acute{b}pq}\right.\nonumber\\
\left.-\frac{1}{\sqrt{30}}\epsilon^{ijpqr}{\cal
P}_{\acute{a}pq}^{\bf T}{\cal
P}_{\acute{b}r}^{kl}-\frac{1}{\sqrt{30}}\epsilon^{klpqr}{\cal
P}_{\acute{a}p}^{ij\bf T}{\cal P}_{\acute{b}qr}\right]
\end{eqnarray}
\begin{eqnarray}
f_{\acute{a}\acute{b}}^{^{(\overline{126})(+)}}\epsilon_{jklmn}{\bf
P}_{\acute{c}\mu}^{kl\bf T}{\bf
P}_{\acute{d}\mu}^{mn}=f_{\acute{a}\acute{b}}^{^{(\overline{126})(+)}}\left[4{\cal
P}_{\acute{c}r}^{pq\bf T}{\cal
P}_{\acute{d}pqj}^r-4\sqrt{\frac{2}{15}}{\cal
P}_{\acute{c}pq}^{\bf T}{\cal
P}_{\acute{d}j}^{pq}\right.\nonumber\\
\left.-\frac{4\sqrt 6}{5}{\cal P}_{\acute{c}}^{p\bf T}{\cal
P}_{\acute{d}pj}\right]
\end{eqnarray}
\begin{eqnarray}
f_{\acute{a}\acute{b}}^{^{(\overline{126})(+)}}{\bf
P}_{\acute{a}\mu}^{ij\bf T}{\bf P}_{\acute{b}j\mu}
=f_{\acute{a}\acute{b}}^{^{(\overline{126})(+)}}\left[-\frac{1}{2\sqrt
{30}}\epsilon^{iklmn}{\cal P}_{\acute{a}lmn}^{p\bf T}{\cal
P}_{\acute{b}kp}-\frac{1}{6\sqrt {2}}\epsilon^{iklmn}{\cal
P}_{\acute{a}lmn}^{p\bf T}{\cal
P}_{\acute{b}kp}^{(S)}\right.\nonumber\\
\left.+\frac{1}{10}\epsilon^{iklmn}{\cal P}_{\acute{a}mn}^{\bf
T}{\cal P}_{\acute{b}kl}+{\cal P}_{\acute{a}l}^{ki\bf T}{\cal
P}_{\acute{b}k}^{l}-\frac{1}{2\sqrt {5}}{\cal P}_{\acute{a}}^{k\bf
T}{\cal P}_{\acute{b}k}^i\right]
\end{eqnarray}
\begin{eqnarray}
\bar{f}_{\acute{c}\acute{d}}^{^{(126)(+)}}{\bf
Q}_{\acute{c}\nu}^{\bf T}{\bf
Q}_{\acute{d}\nu}=\bar{f}_{\acute{c}\acute{d}}^{^{(126)(+)}}\left[\frac{4}{\sqrt{5}
}{\cal Q}_{\acute{c}}^{k\bf T}{\cal Q}_{\acute{d}k}\right]
\end{eqnarray}
\begin{eqnarray}
\bar{f}_{\acute{c}\acute{d}}^{^{(126)(+)}}{\bf
Q}_{\acute{c}\nu}^{\bf T}{\bf
Q}_{\acute{d}ij\nu}=\bar{f}_{\acute{c}\acute{d}}^{^{(126)(+)}}\left[\frac{2}{\sqrt{5}
}{\cal Q}_{\acute{c}k}^{\bf T}{\cal
Q}_{\acute{d}ij}^{k}+\frac{1}{6}\epsilon_{ijklm}{\cal
Q}_{\acute{c}}^{n\bf T}{\cal Q}_{\acute{d}n}^{klm}\right.\nonumber\\
\left.-\frac{1}{\sqrt{30}}\epsilon_{ijklm}{\cal
Q}_{\acute{c}}^{k\bf T}{\cal Q}_{\acute{d}}^{lm}\right]
\end{eqnarray}
\begin{eqnarray}
\bar{f}_{\acute{c}\acute{d}}^{^{(126)(+)}}{\bf
Q}_{\acute{c}\nu}^{i\bf T}{\bf
Q}_{\acute{d}\nu}^{j}=\bar{f}_{\acute{c}\acute{d}}^{^{(126)(+)}}\left[\sqrt{\frac{3}{10}}{\cal
Q}_{\acute{c}k}^{i\bf T}{\cal
Q}_{\acute{d}}^{jk}+\sqrt{\frac{3}{10}}{\cal Q}_{\acute{c}k}^{j\bf
T}{\cal Q}_{\acute{d}}^{ik}\right.\nonumber\\
\left.+\frac{1}{\sqrt{2}}{\cal Q}_{\acute{c}k}^{i\bf T}{\cal
Q}_{(S)\acute{d}}^{jk}+ \frac{1}{\sqrt 2} {\cal Q}_{\acute{c}k}^{j\bf T}{\cal
Q}_{(S)\acute{d}}^{ik}\right]
\end{eqnarray}
\begin{eqnarray}
\bar{f}_{\acute{c}\acute{d}}^{^{(126)(+)}}{\bf
Q}_{\acute{c}\nu}^{\bf T}{\bf
Q}_{\acute{d}\nu}^{j}=\bar{f}_{\acute{c}\acute{d}}^{^{(126)(+)}}\left[
\frac{\sqrt{6}}{5}{\cal Q}_{\acute{c}k}^{\bf T}{\cal
Q}_{\acute{d}}^{jk} +\sqrt{\frac{2}{5}}{\cal Q}_{\acute{c}k}^{\bf
T}{\cal Q}_{(S)\acute{d}}^{jk}+{\cal Q}_{\acute{c}}^{k\bf T}{\cal
Q}_{\acute{d}k}^j\right]
\end{eqnarray}
\begin{eqnarray}
\bar{f}_{\acute{c}\acute{d}}^{^{(126)(+)}}{\bf
Q}_{\acute{c}ij\nu}^{\bf T}{\bf
Q}_{\acute{d}\nu}^{k}=\bar{f}_{\acute{c}\acute{d}}^{^{(126)(+)}}\left[
\frac{1}{2\sqrt{30}}\epsilon_{ijpqr}{\cal Q}_{\acute{c}s}^{pqr\bf
T}{\cal Q}_{\acute{d}}^{ks}
+\frac{1}{6\sqrt{2}}\epsilon_{ijpqr}{\cal Q}_{\acute{c}s}^{pqr\bf
T}{\cal Q}_{(S)\acute{d}}^{ks}\right.\nonumber\\
\left.-\frac{1}{10}\epsilon_{ijpqr}{\cal Q}_{\acute{c}}^{pq\bf
T}{\cal
Q}_{\acute{d}}^{kr}-\frac{1}{2\sqrt{15}}\epsilon_{ijpqr}{\cal
Q}_{\acute{c}}^{pq\bf
T}{\cal Q}_{(S)\acute{d}}^{kr}\right.\nonumber\\
\left.+{\cal Q}_{\acute{c}ij}^{p\bf T}{\cal Q}_{\acute{d}p}^{k}+
\frac{1}{2\sqrt{5}}{\cal Q}_{\acute{c}j}^{\bf T}{\cal
Q}_{\acute{d}i}^{k}\right.\nonumber\\
\left.-\frac{1}{2\sqrt{5}}{\cal Q}_{\acute{c}i}^{\bf T}{\cal
Q}_{\acute{d}j}^{k}\right]
\end{eqnarray}
\begin{eqnarray}
\bar{f}_{\acute{c}\acute{d}}^{^{(126)(+)}}\epsilon^{jklmn}{\bf
Q}_{\acute{c}kl\nu}^{\bf T}{\bf
Q}_{\acute{d}mn\nu}=\bar{f}_{\acute{c}\acute{d}}^{^{(126)(+)}}\left[4{\cal
Q}_{\acute{c}pq}^{r\bf T}{\cal
Q}_{\acute{d}r}^{pqj}-4\sqrt{\frac{2}{15}}{\cal
Q}_{\acute{c}}^{pq\bf T}{\cal Q}_{\acute{d}pq}^{j}
\right.\nonumber\\
\left.-\frac{4\sqrt 6}{5}{\cal Q}_{\acute{c}p}^{\bf T}{\cal
Q}_{\acute{d}}^{pj}\right]
\end{eqnarray}
\begin{eqnarray}
\bar{f}_{\acute{c}\acute{d}}^{^{(126)(+)}}{\bf
Q}_{\acute{c}ij\nu}^{\bf T}{\bf
Q}_{\acute{d}kl\nu}=\bar{f}_{\acute{c}\acute{d}}^{^{(126)(+)}}\left[
\frac{1}{6}\epsilon_{ijpqr}{\cal Q}_{\acute{c}s}^{pqr\bf T}{\cal
Q}_{\acute{d}kl}^{s}+\frac{1}{6}\epsilon_{klpqr}{\cal
Q}_{\acute{c}ij}^{s\bf
T}{\cal Q}_{\acute{d}s}^{pqr}\right.\nonumber\\
\left.+\frac{1}{12\sqrt{5}}\epsilon_{ijpqr}{\cal
Q}_{\acute{c}k}^{pqr\bf T}{\cal
Q}_{\acute{d}l}-\frac{1}{12\sqrt{5}}\epsilon_{klpqr}{\cal
Q}_{\acute{c}i}^{\bf
T}{\cal Q}_{\acute{d}j}^{pqr}\right.\nonumber\\
\left.-\frac{1}{12\sqrt{5}}\epsilon_{ijpqr}{\cal
Q}_{\acute{c}l}^{pqr\bf T}{\cal
Q}_{\acute{d}k}+\frac{1}{12\sqrt{5}}\epsilon_{klpqr}{\cal
Q}_{\acute{c}j}^{\bf
T}{\cal Q}_{\acute{d}i}^{pqr}\right.\nonumber\\
\left.+\frac{1}{10\sqrt{6}}\epsilon_{ijlpq}{\cal
Q}_{\acute{c}}^{pq\bf T}{\cal
Q}_{\acute{d}k}-\frac{1}{10\sqrt{6}}\epsilon_{klipq}{\cal
Q}_{\acute{c}j}^{\bf
T}{\cal Q}_{\acute{d}}^{pq}\right.\nonumber\\
\left.-\frac{1}{10\sqrt{6}}\epsilon_{ijkpq}{\cal
Q}_{\acute{c}}^{pq\bf T}{\cal
Q}_{\acute{d}l}+\frac{1}{10\sqrt{6}}\epsilon_{kljpq}{\cal
Q}_{\acute{c}i}^{\bf
T}{\cal Q}_{\acute{d}}^{pq}\right.\nonumber\\
\left.-\frac{1}{\sqrt{30}}\epsilon_{ijpqr}{\cal
Q}_{\acute{c}}^{pq\bf T}{\cal
Q}_{\acute{d}kl}^{r}-\frac{1}{\sqrt{30}}\epsilon_{klpqr}{\cal
Q}_{\acute{c}ij}^{p\bf T}{\cal Q}_{\acute{d}}^{qr}\right]
\end{eqnarray}
\begin{eqnarray}
\bar{f}_{\acute{c}\acute{d}}^{^{(126)(+)}}{\bf
Q}_{\acute{a}ij\mu}^{\bf T}{\bf
Q}_{\acute{b}\mu}^j=\bar{f}_{\acute{c}\acute{d}}^{^{(126)(+)}}\left[-\frac{1}{2\sqrt
{30}}\epsilon_{iklmn}{\cal Q}_{\acute{c}p}^{lmn\bf T}{\cal
Q}_{\acute{d}}^{kp}-\frac{1}{6\sqrt {2}}\epsilon_{iklmn}{\cal
Q}_{\acute{c}p}^{lmn\bf T}{\cal
Q}_{(S)\acute{d}}^{kp}\right.\nonumber\\
\left.+\frac{1}{10}\epsilon_{iklmn}{\cal Q}_{\acute{c}}^{mn\bf
T}{\cal Q}_{\acute{d}}^{kl}+{\cal Q}_{\acute{c}li}^{k\bf T}{\cal
Q}_{\acute{d}k}^{l}-\frac{1}{2\sqrt {5}}{\cal Q}_{\acute{c}k}^{\bf
T}{\cal Q}_{\acute{d}i}^{k}\right]
\end{eqnarray}

\section{Appendix D: Field normalizations}

\noindent\textbf{(a)}~~\textsc{Normalization of $SU(5)$ components
in 45-plet of $SO(10)$ Higgs} \\

\noindent The 45-plet of $SO(10)$ Higgs $\Phi_{\mu\nu}$ can be
decomposed in $SU(5)$ multiplets as follows

\begin{eqnarray}
\Phi_{c_n\overline c_n}={\mathsf h};~~~\Phi_{c_i\overline
c_j}={\mathsf h}_{j}^i+ \frac{1}{5}\delta_j^i{\mathsf h};~~~
\Phi_{c_ic_j}={\mathsf h}^{ij};~~~\Phi_{\overline c_i\overline
c_j}={\mathsf h}_{ij}
\end{eqnarray}
where ${\mathsf h}$, ${\mathsf h}^{ij}$, ${\mathsf h}_{ij}$ and
${\mathsf h}_{j}^i$ are the 1-plet, 10-plet, $\overline{10}$-plet,
and 24-plet representations of $SU(5)$ respectively To normalize
these $SU(5)$ Higgs fields, we carry out a field redefinition,
\begin{eqnarray}
{\mathsf h}=\sqrt {10} {\mathsf H};~~~{\mathsf h}_{ij}=\sqrt
2{\mathsf H}_{ij}; ~~~ {\mathsf h}^{ij}=\sqrt 2{\mathsf
H}^{ij};~~~{\mathsf h}_{j}^i=\sqrt{2}{\mathsf H}_{j}^i.
\end{eqnarray}
In terms of the normalized fields the kinetic energy of the
 45 plet of Higgs\\
$-\partial_A\Phi_{\mu\nu}\partial^A\Phi_{\mu\nu}^{\dagger}$ takes
the form
\begin{equation}
{\mathsf L}_{kin}^{45-Higgs}=-\partial^A{\mathsf
H}\partial_A{\mathsf H}^\dagger -\frac{1}{2!}
\partial^A{\mathsf H}_{ij}\partial_A{\mathsf H}_{ij\dagger}-\frac{1}{2!}
\partial^A{\mathsf H}^{ij}\partial_A{\mathsf H}^{ij\dagger}
-\partial_A{\mathsf H}_j^i
\partial^A{\mathsf H}_j^{i\dagger}.
\end{equation}\\

\noindent\textbf{(b)}~~\textsc{Normalization of $SU(5)$ components in 45-plet of
$SO(10)$ gauge fields}.\\

\noindent The 45-plet of $SO(10)$ gauge fields  $\Phi_{A\mu\nu}$
can be decomposed in $SU(5)$ multiplets as follows
\begin{eqnarray}
\Phi_{Ac_n\overline c_n}={\mathsf g}_A;~~~\Phi_{Ac_i\overline
c_j}={\mathsf g}_{Aj}^i+ \frac{1}{5}\delta_j^i{\mathsf g}_A;~~~
\Phi_{Ac_ic_j}={\mathsf g}_A^{ij};~~~\Phi_{A\overline c_i\overline
c_j}={\mathsf g}_{Aij}
\end{eqnarray}
To normalize them we make the following redefinitions
\begin{eqnarray}
{\mathsf g}_A=2\sqrt 5 {\mathsf G}_A;~~~{\mathsf g}_{Aij}=\sqrt 2
{\mathsf G} _{Aij};~~~ {\mathsf g}_A^{ij}=\sqrt 2 {\mathsf
G}_A^{ij};~~~{\mathsf g}_{Aj}^i=\sqrt{2} {\mathsf G}_{Aj}^i.
\end{eqnarray}
In terms of the redefined fields the kinetic energy for the
45-plet which is given by $-\frac{1}{4}{\cal F}_{\mu\nu}^
{AB}{\cal F}_{AB\mu\nu}$ takes on the form
\begin{equation}
{\mathsf L}_{kin}^{45-gauge}=-\frac{1}{2}{\cal G}_{AB}{\cal
G}^{AB\dagger}-\frac{1}{2!} \frac{1}{2}{\cal G}^{ABij}{\cal
G}_{AB}^{ij\dagger} -\frac{1}{2!}\frac{1}{2}{\cal G}_j^{ABi}{\cal
G}_{ABi}^{j}
\end{equation}
where ${\cal F}_{\mu\nu}^{AB}$ is the 45 of $SO(10)$ field
strength
tensor.\\
\\
\\
\noindent\textbf{(c)}~~\textsc{Normalization of $SU(5)$ components in 210-plet of
$SO(10)$}\\

\noindent The 210-plet of $SO(10)$ $\Phi_{\mu\nu\rho\sigma}$ has
the following decomposition in $SU(5)$ multiplets

\begin{eqnarray}
\Phi_{c_m\overline c_mc_n\overline c_n}={\mathsf
h};~~~\Phi_{\overline c_i\overline c_j \overline c_k\overline
c_l}=\frac{1}{24}\epsilon_{ijklm}{\mathsf h}^m;~~~
\Phi_{c_ic_jc_kc_l}=\frac{1}{24}\epsilon^{ijklm}{\mathsf h}_{m}\nonumber\\
\Phi_{c_ic_jc_m\overline c_m}={\mathsf h}^{ij};~~~\Phi_{\overline
c_i\overline c_j \overline c_mc_m}={\mathsf
h}_{ij};~~~\Phi_{c_i\overline
c_jc_m\overline c_m}={\mathsf h}_{j}^i+\frac{1}{5}\delta_j^i{\mathsf h}\nonumber\\
\Phi_{c_ic_j\overline c_k\overline c_l}={\mathsf
h}_{kl}^{ij}+\frac{1}{3} \left(\delta_l^i{\mathsf
h}_{k}^j-\delta_k^i{\mathsf h}_{l}^j+ \delta_k^j{\mathsf h}_{l}^i
-\delta_l^j{\mathsf h}_{k}^i\right)+\frac{1}{20}\left(\delta_l^i
\delta_k^j-\delta_k^i\delta_l^j\right){\mathsf h}\nonumber\\
\Phi_{c_ic_jc_k\overline c_l}={\mathsf h}_{l}^{ijk}+\frac{1}{3}
\left(\delta_l^k{\mathsf h}^{ij}-\delta_l^j{\mathsf h}_h^{ik}
+\delta_l^i{\mathsf h}^{jk}\right)\nonumber\\
\Phi_{\overline c_i\overline c_j\overline c_kc_l}={\mathsf
h}_{ijk}^l +\frac{1}{3}\left(\delta_k^l{\mathsf
h}_{ij}-\delta_j^l{\mathsf h}_{ik}+ \delta_i^l{\mathsf
h}_{jk}\right)
\end{eqnarray}
where ${\mathsf h}$, ${\mathsf h}^i$, ${\mathsf h}_{i}$, ${\mathsf
h}^{ij}$, ${\mathsf h}_{ij}$, ${\mathsf h}_{j}^i$, ${\mathsf
h}_{l}^{ijk}$; ${\mathsf h}_{jkl}^i$ and ${\mathsf h}_{kl}^{ij}$
are the 1-plet, 5-plet, $\overline 5$-plet, 10-plet, $\overline
{10}$-plet, 24-plet, 40-plet, $\overline {40}$-plet, and 75-plet
representations of $SU(5)$ respectively. To normalize these fields
we carry out a field redefinition
\begin{eqnarray}
{\mathsf h}=4\sqrt{\frac{5}{3}}{\mathsf H};~~~{\mathsf
h}^i=8\sqrt{6}{\mathsf H}^i;~~~{\mathsf h}_{i}
=8\sqrt{6}{\mathsf H}_{i}\nonumber\\
{\mathsf h}^{ij}=\sqrt{2}{\mathsf H}^{ij};~~~{\mathsf
h}_{ij}=\sqrt{2}{\mathsf H}_{ij};
~~~{\mathsf h}_{j}^i=\sqrt{2}{\mathsf H}_{j}^i\nonumber\\
{\mathsf h}_{l}^{ijk}=\sqrt{\frac{2}{3}}{\mathsf
H}_{l}^{ijk};~~~{\mathsf h}_{jkl}^i =\sqrt{\frac{2}{3}}{\mathsf
H}_{jkl}^i;~~~{\mathsf h}_{kl}^{ij} ={\frac{2}{\sqrt3}}{\mathsf
H}_{kl}^{ij}.
\end{eqnarray}
Now the kinetic energy for the 210 dimensional Higgs field is
$-\partial_A\Phi_{\mu\nu\rho\lambda}\partial^A
\Phi_{\mu\nu\rho\lambda}^{\dagger}$ which in terms of the
redefined fields takes the form
\begin{eqnarray}
{\mathsf L}_{kin}^{210-Higgs}=-\partial_A{\mathsf H}\partial^A
{\mathsf H}^{\dagger} -\partial_A{\mathsf H}^i\partial^A{\mathsf
H}^{i\dagger}-\partial_A{\mathsf H}_i\partial^A{\mathsf
H}_{i}^{\dagger}\nonumber\\
-\frac{1}{2!}\partial_A{\mathsf H}_{ij}
\partial^A{\mathsf H}_{ij}^{\dagger}-\frac{1}{2!}\partial_A{\mathsf H}^{ij}
\partial^A{\mathsf H}^{ij\dagger}
-\partial_A{\mathsf H}_j^i\partial^A{\mathsf
H}_j^{i\dagger}\nonumber\\
 -\frac{1}{3!}\partial_A{\mathsf
H}^l_{ijk}
\partial^A{\mathsf H}^{l\dagger}_{ijk}-\frac{1}{3!}\partial_A{\mathsf
H}_l^{ijk}
\partial^A{\mathsf H}_l^{ijk\dagger}
-\frac{1}{2!}\frac{1}{2!}
\partial_A{\mathsf H}_{kl}^{ij}\partial^A{\mathsf H}_{kl}^{ij\dagger}.
\end{eqnarray}\\

\noindent\textbf{(d)}~~\textsc{Normalization of  $SU(5)$ components in 10-plet of
$SO(10)$}\\

\noindent The 10-plet of $SO(10)$ $\Phi_{\mu}$ can  be decomposed
in $SU(5)$ components as follows
\begin{eqnarray}
\Phi_{\bar c_i}={\mathsf h}_{i};~~~~~\Phi_{c_i}={\mathsf h}^{i}
\end{eqnarray}

The tensors are already in their irreducible form and one can
identify $\Phi_{c_i}$ with the 5 plet of Higgs and $\Phi_{\bar
c_i}$ with the $\bar 5$ plet of Higgs. To normalize the fields we
define
\begin{eqnarray}
{\mathsf h}_{i}=\frac{1}{\sqrt{2}}{\mathsf H}_{i};~~~~~{\mathsf
h}^{i}=\frac{1}{\sqrt{2}}{\mathsf H}^{i}
\end{eqnarray}
Now the kinetic energy for the 10 dimensional Higgs
field is $-\partial_A\Phi_{\mu}\partial^A \Phi_{\mu}^{\dagger}$
which in terms of the redefined fields takes the form
\begin{eqnarray}
{\mathsf L}_{kin}^{10-Higgs}= -\partial_{A}{\mathsf
H}_{i}\partial^{A}{\mathsf H}_{i}^{\dagger}
 -\partial_{A}{\mathsf H}^i\partial^{A}{\mathsf H}^{i\dagger} .
\end{eqnarray}\\

\noindent\textbf{(e)}~~\textsc{Normalization of $SU(5)$ components in 120-plet of
$SO(10)$}\\

\noindent
The 120-plet of $SO(10)$ $\Phi_{\mu\nu\rho}$ can be decomposed in $SU(5)$ components
as follows

\begin{eqnarray}
\Phi_{c_ic_j\bar c_k}={\mathsf
h}^{ij}_k+\frac{1}{4}\left(\delta^i_k {\mathsf h}^j- \delta^j_k
{\mathsf h}^i\right),~~~ \Phi_{c_i\bar c_j\bar c_k}={\mathsf
h}^{i}_{jk}+\frac{1}{4}\left(\delta^i_j {\mathsf h}_k-
\delta^i_k {\mathsf h}_j\right)\nonumber\\
\Phi_{c_ic_jc_k}=\epsilon^{ijklm}{\mathsf h}_{lm},~~~ \Phi_{\bar
c_i\bar c_j\bar c_k}=\epsilon_{ijklm}{\mathsf h}^{lm},
~~~\Phi_{\bar c_nc_n c_i}={\mathsf h}^i, \Phi_{\bar c_nc_n \bar
c_i}={\mathsf h}_i
\end{eqnarray}
 where ${\mathsf h}_{i}$, ${\mathsf h}^{i}$, ${\mathsf h}_{ij}$, ${\mathsf h}^{ij}$, ${\mathsf h}^{ij}_k$, ${\mathsf h}^i_{jk}$
 are the $\bar{5}$, 5, $\overline{10}$, 10, 45 and $\overline{45}$ plet representations of $SU(5)$.
To normalize them we make the following redefinition of fields
 \begin{eqnarray}
{\mathsf h}^{i}=\frac{4}{\sqrt 3} {\mathsf H}^{i},~~~ {\mathsf
h}^{ij}=\frac{1}{\sqrt 3} {\mathsf H}^{ij},~~~
{\mathsf h}^{ij}_k=\frac{2}{\sqrt 3} {\mathsf H}^{ij}_k\nonumber\\
{\mathsf h}_{i}=\frac{4}{\sqrt 3} {\mathsf H}_{i},~~~ {\mathsf
h}_{ij}=\frac{1}{\sqrt 3} {\mathsf H}_{ij},~~~ {\mathsf
h}^{i}_{jk}=\frac{2}{\sqrt 3} {\mathsf H}^{i}_{jk}
\end{eqnarray}
In terms of the redefined fields the kinetic energy term for the
120 multiplet which is given by
$-\partial_{A}\Phi_{\mu\nu\lambda}$
$\partial^{A}\Phi_{\mu\nu\lambda}^{\dagger}$ takes on the form
\begin{eqnarray}
{\mathsf L}_{kin}^{120-Higgs}=-\frac{1}{2!}\partial_{A}{\mathsf H}
^{ij}
\partial^{A} {\mathsf H}^{ij\dagger}
-\frac{1}{2!}\partial_{A} {\mathsf H}_{ij}
\partial^{A} {\mathsf H}_{ij}^{\dagger}
-\frac{1}{2!}\partial_{A} {\mathsf H}^{ij}_k
\partial^{A}{\mathsf H}^{ij\dagger}_k  \nonumber\\
 -\frac{1}{2!}\partial_{A} {\mathsf H}_{jk}^i
\partial^{A}{\mathsf H}_{jk}^{i\dagger}
-\partial_{A} {\mathsf H}^{i}
\partial^{A} {\mathsf H}^{i\dagger}
-\partial_{A} {\mathsf H}_{i}
\partial^{A} {\mathsf H}_{i}^{\dagger}
\end{eqnarray}\\

\noindent\textbf{(f)}~~\textsc{Normalization of $SU(5)$ components in
$126$ and $\overline{126}$ -plets of $SO(10)$}\\

\noindent
To deal with the $126$ and $\overline{126}$ -plets  $\Phi_{\mu\nu\rho\lambda\sigma}$ and
$\overline{\Phi}_{\mu\nu\rho\lambda\sigma}$,  we first introduce the full 252-dimensional tensor,
 $\Xi_{\mu\nu\lambda\rho\sigma}$ which can be
be decomposed as $\Xi_{\mu\nu\lambda\rho\sigma}$=$\overline
{\Phi}_{\mu\nu\lambda\rho\sigma}$+
$\Phi_{\mu\nu\lambda\rho\sigma}$, where
\begin{equation}
\left(\matrix{\overline{\Phi}_{\mu\nu\lambda\rho\sigma}\cr
 \Phi_{\mu\nu\lambda\rho\sigma}}\right)=
   \frac{1}{2}\left(\delta_{\mu\alpha}
\delta_{\nu\beta}\delta_{\rho\gamma}\delta_{\lambda\delta}\delta_{\sigma\theta}
\pm
\frac{i}{5!}\epsilon_{\mu\nu\rho\lambda\sigma\alpha\beta\gamma\delta\theta}\right)
\Xi_{\alpha\beta\gamma\delta\theta}
\end{equation}
and where the $\overline{\Phi}_{\mu\nu\lambda\rho\sigma}$ is the
${\overline{126}}$ plet
 and $\Phi_{\mu\nu\lambda\rho\sigma}$ is the ${126}$ plet representation.
 The decomposition of these in $SU(5)$ components is then given by

\begin{eqnarray}
 \Xi_{c_ic_jc_k\bar c_nc_n}={\mathsf
 h}^{ijk};~~~~~~\Xi_{\bar c_i\bar c_j\bar c_kc_n\bar c_n}={\mathsf
 h}_{ijk}\nonumber\\
\Xi_{c_i\bar c_nc_n\bar c_pc_p}={\mathsf h}^{i};~~~~~~\Xi_{\bar
c_i\bar c_nc_n\bar c_pc_p}={\mathsf h}_{i}\nonumber\\
\Xi_{c_i\bar c_j\bar c_k\bar c_nc_n}={\mathsf
 h}^{i}_{jk}+\frac{1}{4}\left(\delta^i_k{\mathsf h}_j-\delta^i_j{\mathsf h}_k\right)\nonumber\\
\Xi_{\bar c_i c_j c_k c_n\bar c_n}={\mathsf
 h}_{i}^{jk}+\frac{1}{4}\left(\delta_i^k{\mathsf h}^j-\delta_i^j{\mathsf h}^k\right)\nonumber\\
\Xi_{c_ic_jc_kc_l\bar c_m}={\mathsf h}^{ijkl}_m+\frac{1}{2}
\left(\delta^i_m{\mathsf h}^{jkl}-\delta^j_m{\mathsf h}^{ikl}+
\delta^k_m{\mathsf h}^{ijl}-\delta^l_m{\mathsf h}^{ijk} \right)\nonumber\\
\Xi_{c_ic_jc_k\bar c_l\bar c_m}={\mathsf h}^{ijk}_{lm}+\frac{1}{2}
\left(\delta^i_l{\mathsf h}^{jk}_m -\delta^j_l{\mathsf h}^{ik}_m
+\delta^k_l{\mathsf h}^{ij}_m -\delta^i_m{\mathsf h}^{jk}_l
+\delta^j_m{\mathsf h}^{ik}_l -\delta^k_m{\mathsf h}^{ij}_l\right)\nonumber\\
+\frac{1}{12}\left(\delta^i_l\delta^j_m{\mathsf
h}^k-\delta^j_l\delta^i_m{\mathsf h}^k
-\delta^i_l\delta^k_m{\mathsf h}^j+\delta^k_l\delta^i_m{\mathsf
h}^j
+\delta^j_l\delta^k_m{\mathsf h}^i-\delta^k_l\delta^j_m{\mathsf h}^i\right)\nonumber\\
\Xi_{c_ic_j\bar c_k\bar c_l\bar c_m}={\mathsf
h}^{ij}_{klm}+\frac{1}{2} \left(\delta^i_k{\mathsf h}^{j}_{lm}
-\delta^i_l{\mathsf h}^{j}_{km} +\delta^i_m{\mathsf h}^{j}_{kl}
-\delta^j_k{\mathsf h}^{i}_{lm}
+\delta^j_l{\mathsf h}^{i}_{km} -\delta^j_m{\mathsf h}^{i}_{kl}\right)\nonumber\\
+\frac{1}{12}\left(\delta^i_k\delta^j_l{\mathsf
h}_m-\delta^i_k\delta^j_m{\mathsf h}_l
-\delta^i_l\delta^j_k{\mathsf h}_m+\delta^i_l\delta^j_m{\mathsf
h}_l
+\delta^i_m\delta^j_k{\mathsf h}_l-\delta^i_m\delta^j_l{\mathsf h}_k\right)\nonumber\\
\Xi_{c_i\bar c_j\bar c_k\bar c_l\bar c_m}={\mathsf h}^i_{jklm}+
\frac{1}{2} \left(\delta^i_j{\mathsf h}_{klm}-\delta^i_k{\mathsf
h}_{jlm}+
\delta^i_l{\mathsf h}_{jkm}-\delta^i_m{\mathsf h}_{jkl} \right)\nonumber\\
\Xi_{c_ic_jc_kc_lc_m}=\epsilon^{ijklm} {\mathsf h},~~~~~~
\Xi_{\bar c_i\bar c_j\bar c_k\bar c_l\bar c_m}= \epsilon_{ijklm}
\bar {\mathsf h}
\end{eqnarray}
The fields that appear above are not yet properly normalized.
 To normalize the fields we carry out a field redefinition
 so that
\begin{eqnarray}
{\mathsf h}=\frac{2}{\sqrt {15}}{\mathsf H},~~~\overline{{\mathsf
h}}=\frac{2}{\sqrt {15}}\overline{{\mathsf H}},~~~ {\mathsf
h}^i=4\sqrt{\frac{ 2}{5}}{\mathsf H}^i,~~~{\mathsf
h}_i=4\sqrt{\frac{ 2}{5}}{\mathsf H}_i\nonumber\\
 {\mathsf
h}^{ijk}=\sqrt{\frac{ 2}{15}} \epsilon^{ijklm}{\mathsf
H}_{lm},~~~{\mathsf h}_{ijk}=\sqrt{\frac{ 2}{15}}
\epsilon_{ijklm}{\mathsf H}^{lm}\nonumber\\
 {\mathsf h}^i_{jklm}=
\sqrt{\frac{2}{15}}\epsilon_{jklmn}{\mathsf
H}^{ni}_{(S)},~~~{\mathsf h}_i^{jklm}=
\sqrt{\frac{2}{15}}\epsilon^{jklmn}{\mathsf
H}_{ni}^{(S)}\nonumber\\
 {\mathsf h}^i_{jk}=2\sqrt{\frac{2}{15}} {\mathsf
 H}^i_{jk},~~~{\mathsf h}_i^{jk}=2\sqrt{\frac{2}{15}} {\mathsf
 H}_i^{jk}\nonumber\\
  {\mathsf h}^{ijk}_{lm}=\frac{2}{\sqrt {15}} {\mathsf
  H}^{ijk}_{lm},~~~{\mathsf h}_{ijk}^{lm}=\frac{2}{\sqrt {15}} {\mathsf H}_{ijk}^{lm}
 \end{eqnarray}
The  kinetic energy for the $252$ plet field
$-\partial_{A}\Xi_{\mu\nu\lambda\rho\sigma}
\partial^{A}\Xi_{\mu\nu\lambda\rho\sigma}^{\dagger}$
in terms of the normalized fields is then given by
 \begin{eqnarray}
L_{kin}^{252-Higgs}=-\partial_{A} \overline{{\mathsf
H}}\partial^{A} \overline{{\mathsf H}}^{\dagger}-\partial_{A}
{\mathsf H}\partial^{A} {\mathsf H}^{\dagger}
 - \partial_{A} {\mathsf H}_{i}\partial^{A} {\mathsf
 H}^{\dagger}_i
 - \partial_{A} {\mathsf H}^{i}\partial^{A} {\mathsf
 H}^{i\dagger}\nonumber\\
- \frac{1}{2!}\partial_{A} {\mathsf H}^{ij}
\partial^{A} {\mathsf H}^{ij\dagger}
- \frac{1}{2!}\partial_{A} {\mathsf H}_{ij}
\partial^{A} {\mathsf H}_{ij}^{\dagger}
- \frac{1}{2!}\partial_{A} {\mathsf H}_{ij}^{(S)}
\partial^{A} {\mathsf H}^{(S)\dagger}_{ij}\nonumber\\
- \frac{1}{2!}\partial_{A} {\mathsf H}^{ij}_{(S)}
\partial^{A} {\mathsf H}^{ij\dagger}_{(S)}
-\frac{1}{2!}\partial_{A} {\mathsf H}^{jk}_i
\partial^{A} {\mathsf H}_{i}^{jk\dagger}
-\frac{1}{2!}\partial_{A} {\mathsf H}_{jk}^i
\partial^{A} {\mathsf H}_{jk}^{i\dagger}\nonumber\\
-\frac{1}{3!2!}\partial_{A} {\mathsf H}_{ijk}^{lm}
\partial^{A} {\mathsf H}^{lm\dagger}_{ijk}
-\frac{1}{3!2!}\partial_{A} {\mathsf H}^{ijk}_{lm}
\partial^{A} {\mathsf H}^{ijk\dagger}_{lm}
\end{eqnarray}
where $\overline{{\mathsf H}}({\mathsf H}), {\mathsf H}_{i},
{\mathsf H}^{i}, {\mathsf H}^{ij}, {\mathsf H}_{ij}, {\mathsf
H}_{ij}^{(S)}, {\mathsf H}^{ij}_{(S)}, {\mathsf H}^{jk}_i,
{\mathsf H}_{jk}^i, {\mathsf H}_{ijk}^{lm}, {\mathsf
H}^{ijk}_{lm}$ are the $1, \bar{5}, 5, 10, \overline{10},\\
\overline{15}, 15, 45, \overline{45}, \overline{50}, 50$ plet
representations of $SU(5)$.

\section{Appendix E: A Simplification of Quartic Couplings}
The quartic couplings  discussed in Secs.(5-6) are obtained by integrating out
the intermediate fields which belong to the set of tensor representations
1, 45, 210, 10, 120, $126+\overline{126}$.
The analysis given in the paper is quite general allowing for an arbitrary number of
such intermediate tensor set. The results, however, can be simplified if one assumes
just a single tensor field for each  term in the set listed above. In this case
the couplings show a factorization.
 This case can be gotten from
the analysis of the paper by the following simple algorithm of
replacements

\begin{eqnarray}
\bar{\lambda}_{\acute{a}\acute{b},\acute{c}\acute{d}}^{^{(.)}}~~\rightarrow~~-\frac{1}{4}\frac{\bar{h}_{\acute{a}\acute{b}}
^{^{(.)}}\bar{h}_{\acute{c}\acute{d}}^{^{(.)}}}{{\cal M}};~~~~
\lambda_{\acute{a}\acute{b},\acute{c}\acute{d}}^{^{(.)}}~~\rightarrow~~-\frac{1}{4}\frac{h_{\acute{a}\acute{b}}^{^{(.)}}
h_{\acute{c}\acute{d}}^{^{(.)}}}{{\cal M}}\nonumber\\
\zeta_{\acute{a}\acute{b},\acute{c}\acute{d}}^{^{(.)}}~~\rightarrow~~\frac{1}{2}\frac{f_{\acute{a}\acute{b}}
^{^{(.)}}\bar{h}_{\acute{c}\acute{d}}^{^{(.)}}}{{\cal M}};~~~~
\xi_{\acute{a}\acute{b},\acute{c}\acute{d}}^{^{(.)}}~~\rightarrow~~\frac{1}{2}\frac{f_{\acute{a}\acute{b}}
^{^{(.)}}h_{\acute{c}\acute{d}}^{^{(.)}}}{{\cal M}}\nonumber\\
\theta_{\acute{a}\acute{b},\acute{c}\acute{d}}^{^{(.)}}~~\rightarrow~~\frac{1}{2}\frac{h_{\acute{a}\acute{b}}
^{^{(.)}}\bar{h}_{\acute{c}\acute{d}}^{^{(.)}}}{{\cal M}};~~~~
\kappa_{\acute{a}\acute{b},\acute{c}\acute{d}}^{^{(.)}}~~\rightarrow~~-\frac{h_{\acute{a}\acute{b}}
^{^{(.)}}\bar{h}_{\acute{c}\acute{d}}^{^{(.)}}}{{\cal
M}}\nonumber\\
\varrho_{\acute{a}\acute{b},\acute{c}\acute{d}}^{^{(.)}}~~\rightarrow~~-\frac{f_{\acute{a}\acute{b}}
^{^{(.)}}\bar{h}_{\acute{c}\acute{d}}^{^{(.)}}}{{\cal M}};~~~~
\varsigma_{\acute{a}\acute{b},\acute{c}\acute{d}}^{^{(.)}}
~~\rightarrow~~-\frac{h_{\acute{a}\acute{b}}^{^{(.)}}\bar{f}_{\acute{c}\acute{d}}^{^{(.)}}}{{\cal
M}}
\end{eqnarray}

\section{Appendix F:  The Technique to Evaluate SO(10) Vector-Spinor Couplings }

In this Appendix we illustrate   the technique to
evaluate SO(10) vector-spinor couplings. For that purpose,  we
choose a simple example of the matrix element 
$<\Upsilon^{*}_{(+)\mu}|B|\Upsilon_{(+)\mu}>$. Using Eqs. (5), (9)
and (10) one can write
\begin{eqnarray}
<\Upsilon^{*}_{(+)\mu}|B|\Upsilon_{(+)\mu}>=-i{\bf Q}_{\mu}{\bf
P}_{\mu}<0|b_5b_4b_3b_2b_1\prod_{s=1}^5
(b_s-b_s^{\dagger})|0>\nonumber\\
-\frac{i}{24}\epsilon^{ijklm}{\bf Q}_{ij\mu}{\bf P}_{\mu}^{pq}
<0|b_mb_lb_k\prod_{s=1}^5
(b_s-b_s^{\dagger})b_p^{\dagger}b_q^{\dagger}|0>\nonumber\\
-\frac{i}{24}\epsilon^{ijklm}{\bf Q}_{\mu}^r{\bf P}_{i\mu}^{pq}
<0|b_r\prod_{s=1}^5
(b_s-b_s^{\dagger})b_j^{\dagger}b_k^{\dagger}b_l^{\dagger}b_m^{\dagger}|0>
\end{eqnarray}
Simplifying we get,
\begin{equation}
<\Upsilon^{*}_{(+)\mu}|B|\Upsilon_{(+)\mu}>=i\left[{\bf
Q}_{\mu}{\bf P}_{\mu}-\frac{1}{2}{\bf Q}_{ij\mu}{\bf
P}_{\mu}^{ij}+{\bf Q}_{\mu}^i{\bf P}_{i\mu}^{pq}\right]
\end{equation}
Using  the  Basic Theorem we can expand the terms in Eq. (208) as
\begin{eqnarray}
{\bf Q}_{\mu}{\bf P}_{\mu}={\bf Q}_{ c_i}{\bf P}_{\bar c_i}+{\bf
Q}_{\bar c_i}{\bf P}_{ c_i}\nonumber\\
{\bf Q}_{ij\mu}{\bf P}_{\mu}^{ij}={\bf Q}_{ijc_k}{\bf P}_{\bar
c_k}^{ij}+{\bf
Q}_{ij\bar c_k}{\bf P}_{ c_k}^{ij}\nonumber\\
{\bf Q}_{\mu}^i{\bf P}_{i\mu}={\bf Q}_{ c_j}^i{\bf P}_{i\bar
c_j}+{\bf Q}_{\bar c_j}^i{\bf P}_{i c_j}
\end{eqnarray}
Further, 
using Eq.(13) directly or the  third equation in Eq.(203)
\begin{equation}
{\bf Q}_{\mu}{\bf P}_{\mu}={\bf Q}^i{\bf P}_{i}+{\bf Q}_{i}{\bf
P}^i
\end{equation}
\begin{equation}
{\bf Q}_{ijc_k}={\bf S}^k_{[ij]}={\bf
Q}^k_{ij}+\frac{1}{4}\left({\delta^k_j\widehat {\bf
Q}}_i-{\delta^k_{i}\widehat  {\bf Q}}_j\right)={\bf
Q}^k_{ij}+\frac{1}{4}\left({\delta^k_i{\bf Q}}_j-{\delta^k_{j}
{\bf Q}}_i\right)
\end{equation}
where in the last step we have used Eq.(16).
Similarly,
\begin{equation}
{\bf P}_{\bar c_k}^{ij}={\bf
P}_k^{ij}+\frac{1}{4}\left({\delta_k^i{\bf P}}^j-{\delta_k^{j}
{\bf P}}^i\right)
\end{equation}
Thus we have 
\begin{equation}
{\bf Q}_{ijc_k}{\bf P}_{\bar c_k}^{ij}={\bf Q}^k_{ij}{\bf
P}_k^{ij}+\frac{1}{2}{\bf Q}_i{\bf P}^i
\end{equation}
where we have used the fact that ${\bf Q}^k_{ij}$ and ${\bf
P}_k^{ij}$ are  traceless tensors.
 Again using Eq.(13) directly or the 
fourth equation in Eq.(193) we can write
\begin{eqnarray}
{\bf Q}_{ij\bar c_k}={\bf S}_{[ij]k}=\epsilon_{ijlmn}{\bf
S}^{'lmn}_k=\epsilon_{ijlmn}\left[{\bf
Q}^{lmn}_k+\frac{1}{3}\left(\delta^n_k{\widehat{\bf
Q}}^{lm}-\delta^m_k{\widehat{\bf Q}}^{ln}+\delta^l_k{\widehat{\bf
Q}}^{mn}\right)\right]\nonumber\\
=\epsilon_{ijlmn}{\bf Q}^{lmn}_k+\epsilon_{ijklm}{\widehat{\bf
Q}}^{lm}=\epsilon_{ijlmn}{\bf
Q}^{lmn}_k-\frac{1}{6}\epsilon_{ijklm}{{\bf Q}}^{lm}
\end{eqnarray}
where again in the last step we have used Eq. (16). Similarly,
\begin{equation}
{\bf P}_{ c_k}^{ij}=\epsilon^{ijlmn}{\bf
P}_{lmn}^k-\frac{1}{6}\epsilon^{ijklm}{{\bf P}}_{lm}
\end{equation}
Computing the product ${\bf Q}_{ij\bar c_k}{\bf P}_{ c_k}^{ij}$ we
get
\begin{equation}
{\bf Q}_{ij\bar c_k}{\bf P}_{ c_k}^{ij}=12{\bf Q}^{lmn}_k{\bf
P}_{lmn}^k+\frac{1}{3}{{\bf Q}}^{lm}{{\bf P}}_{lm}
\end{equation}
Here we have used the results:
\begin{eqnarray}
\epsilon_{ijlmn}\epsilon^{ijkpq}{\bf Q}^{lmn}_k{\bf
P}_{pq}=0=\epsilon_{ijklm}\epsilon^{ijpqr}{\bf Q}^{lm}{\bf
P}_{pqr}^k\nonumber\\
\epsilon_{ijklm}\epsilon^{ijpqr}{\bf Q}^{lmn}_k{\bf
P}_{pqr}^k=12{\bf Q}^{lmn}_k{\bf P}_{lmn}^k\nonumber\\
\epsilon_{ijklm}\epsilon^{ijkpq}{\bf Q}^{lm}{\bf P}_{pq}=12{\bf
Q}^{lm}{\bf P}_{lm}
\end{eqnarray}
The first of these equations in Eq. (207) follow from the
tracelessness of ${\bf Q}^{lmn}_k$ and ${\bf P}_{ijk}^l$.
Further, on using Eq.(13)
\begin{eqnarray}
{\bf Q}_{ c_j}^i={\bf S}^{ij}=\frac{1}{2}\left({\bf S}^{[ij]}+{\bf
S}^{\{ij\}}\right)\nonumber\\
=\frac{1}{2}\left({\bf Q}^{ij}+{\bf
Q}^{ij}_{(S)}\right)
\end{eqnarray}
 and similarly,
\begin{equation}
{\bf P}_{i\bar c_j}=\frac{1}{2}\left({\bf P}_{ij}+{\bf
P}_{ij}^{(S)}\right)
\end{equation}
which gives,
\begin{equation}
{\bf Q}_{ c_j}^i{\bf P}_{i\bar c_j}=\frac{1}{4}{\bf Q}^{ij}{\bf
P}_{ij}+\frac{1}{4}{\bf Q}^{ij}_{(S)}{\bf P}_{ij}^{(S)}
\end{equation}
Note that the cross terms do no couple in Eq.(220) as one is
antisymmetric and the other is a symmetric tensor in the exchange
of indices $i$ and $j$.
Finally, on using Eq. (13) once again
\begin{eqnarray}
{\bf Q}_{\bar c_j}^i{\bf P}_{i c_j}={\bf S}^{i}_j{\bf
R}^{j}_i=\left({\bf Q}^{i}_j+\frac{1}{5}\delta^i_j{\widehat {\bf
Q}}\right)\left({\bf P}^{j}_i+\frac{1}{5}\delta^j_i{\widehat
{\bf P}}\right)\nonumber\\
={\bf Q}^{i}_j{\bf P}^{j}_i
\end{eqnarray}
In the last step we have used Eq.(16).
Substituting Eqs. (210), (213), (216), (220) and (221) in Eq.(208)
we get,
\begin{eqnarray}
<\Upsilon^{*}_{(+)\mu}|B|\Upsilon_{(+)\mu}>=\frac{3i}{4}{\bf
Q}^i{\bf P}_{i}+i{\bf Q}_{i}{\bf P}^i\nonumber\\
+\frac{i}{12}{\bf Q}^{ij}{\bf P}_{ij}+\frac{i}{4}{\bf
Q}^{ij}_{(S)}{\bf P}_{ij}^{(S)}\nonumber\\
+i{\bf Q}^{i}_j{\bf P}_{i}^j-6i{\bf Q}^{ijk}_l{\bf
P}_{ijk}^l\nonumber\\
-\frac{i}{2}{\bf Q}^{ij}_k{\bf P}_{k}^{ij}
\end{eqnarray}
One can now use normalized fields exhibited in Eqs. (19) and (20).


\end{document}